\documentclass[aps,prd,twocolumn,amsmath,amssymb,nofootinbib]{revtex4-2}

\usepackage{amsmath,amssymb}
\usepackage{graphicx}
\usepackage[hidelinks]{hyperref}
\usepackage{xcolor}
\newcommand{\revblue}[1]{#1}
\colorlet{BLUE}{blue}
\newcommand{\rev}[1]{#1}
\newcommand{\revgreen}[1]{#1}
\usepackage{tikz}
\usetikzlibrary{arrows.meta,positioning,decorations.pathmorphing,decorations.markings}

\newcommand{\Tr}{\mathrm{Tr}}

\newcommand{\e}{\mathrm{e}}
\newcommand{\dd}{\mathrm{d}}
\newcommand{\ii}{\mathrm{i}}

\newcommand{\slashD}{\slashed{D}}
\newcommand{\slashed}[1]{\not\!#1}

\begin{document}

\title{Minimum Virtual Proper Time and Finite Mass--Charge Matching in QED}

\author{Mustafa Bakr}
\email{mustafa.bakr@physics.ox.ac.uk}
\affiliation{Clarendon Laboratory, Department of Physics, University of Oxford, Oxford OX1 3PU, UK}
\affiliation{Centre for Quantum Technologies, National University of Singapore, Singapore}

\begin{abstract}
We formulate four-dimensional finite-proper-time QED from a gauge-covariant source functional with a retained spectral lower boundary $s_0=\Lambda^{-2}>0$ and finite physical duration $\mathcal T$. Physically, the basic $s_0$ operation is a truncated Laplace transform: the arbitrarily small-proper-time part of the spectral integral, which carries the ultraviolet singularity, is excluded without imposing a cutoff on momentum space, while finite $\mathcal T$ expresses that a physical observable is prepared and detected during a finite experiment rather than by an infinite-time measurement. The retained $s_0$ belongs to the defining source construction and controls the correlated finite corrections after mass and charge matching. The charged interval and circle sectors are generated by the same covariant heat semigroup, while the photon covariance carries the same retained lower bound at the source level. At every fixed perturbative order, gauge covariance gives the Ward--Takahashi hierarchy and an assembled graph-level quadratic-form bound establishes ultraviolet finiteness. The physical electron, positron, and transverse-photon poles retain their canonical residues, and \revblue{the fixed-order cutting relation uses the matched daughter domains on the stated regulated continuation domain.} Mass and charge renormalisation are replaced by finite matching, while independent observables retain correlated $s_0$ dependence. \revblue{At fixed $s_0>0$, the Euclidean source has a nonperturbative definition. Its physical standing-wave response is obtained order by order from the continued FPT amplitudes with the same finite-duration channel data. An exact all-coupling physical boundary realisation requires additional channel-norm control.} The limits $s_0\to0^+$, $\mathcal T\to\infty$, and exact all-coupling free-space exhaustion are separate limiting questions and are not used to define the theory.
\end{abstract}

\maketitle

\section{Introduction}
\label{sec:intro}

Renormalised quantum electrodynamics is one of the most successful physical theories ever constructed. Its agreement with precision measurements, such as the anomalous magnetic moment of the electron, is extraordinary. Nevertheless, the perturbative formulation contains ultraviolet (UV) divergences which are removed by renormalisation. Historically, this was viewed in two rather different ways.

In the Feynman--Dyson programme the divergences are intermediate artifacts in an $S$-matrix calculation and disappear after expressing the bare parameters in terms of measured quantities~\cite{Feynman1949,Dyson1949}. In Schwinger's proper-time formulation, building on Fock's earlier representation~\cite{Fock1937}, the same divergences appear geometrically as the lower-endpoint singularity of the heat-kernel integral~\cite{Schwinger1951}.

In Wilson's later language, the relation between microscopic and measured parameters is a statement about scale dependence and matching rather than a purely formal subtraction~\cite{Wilson1971,Wilson1975,Polchinski1984}.

\revblue{The operator-first viewpoint was motivated from earlier work on singular electromagnetic boundary problems. In singular azimuthal fields and in spherical cavities, the local solutions of the field equations and the states admitted by the full operator domain are not the same set, and the physical spectrum belongs to the latter~\cite{BakrAmariAzimuth2023,BakrZhangAmariSphere2026,BakrAmariWedge2025,BakrAmariAngular2025}. In circuit QED, boundary formulations of readout and of spatial crosstalk relate the measured response to the complete mode structure and to its coupling to external channels~\cite{BakrBoundaryReadout2025,BakrWopalenski2026,AlghadeerCrosstalk2026}; programmable anharmonic phase gates separate the implemented response from its finite-width, post-selected measurement, and path signatures retain the ordering information of a finite readout record~\cite{FontaineAnharmonic2026,CaoPathSignature2025}. In ultrastrong-coupling cavity QED, representation changes, truncations and measured field operators must be treated together or the predictions become gauge dependent~\cite{DiStefanoStassiGauge2019,StassiOutput2016}. The common lesson is that the domain, the complete source and the measurement map are specified first and approximations are made afterwards. None of these results derives the retained proper-time boundary or supplies experimental evidence for it. Here that boundary is a postulate, and a change of representation is allowed only if it leaves the source-defined physical response unchanged.}

The present work develops a finite-proper-time formulation of QED. The central postulate is that every independent complete virtual history has a nonzero lower bound on its Schwinger proper-time modulus,
\begin{equation}
T\geq s_{0}=\Lambda^{-2}.
\label{eq:central_postulate}
\end{equation}
$T$ is the invariant proper-time parameter of the spectral representation. For a positive covariant operator $P$, the ordinary Laplace representation
\begin{equation}
P^{-1}=\int_{0}^{\infty}\dd s\,\e^{-sP}
\end{equation}
is replaced by the truncated transform
\begin{equation}
R_{s_0}(P)=\int_{s_0}^{\infty}\dd s\,\e^{-sP}
=\e^{-s_0P}P^{-1}.
\label{eq:intro_truncated_Laplace}
\end{equation}
The change is spectral rather than a momentum cutoff: all momenta remain integrated over, while only the inadmissibly short part $0<s<s_0$ of the proper-time domain is removed. Because the lower endpoint is imposed on the covariant operator before functional differentiation, the construction introduces no preferred frame and retains gauge covariance. At a physical pole, $\lim_{\lambda\to0}\lambda R_{s_0}(\lambda)=1$, so the pole position and unit residue are also unchanged. 

The physical motivation is therefore direct. In the proper-time representation the ultraviolet singularity is concentrated at the lower endpoint $s=0$. FPT removes that region from the admissible virtual spectrum. The usual UV divergences are consequently interpreted as the result of extending virtual histories to arbitrarily short proper time. This idea is close in mathematical form to the proper-time and heat-kernel representations of charged particles in background fields~\cite{Schwinger1951,Fradkin1966}, but differs in physical status: in the standard use of proper time the lower endpoint is removed after regularisation, whereas here it remains finite and appears in observable matching relations.

The finite amplitudes can of course be expanded at low external invariants in an EFT-like derivative series, but that series is a representation of the finite FPT amplitudes, not their definition. Once the ordinary measured QED parameters are fixed, the endpoint-dependent coefficients of that series are correlated functions of the same retained $s_0$ inherited from the source construction; they are not independent Wilson coefficients assigned separately to successive operators. Promoting them to independent constants would therefore define a different effective theory and would discard the cross-observable content of FPT QED.

This distinction is experimentally testable. After the measured pole mass and low-energy charge are fixed, Sec.~\ref{subsec:nonredundancy} proves that the one-loop Pauli form factor is a continuous, strictly monotonic function of $s_0$ and therefore fixes a unique endpoint within the minimal FPT theory. The same calibrated $s_0$ then enters the momentum-subtraction charge flow and weak-field four-photon coefficient, which are independent predictions rather than additional matching parameters. Thus one endpoint-sensitive observable calibrates the spectral boundary while further observables overconstrain the same one-parameter trajectory. This is stronger than ultraviolet finiteness alone: a regulator may render amplitudes finite while remaining arbitrary, whereas here the retained endpoint is physically calibratable and cross-observable consistency is falsifiable.

The proposal is also related to the Efimov--Krasnikov--Tomboulis and weakly nonlocal field-theory programmes, where entire functions soften high-energy behaviour~\cite{Efimov1967,Efimov1972,Krasnikov1987,Tomboulis1997,ModestoPivaRachwal2016}, to finite nonlocal QED~\cite{Moffat1990,EvensMoffatKleppeWoodard1991}, and to additional-pole and higher-derivative ultraviolet constructions~\cite{LeeWick1969,Slavnov1972}. Short-proper-time suppression has also been connected to a zero-point length through a smooth modification of the proper-time weight rather than a sharp endpoint~\cite{Padmanabhan1997}. A mathematically different finite-QED programme is causal perturbation theory, where singular products are defined by causal splitting and finite normalization rather than by retaining a microscopic proper-time scale~\cite{Scharf1995}. The distinction of FPT QED is therefore not the use of proper time or an entire function by itself. It is that the lower endpoint is imposed on the spectral domain of a complete covariant virtual history before perturbative decomposition, and the propagators, vertices, contact terms, and matching relations are derived from that common construction.

At the operator level, functional differentiation partitions a pre-existing charged interval or circle into non-negative segments without assigning an independent endpoint to each segment. After local interactions are sewn into a graph, the relevant closed virtual momentum histories are identified from the assembled topology. Their primitive supports are the graph circuits. The photon covariance is fixed independently by the same lower endpoint in its own spectral integral, as specified below. Thus operator ancestry determines how an edge is partitioned, while graph topology determines which collections of edge totals form closed virtual circulations.

The open charged line and the closed fermion loop are also not defined by separate finite-proper-time prescriptions. They are the interval and circle sectors of the same covariant heat semigroup generated by $\Delta_A$, as shown in Fig.~\ref{fig:open_closed_topology}. The different boundary conditions produce different spectral functions of the common operator, while an exact mass-derivative identity ties the two sectors together. The propagator, determinant, vertices and contact terms belong to one gauge-covariant construction.

\begin{figure*}[t]
\centering
\resizebox{0.8\textwidth}{!}{%
\begin{tikzpicture}[
>=Latex,
node distance=16mm and 14mm,
every node/.style={font=\small},
box/.style={draw=black, rounded corners, thick, text=black, align=center, inner sep=6pt},
arr/.style={->, thick, draw=black},
darr/.style={->, thick, dashed, draw=black}
]
\node[box, text width=0.23\textwidth] (gen) {Single heat semigroup\\[1mm]
$U_A(T)=\e^{-T\Delta_A}$\\[1mm]
$T\ge s_0$};

\node[box, below left=of gen, text width=0.28\textwidth] (open) {Open sector\\[1mm]
interval with fixed endpoints\\[1mm]
$S_\Lambda[A]=(m-\slashD_A)\displaystyle\int_{s_0}^{\infty}\dd T\,U_A(T)$};

\node[box, below right=of gen, text width=0.28\textwidth] (closed) {Closed sector\\[1mm]
closed loop with cyclic origin removed\\[1mm]
$\Gamma_{1,\Lambda}[A]-\Gamma_{1,\Lambda}[0]=\frac12\displaystyle\int_{s_0}^{\infty}\frac{\dd T}{T}\,\Tr\!\left[U_A(T)-U_0(T)\right]$};

\draw[arr] (gen) -- node[left, text=black] {fixed endpoints} (open);
\draw[arr] (gen) -- node[right, text=black] {trace and divide by $T$} (closed);
\draw[darr] (closed) -- node[below, text=black, align=center] {mass derivative $\partial_m$\\trace identity} (open);
\end{tikzpicture}%
}
\caption{Open and closed charged histories from a single covariant proper-time construction. Both sectors are generated by the same heat semigroup $U_A(T)=\e^{-T\Delta_A}$ with the same retained lower endpoint $s_0$. The open sector is an interval with fixed external endpoints, whereas the closed sector is a circle with the cyclic $\dd T/T$ measure. They are therefore not independent modifications of propagators and loops. The mass derivative removes the cyclic factor and gives the exact open--closed relation derived in Sec.~\ref{sec:definition}.}
\label{fig:open_closed_topology}
\end{figure*}

In standard Pauli--Villars and dimensional regularisation schemes the regulator is removed after subtraction. In weakly nonlocal constructions the high-energy scale may be retained, but the propagator and vertex structure follows from the chosen nonlocal action. In the present formulation, the closed-loop determinant, the open fermion-line propagator, the one-photon vertex, the two-photon contact kernel, and the electron self-energy are all obtained from covariant heat-kernel objects by functional differentiation. The Ward--Takahashi identity then follows from gauge covariance~\cite{BallChiu1980}. The transverse part and contact terms are generated by the same functional derivative.

The distinction between the present formulation and the dressed-action approach merits emphasis. In the standard nonlocal programme~\cite{Efimov1967,Tomboulis1997,ModestoPivaRachwal2016,Moffat1990}, the classical action is modified by dressing the kinetic operator with an entire function. Functional differentiation of that nonlocal kinetic term then generates interaction vertices whose high-momentum behaviour can compensate the propagator damping. Ultraviolet convergence therefore depends on the balance between the dressed propagators and the induced vertices.

In the present work, the theory is defined not by a dressed classical action but by a gauge-covariant heat-kernel generating functional with the physical proper-time endpoint $s_{0}=\Lambda^{-2}$. Photon insertions on a charged worldline are local operations on one parent heat kernel. The ultraviolet estimate must therefore be made after the complete gauge-covariant graph has been assembled.

The resulting connected Euclidean amplitudes are ultraviolet finite at every fixed perturbative order and fixed infrared prescription. After the open-line insertions, contact terms, closed charged sectors, and photon contractions are assembled, the proper-time exponent defines a strictly positive quadratic form in every nonzero simultaneous loop-momentum direction. The resulting Gaussian coercivity controls overlapping as well as overall ultraviolet limits.

Finiteness changes the role of renormalisation. The measured electron mass and electric charge are imposed as physical matching conditions, but no divergent counterterm is required. These conditions determine the microscopic parameters at fixed $s_{0}$. Independent observables therefore retain correlated dependence on the same $s_{0}$. At one loop the anomalous magnetic moment receives a calculable $O(m^{2}/\Lambda^{2})$ correction, while the vacuum polarisation, static exchange, and weak-field photon amplitudes provide independent finite-proper-time observables.

The retained endpoint modifies virtual propagation without introducing additional particle poles. The free propagators retain canonical unit residues and contain no additional poles. More generally, the differentiated open-line kernels can be written as ordinary pole factors multiplied by jointly entire complete-history factors. These factors reduce to unity when the corresponding cut denominators are placed on shell. Generalised amputation therefore removes the endpoint factor from real external particles, while the physical thresholds remain those of the ordinary electron, positron, and photon states.

Lorentzian amplitudes are defined at fixed order by analytic continuation of the Euclidean correlators in the external invariants with $s_{0}$ held fixed. The finite-proper-time factors are entire in the virtual denominators, while the physical discontinuities arise from the ordinary propagator poles. The resulting fixed-order amplitudes therefore obey the standard cutting relations with modified off-shell virtual propagation.

The infrared problem is separate from the short-proper-time boundary. The photon remains massless and its physical pole is unchanged. A real observable is prepared and detected during a finite interval $\mathcal T=t_f-t_i$; it does not require an apparatus to wait for the idealised limit $t_i\to-\infty$, $t_f\to+\infty$. The finite endpoints replace the usual soft factor $1/(p\!\cdot\!k)$ by the bounded difference quotient $[\e^{i(k\cdot u)\mathcal T}-1]/(k\!\cdot\!u)$, so the zero-energy endpoint is integrable for massive charged states. Standard QED soft-photon factorisation then separates this universal finite-duration contribution from an infrared-finite hard remainder~\cite{YennieFrautschiSuura1961,Weinberg1965}. Thus $s_{0}$ controls the lower boundary of virtual proper time, whereas $\mathcal T$ is the finite duration over which a physical preparation and detection are actually defined.

On reflection positivity, the raw finite-$s_{0}$ Euclidean covariance does not satisfy the strict Osterwalder--Schrader condition and is therefore not used as the physical state metric. Physical scattering states are instead identified with the ordinary positive electron, positron, and transverse-photon pole states. \revblue{On the physical channel domain, the continued source amplitudes define a standing-wave response order by order. Its Hermiticity follows from the fixed-order cutting relation, and the reaction representation reproduces those same amplitudes.} Taking the spatial boundary to infinity in a regular way recovers the free-space perturbative amplitudes coefficient by coefficient.

The physical ordering is therefore
\begin{equation}
\boxed{\begin{gathered}
\text{complete virtual histories at }s_0>0\\[-1pt]
+\ \text{finite-duration physical channels at }\mathcal T<\infty
\\[2pt]
\Downarrow\\[-1pt]
\text{Euclidean spectral calculus for virtual amplitudes}
\\[2pt]
\Downarrow\\[-1pt]
\mathcal M_E^{\rm comp}\longrightarrow\mathcal M_{\rm phys}^{\pm}
\longrightarrow\mathcal H_{\rm phys} .
\end{gathered}}
\label{eq:physical_scattering_ordering}
\end{equation}
The Euclidean operator calculus thus computes the virtual histories connecting physical preparation and detection channels. This scattering construction also makes the non-EFT distinction concrete. Expanding \(\mathcal M_{\rm phys}^{\pm}\) at low invariants can produce an infinite local operator series, but the coefficients are derived from the same finite-$s_0$ amplitude. After the ordinary QED parameters and $s_0$ have been fixed, that entire endpoint-dependent tower is predicted. Replacing its correlated coefficients by independent constants would be an EFT reparametrisation of the low-energy amplitudes, not the defining FPT theory.

The physical theory is defined with a retained spectral lower boundary and finite physical duration,
\[
s_0>0,\qquad 0<\mathcal T<\infty,
\]
on a specified self-adjoint physical domain. \revblue{At these values the exact Euclidean source is defined, while the electron, positron and transverse-photon pole measures specify the physical channel test space. The physical reaction representation is derived coefficient by coefficient below.} Its fixed-order perturbative expansion is ultraviolet finite and gives the free-space matching, form-factor, cutting, and scattering quantities calculated below. Its retained endpoint is inherited by the source construction and is calibratable. The formal boundary removal $s_0\to0^+$, the idealised infinite-duration limit $\mathcal T\to\infty$, and taking the spatial boundary to infinity are distinct limiting comparisons; none is used to define the finite-resolution theory. Osterwalder--Schrader reconstruction of the raw Euclidean covariance is likewise not assumed.

\section{Unified Finite-Proper-Time Path-Integral Definition}
\label{sec:definition}
The theory is defined first through spectral operator functions. Proper-time and worldline representations make their geometry explicit but are not an independent set of diagrammatic rules. At one-history level the charged open interval and closed circle retain the same lower spectral endpoint $s_0$. Functional differentiation partitions that pre-existing operator history into non-negative segments without assigning an independent endpoint to each segment. Once interaction vertices are sewn, those segments become edges of an assembled graph and the closed virtual momentum histories are identified by its primitive circuits. The photon sector is fixed by its own spectral covariance below. In particular, the construction is not obtained by damping every fermion propagator segment separately.

The generating functional is
\begin{equation}
Z[J,\bar\eta,\eta]=\mathcal N\!\int\!\dd\mu_C(A)\,\e^{-\Gamma_{1,\Lambda}[A]}
\exp\!\Big(\bar\eta S_\Lambda[A]\eta+J\!\cdot\!A\Big),
\label{eq:main_unified_Z}
\end{equation}
where the gauge-fixed photon measure has covariance $\e^{-s_0k^2}/k^2$, $\Gamma_{1,\Lambda}[A]$ is the vacuum-subtracted closed charged functional, and $S_\Lambda[A]$ is the open charged kernel. Correlation functions are obtained by functional differentiation and Gaussian contraction. The open and closed sectors are not the inverse and determinant of one modified Grassmann covariance. They are two heat-kernel objects associated with the same covariant second-order operator, with interval and circle worldline boundary conditions respectively.

For a real Euclidean background let
\begin{equation}
\Delta_A=(m-\slashD_A)(m+\slashD_A)
=-D_A^2+m^2+\frac e2\sigma^{\mu\nu}F_{\mu\nu}.
\end{equation}
The two charged sectors are
\begin{align}
S_\Lambda[A]
&=(m-\slashD_A)\int_{s_0}^{\infty}\dd T\,\e^{-T\Delta_A},\nonumber\\
\Gamma_{1,\Lambda}[A]-\Gamma_{1,\Lambda}[0]
&=\frac12\int_{s_0}^{\infty}\frac{\dd T}{T}\,
\Tr\!\left(\e^{-T\Delta_A}-\e^{-T\Delta_0}\right).
\end{align}
Although these are different functions of $\Delta_A$, they are locked by the exact identity
\begin{equation}
\frac{\partial}{\partial m}\Big(\Gamma_{1,\Lambda}[A]-\Gamma_{1,\Lambda}[0]\Big)
=-\Tr\Big(S_\Lambda[A]-S_\Lambda[0]\Big).
\label{eq:main_mass_lock}
\end{equation}
Differentiating the heat kernel brings down $-2mT$, which cancels the cyclic $1/T$ measure of the closed loop, while the Dirac-odd trace vanishes. Closing an open charged history therefore reproduces the mass derivative of the closed functional. 

Gauge covariance is preserved for every fixed $s_0$. Under $A_\mu\to A_\mu+\partial_\mu\chi$,
\begin{align}
S_\Lambda[A+\partial\chi](x,y)
&=\e^{\ii e[\chi(x)-\chi(y)]}S_\Lambda[A](x,y),\nonumber\\
\Gamma_{1,\Lambda}[A+\partial\chi]
&=\Gamma_{1,\Lambda}[A].
\end{align}
One functional derivative gives
\revblue{\begin{equation}
\partial_\mu^z\frac{\delta S_\Lambda(x,y)}{\delta A_\mu(z)}
=-\ii e\,[\delta(z-x)-\delta(z-y)]S_\Lambda(x,y),
\label{eq:main_kernel_WT}
\end{equation}}
and repeated differentiation gives the complete Ward--Takahashi hierarchy on open lines. Gauge invariance of the closed functional gives transversality of every closed-loop photon function. These statements hold before any diagram is isolated.

The lower endpoint also fixes the internal photon covariance. For a positive Euclidean kinetic operator $P$,
\begin{equation}
\int_{s_0}^{\infty}\dd s\,\e^{-sP}=\e^{-s_0P}P^{-1},
\end{equation}
so in the unit covariant gauge
\begin{equation}
D_{\mu\nu,\Lambda}(k)=\delta_{\mu\nu}\frac{\e^{-s_0k_E^2}}{k_E^2}.
\label{eq:main_photon_covariance}
\end{equation}
The Gaussian factor is a consequence of the retained proper-time endpoint.

The operator and graph statements must be kept distinct. Insertions on an open charged kernel partition one pre-existing interval, so the daughter parameters are non-negative and only their inherited total is constrained at that stage. After sewing, an edge total may participate in one or more primitive graph circuits, and it is those closed circulations that organise the graph-level ultraviolet and cutting analysis. The photon covariance contributes its own parameter $\sigma\ge s_0$ because Eq.~\eqref{eq:main_photon_covariance} is part of the source functional. After external charged legs are amputated on shell, their operator-history constraint disappears; no independent endpoint is assigned to the remaining charged segment merely because it is an internal propagator. Figure~\ref{fig:operator_to_circuit} summarises this order of operations.

\begin{figure*}[t]
\centering
\resizebox{0.97\textwidth}{!}{%
\begin{tikzpicture}[
>=Latex,
every node/.style={font=\small},
fermion/.style={line width=1.45pt},
photon/.style={decorate,decoration={snake,amplitude=1.2pt,segment length=5.5pt},line width=0.9pt},
block/.style={draw,rounded corners,thick,minimum width=1.55cm,minimum height=0.9cm,align=center}
]
\node[font=\bfseries] at (0,1.55) {(a) Operator insertion};
\draw[fermion] (-3.2,0.35)--(3.2,0.35);\draw[->,line width=1.0pt] (-0.35,0.35)--(0.35,0.35);
\fill (-1.05,0.35) circle (2.2pt);
\fill (1.05,0.35) circle (2.2pt);
\node[below=7pt] at (-2.15,0.35) {$u_0\ge0$};
\node[below=7pt] at (0,0.35) {$u_1\ge0$};
\node[below=7pt] at (2.15,0.35) {$u_2\ge0$};
\draw[thick] (-3.2,-0.62)--(3.2,-0.62);\draw[thick] (-3.2,-0.72)--(-3.2,-0.52);\draw[thick] (3.2,-0.72)--(3.2,-0.52);
\node[below=8pt] at (0,-0.62) {$T=u_0+u_1+u_2\ge s_0$};

\draw[->,very thick] (3.7,0.2)--(5.0,0.2);
\node[above] at (4.35,0.2) {sewing};

\node[font=\bfseries] at (8.25,1.78) {(b) One-loop QED circuit};
\coordinate (vL) at (6.4,0.3);
\coordinate (vR) at (10.1,0.3);
\draw[fermion,->] (5.35,0.3)--(vL);
\draw[fermion,->] (vL)--(vR);
\draw[fermion,->] (vR)--(11.15,0.3);
\draw[photon] (vL) to[bend left=52] (vR);
\fill (vL) circle (2.3pt); \fill (vR) circle (2.3pt);
\node[below=8pt] at (8.25,0.3) {$t\ge0$};
\node[anchor=west] at (9.35,1.30) {$\sigma\ge s_0$};
\node at (8.25,-0.92) {$t+\sigma\ge s_0$};

\node[font=\bfseries] at (14.8,1.78) {(c) Self-energy insertion};
\coordinate (cL) at (12.35,0.3);
\coordinate (c1) at (14.05,0.3);
\coordinate (c2) at (15.55,0.3);
\coordinate (cR) at (17.25,0.3);
\draw[fermion,->] (cL)--(c1);
\draw[fermion,->] (c2)--(cR);
\draw[photon] (cL) to[bend left=42] (cR);
\coordinate (SigmaW) at (14.05,0.3);
\coordinate (SigmaE) at (15.55,0.3);
\draw[thick] (14.8,0.3) ellipse[x radius=0.75cm,y radius=0.42cm];
\node at (14.8,0.3) {$\Sigma_{\rm 1PI}$};
\draw[fermion] (c1)--(SigmaW);
\draw[fermion] (SigmaE)--(c2);
\fill (cL) circle (2.2pt); \fill (cR) circle (2.2pt);
\draw[dashed,thick,rounded corners] (14.00,-0.13) rectangle (15.60,0.73);
\end{tikzpicture}%
}
\caption{From operator differentiation to the graph-level finite-proper-time domain. (a) Functional differentiation partitions one charged spectral history into non-negative daughter intervals with the inherited condition $T=u_0+u_1+u_2\ge s_0$; no daughter interval receives an independent endpoint. (b) After sewing, the one-loop electron self-energy contains an internal charged segment $t\ge0$ and a photon parameter $\sigma\ge s_0$ from the photon covariance, so the closed circulation obeys $t+\sigma\ge s_0$. The external charged-history constraint disappears under on-shell amputation. (c) A conventional 1PI self-energy insertion, denoted by the blob $\Sigma_{\rm 1PI}$, embedded in a larger closed circuit. \revblue{If a non-redundant charged parent circle crosses the insertion boundary, its proper-time variables remain correlated with the surrounding circle. The insertion cannot then be replaced by a parent-independent momentum-only dressed propagator before that shared domain is retained.}}
\label{fig:operator_to_circuit}
\end{figure*}

The operator identities, finite-volume source relations, divided-difference calculus, explicit derivatives through fourth order, and the detailed sewing criterion used later are collected in Appendix~\ref{sec:formulations}. 

\section{Closed Charged Histories: Gauge-Covariant Determinant}
\label{sec:closed}
Throughout, the closed-loop functional denotes the vacuum-subtracted trace, $\Gamma_{1,\Lambda}[A]-\Gamma_{1,\Lambda}[0]$, in finite spacetime volume. The field-independent term is otherwise infinite through the volume factor even at finite $s_0$. With the Euclidean conventions introduced in Sec.~\ref{sec:definition}, define $F_{\mu\nu}=\partial_\mu A_\nu-\partial_\nu A_\mu$ and $\sigma^{\mu\nu}=\frac{\ii}{2}[\gamma^\mu,\gamma^\nu]$. The squared Dirac operator is
\begin{equation}
\Delta_A=(m-\slashD_A)(m+\slashD_A)
=-D^2+m^2+\frac{e}{2}\sigma^{\mu\nu}F_{\mu\nu}.
\label{eq:DeltaA}
\end{equation}
For $A=0$, $\Delta_0=p_E^2+m^2$ in momentum space, where $p_E$ denotes Euclidean momentum. The one-loop effective action is the vacuum-subtracted proper-time trace defined in Sec.~\ref{sec:definition}, with its integration starting at $s_0>0$; no $s_0\to0$ limit is taken.

The finite endpoint does not disturb gauge invariance. Under $A_\mu\to A_\mu+\partial_\mu\chi$, the covariant derivative transforms as $D_\mu\to U D_\mu U^{-1}$ with $U(x)=\e^{\ii e\chi(x)}$, and consequently $\Delta_A\to U\Delta_AU^{-1}$. The heat kernel undergoes the same similarity transformation, while cyclicity of the functional trace makes $\Tr\,\e^{-s\Delta_A}$ invariant at each fixed $s$. Restricting the final integration to $s\ge s_0$ therefore leaves $\Gamma_{1,\Lambda}[A]$ gauge invariant.

Gauge invariance gives transversality of the vacuum polarisation tensor, $\Pi_{\mu\nu}(q)=(q^2\delta_{\mu\nu}-q_\mu q_\nu)\Pi(q^2)$. In terms of the fine-structure constant $\alpha$, and writing $q_E$ for the Euclidean external photon momentum, the scalar polarisation is
\begin{equation}
\Pi(q_E^2)
=-\frac{2\alpha}{\pi}\int_0^1\dd x\,x(1-x)
\,\Gamma\!\left(0,\frac{m^2+x(1-x)q_E^2}{\Lambda^2}\right),
\label{eq:Pi_gamma}
\end{equation}
where $\Gamma(0,z)=\int_z^\infty \dd t\,\e^{-t}/t$. Since $\Gamma(0,z)$ is positive and monotonically decreasing for $z>0$, and $m^2+x(1-x)q_E^2\ge m^2$ for Euclidean $q_E^2\ge0$, the magnitude of the integral is maximised at $q_E^2=0$.
\begin{equation}
|\Pi(q_E^2)|\le -\Pi(0)
=\frac{\alpha}{3\pi}\Gamma\!\left(0,\frac{m^2}{\Lambda^2}\right).
\label{eq:Pi_bound}
\end{equation}

For $q_E^2,m^2\ll\Lambda^2$, $\Gamma(0,z)\simeq-\gamma_E-\ln z$, where $\gamma_E$ is the Euler--Mascheroni constant. This reproduces the standard renormalised QED vacuum polarisation up to a finite matching constant and corrections of order $\max(q_E^2,m^2)/\Lambda^2$.

For $q_E^2\gg \Lambda^2$, the integral is dominated by the endpoint regions $x\simeq0$ and $x\simeq1$. Near $x=0$, putting $y=q_E^2x/\Lambda^2$ and using $\int_0^\infty\dd y\,y\Gamma(0,y)=\frac12$ gives
\begin{equation}
\Pi(q_E^2)\sim -\frac{2\alpha}{\pi}\frac{\Lambda^4}{q_E^4}
\left[1+O\!\left(\frac{m^2}{\Lambda^2}\right)\right],
\qquad q_E^2\gg\Lambda^2.
\label{eq:Pi_asymptotic}
\end{equation}
The leading high-energy tail is polynomial and contains no leading factor of $\ln(\Lambda^2/m^2)$.

We subtract at zero momentum, $\Pi_R(q_E^2)=\Pi(q_E^2)-\Pi(0)$, so that $\Pi_R(0)=0$ and the measured low-energy charge is the input parameter. We denote the corresponding Thomson-limit fine-structure constant by $\alpha_{\rm phys}$. With this convention the Euclidean exchange strength is
\begin{equation}
\alpha_{\rm exch}(q_E^2)=\frac{\alpha_{\rm phys}}{e^{q_E^2/\Lambda^2}-\Pi_R(q_E^2)}.
\label{eq:alpha_eff_ren}
\end{equation}
Because $|\Pi_R(q_E^2)|\le|\Pi(0)|$, the displayed one-loop denominator has no spacelike zero whenever $|\Pi(0)|<1$. At $q_E^2\gg\Lambda^2$ the direct endpoint factor dominates and $\alpha_{\rm exch}$ is exponentially suppressed. This is a one-loop statement~\cite{Landau1955}.

\section{Open Charged Histories and the Derived Vertex}
\label{sec:open}
The open fermion line must be defined by a gauge-covariant kernel, not by damping a free propagator in isolation. We write
\begin{equation}
S_\Lambda[A](x,y)=
(m-\slashD_x)
\int_{s_0}^{\infty}\dd T\,\e^{-m^2T}\,
\langle x|\e^{-T H_A}|y\rangle,
\label{eq:open_kernel}
\end{equation}
where $H_A=-D^2+\frac{e}{2}\sigma^{\mu\nu}F_{\mu\nu}$. The prefactor $m-\slashD_x$ converts the heat kernel of the squared Dirac operator into the Dirac propagator. At $A=0$, writing $N(p)=m-\ii\slashed p$, $a_p=p^2+m^2$, and $R(a)=\e^{-s_0a}/a$,
\begin{multline}
S_\Lambda(p)=\frac{m-\ii\slashed p}{p^2+m^2}\,\e^{-s_0(p^2+m^2)},
\qquad \\
S_\Lambda^{-1}(p)=\e^{s_0(p^2+m^2)}(m+\ii\slashed p).
\label{eq:S_inverse}
\end{multline}

The open kernel transforms covariantly according to
$S_\Lambda[A^\chi](x,y)=\e^{\ii e\chi(x)}S_\Lambda[A](x,y)\e^{-\ii e\chi(y)}$. Differentiation first with respect to $\chi$ and then with respect to $A_\mu$ gives the finite-$\Lambda$ Ward identity. For the one-photon form, define
$G_\Lambda^\mu(p',p)=\delta S_\Lambda[A](p',p)/\delta A_\mu(q)|_{A=0}$ with $q=p'-p$. Applying
$\delta\e^{-T H_A}=-\int_0^T\dd \tau\,\e^{-(T-\tau)H_A}(\delta H_A)\e^{-\tau H_A}$ shows that the one-photon heat-kernel insertion at $A=0$ is
\begin{equation}
V_1^\mu(p',p)=e\left[(p'+p)^\mu-\ii\sigma^{\nu\mu}(p'-p)_\nu\right],
\label{eq:V1}
\end{equation}
and the prefactor contribution from differentiating $m-\slashD_x$ is $E_1^\mu=\ii e\gamma^\mu$. The first divided-difference kernel is
\begin{equation}
I(a',a)=\int_{s_0}^{\infty}\dd T\int_0^T\dd\tau\,
\e^{-(T-\tau)a'}\e^{-\tau a}
=\frac{R(a)-R(a')}{a'-a}.
\label{eq:I_def}
\end{equation}
The unamputated one-photon kernel is therefore
\begin{equation}
G_\Lambda^\mu(p',p)=
E_1^\mu R(a_p)
+N(p')V_1^\mu(p',p)I(a_{p'},a_p).
\label{eq:G_mu_explicit}
\end{equation}
With $\sigma^{\mu\nu}=\tfrac{\ii}{2}[\gamma^\mu,\gamma^\nu]$, the spin term in $V_1$ has the minus sign shown in Eq.~\eqref{eq:V1}, fixed by the squared Dirac operator. Its contraction with $q_\mu$ vanishes by antisymmetry. The appearance of $\sigma^{\mu\nu}$ in this auxiliary heat-kernel insertion does not by itself represent a tree-level anomalous magnetic moment: the physical vertex is obtained only after the endpoint and heat-kernel pieces are combined. The one-loop Pauli form factor is derived from the complete third derivative in Sec.~\ref{subsec:g2}.

Contracting with $q_\mu$ and using $q_\mu V_1^\mu=e(a_{p'}-a_p)$ and $(a_{p'}-a_p)I(a_{p'},a_p)=R(a_p)-R(a_{p'})$ reproduces the unamputated Ward identity $q_\mu G_\Lambda^\mu=e[S_\Lambda(p)-S_\Lambda(p')]$. The amputated vertex $\Gamma_\Lambda^\mu=S_\Lambda^{-1}(p')G_\Lambda^\mu S_\Lambda^{-1}(p)$ therefore satisfies
\begin{equation}
q_\mu\Gamma_\Lambda^\mu(p',p)=e\left[S_\Lambda^{-1}(p')-S_\Lambda^{-1}(p)\right].
\label{eq:WTI}
\end{equation}
This identity is derived from gauge covariance of the kernel; no vertex ansatz is imposed. Write the inverse propagator as $S_\Lambda^{-1}(p)=\ii A(p^2)\slashed p+B(p^2)$, with $A(p^2)=B(p^2)/m=\e^{s_0(p^2+m^2)}$. A kinematically regular longitudinal solution of the Ward identity~\eqref{eq:WTI} is~\cite{BallChiu1980}
\begin{align}
\Gamma_{\Lambda,L}^{\mu}(p',p)
=e\Bigg\{&\frac{A'+A}{2}\,\ii\gamma^\mu
+\frac{(p'+p)^\mu}{p'^2-p^2}\nonumber\\
&\times\left[
\frac{A'-A}{2}\,\ii(\slashed p'+\slashed p)
+(B'-B)\right]\Bigg\}.
\end{align}
where $A'=A(p'^2)$ and $B'=B(p'^2)$. The vertex obtained from the functional derivative decomposes exactly as $\Gamma_\Lambda^\mu=\Gamma_{\Lambda,L}^\mu+\Gamma_{\Lambda,T}^\mu$, with $q_\mu\Gamma_{\Lambda,T}^\mu=0$. The Ward identity fixes the contraction $q_\mu\Gamma_\Lambda^\mu$; the transverse part is supplied by the derived kernel rather than by the identity alone.

The full open-line derivative can be written before any pole limit is taken. Write
\begin{equation}
M_A=m-\slashD_A,\qquad S_\Lambda[A]=M_A R_{s_0}(\Delta_A).
\end{equation}
For a labelled photon derivative $\delta_i\equiv\delta/\delta A_{\mu_i}(q_i)$ define
\begin{equation}
E_i=\delta_iM_A,\qquad V_i=-\delta_i\Delta_A,\qquad V_{ij}=-\delta_i\delta_j\Delta_A.
\label{eq:complete_derivative_vertices}
\end{equation}
The two-photon contact contained in $V_{ij}$ is a standard feature of the second-order representation of spinor QED. It arises from the $A^2$ term in $-D_A^2$ after the first-order Dirac propagator is factorised through $\Delta_A$; it is not an additional fundamental interaction of the first-order Dirac Lagrangian. In the second-order representation it must be retained together with the one-photon and Dirac-prefactor contributions for equivalence with ordinary spinor QED~\cite{Schubert2001,Ahmadiniaz2020,AngelesMartinez2012}.
Because the prefactor $M_A$ is linear in $A$ and the operator $\Delta_A$ is at most quadratic, no second derivative of $M_A$ and no third derivative of $\Delta_A$ can occur, so
\begin{equation}
\delta_i\delta_jM_A=0,\qquad \delta_{i_1}\delta_{i_2}\delta_{i_3}\Delta_A=0.
\label{eq:derivative_truncation}
\end{equation}
Let $\mathfrak P_{1,2}(I)$ denote the partitions of the labelled set $I$ into singleton and pair blocks, with $V_B=V_i$ for $B=\{i\}$ and $V_B=V_{ij}$ for $B=\{i,j\}$. For a list of $r$ such blocks define
\begin{multline}
\mathfrak D_{s_0}[B_1,\ldots,B_r]
=\sum_{\sigma\in S_r}
\int_{\substack{u_0,\ldots,u_r\ge0\\u_0+\cdots+u_r\ge s_0}}
\prod_{j=0}^{r}\dd u_j\\
\times e^{-u_0\Delta_A}V_{B_{\sigma(1)}}e^{-u_1\Delta_A}\cdots
V_{B_{\sigma(r)}}e^{-u_r\Delta_A}.
\label{eq:complete_Duhamel_block}
\end{multline}
where $S_r$ is the permutation group of the blocks, so $\mathfrak D_{s_0}$ already contains every ordering of the listed insertions. With the empty list defined by $\mathfrak D_{s_0}[\varnothing]=R_{s_0}(\Delta_A)$, repeated application of the product rule gives
\begin{multline}
\delta_I S_\Lambda[A]
=M_A\sum_{\pi\in\mathfrak P_{1,2}(I)}\mathfrak D_{s_0}[\pi]\\
+\sum_{r\in I}E_r
\sum_{\pi\in\mathfrak P_{1,2}(I\setminus\{r\})}
\mathfrak D_{s_0}[\pi].
\label{eq:complete_derivative_master}
\end{multline}
Equation~\eqref{eq:complete_derivative_master} contains the ordered bulk insertions, the two-photon contacts, and the prefactor contributions in one expression. The proper-time subintervals satisfy $u_j\ge0$, while the lower bound applies only to their sum. This follows directly from the integral form of the product rule
\begin{equation}
\delta e^{-T\Delta_A}=-\int_0^T\dd\tau\,
e^{-(T-\tau)\Delta_A}(\delta\Delta_A)e^{-\tau\Delta_A}.
\end{equation}
No higher class can appear because of Eq.~\eqref{eq:derivative_truncation}.

\revgreen{The reader can keep the following picture of what the construction changes and what it leaves alone. The graphs, their symmetry factors, the Dirac numerators, the photon polarisation tensors and the gauge-fixing structure are those of ordinary QED in its second-order form, with the linear insertions, the seagull and the Pauli term as its vertex set. The retained endpoint acts only on the proper-time weights: the scalar domain integrals $I(a_1,\ldots,a_n)$ of Eq.~\eqref{eq:complete_Duhamel_block} replace $\prod_ja_j^{-1}$, while every tensor factor is unchanged, exactly as the circuit domain replaces the local Schwinger domain in the scalar theory~\cite{BakrZhang2026} without touching its combinatorics. The one structural difference between the two theories is that the scalar field is both the propagating and the inserted field, so an insertion only partitions the proper time of the history it sits on, whereas in QED the photon is a second field: a photon insertion on a charged line likewise only partitions that line, but an internal photon is a complete history of its own and therefore carries the endpoint. The criterion is not that the fields differ but that the photon has its own Gaussian measure in the exact source, dressed by the charged-sector weight; a field integrated over with its own covariance is a parent and every one of its internal lines carries the endpoint, a self-interacting field is not and its histories are the circuits of its own network, and in both cases insertions partition and never add an endpoint. A complex scalar with a $\phi\phi\phi^*\phi^*$ vertex has one network and only circuits; two charged scalars coupled by $\phi\phi\chi^*\chi^*$ have two networks, both circuit carriers and no parent; scalar QED has the photon as parent and the charged loops as circuits, exactly as here. This is why the Ward--Takahashi hierarchy is exact and why the amplitudes nevertheless differ from the local ones. The same $s_0$ enters every complete history, closed charged circles and photons alike, because it is a constant of the theory and not a property of a species: at the scale $\sqrt{s_0}$ the electron appears as an extended charge, the resolved Coulomb potential of Eq.~\eqref{eq:erf_potential} being that of a Gaussian charge distribution of width $\sqrt{2s_0}$, and the photon shows the same length in the damping of its propagation. Gauge invariance does not require the photon to carry the endpoint, since the hierarchy holds for any positive photon covariance; the single value is a postulate, and it is tested by the matching, because the anomalous moment is controlled by the photon's endpoint and the vacuum-polarisation tail by the charged circle's, so one number must fit both.}

\subsection{Complete open-line derivatives and pole reduction}
The second, third, and fourth derivatives are written before any pole limit is taken. We keep the ordered block functional $\mathfrak D_{s_0}[\boldsymbol V]$ defined in Eq.~\eqref{eq:complete_Duhamel_block}; its number of arguments is the number of insertion blocks, so no derivative-like superscript is introduced for this bookkeeping.

\paragraph{Second derivative.}
The second derivative is
\begin{align}
\delta_1\delta_2S_\Lambda
={}&M_A\Big[\mathfrak D_{s_0}[V_1,V_2]+\mathfrak D_{s_0}[V_{12}]\Big]
\nonumber\\
&+E_1\mathfrak D_{s_0}[V_2]+E_2\mathfrak D_{s_0}[V_1].
\label{eq:second_derivative_complete_global}
\end{align}
The first term contains the two orderings $V_1V_2$ and $V_2V_1$, the second is the contact insertion $V_{12}$, and the remaining two are the prefactor contributions. Thus Eq.~\eqref{eq:second_derivative_complete_global} contains five ordered terms, whose unit-covariant-gauge contraction is evaluated directly in Appendix~\ref{app:self_energy_derivation}.

\paragraph{Third derivative.}
A third derivative gives
\begin{align}
\delta_1\delta_2\delta_3S_\Lambda
={}&M_A \mathfrak D_{s_0}[V_1,V_2,V_3]
\nonumber\\
&+M_A \mathfrak D_{s_0}[V_{12},V_3]
+M_A \mathfrak D_{s_0}[V_{13},V_2]
\nonumber\\
&+M_A \mathfrak D_{s_0}[V_{23},V_1]
\nonumber\\
&+E_1\mathfrak D_{s_0}[V_2,V_3]+E_1\mathfrak D_{s_0}[V_{23}]
\nonumber\\
&+E_2\mathfrak D_{s_0}[V_1,V_3]+E_2\mathfrak D_{s_0}[V_{13}]
\nonumber\\
&+E_3\mathfrak D_{s_0}[V_1,V_2]+E_3\mathfrak D_{s_0}[V_{12}].
\label{eq:third_derivative_complete_global}
\end{align}
The four classes contain six $V_1^3$ orderings, six $V_2V_1$ orderings, six $EV_1^2$ orderings, and three $EV_2$ orderings, giving 21 labelled terms in all.

\paragraph{Fourth derivative.}
The fourth derivative is the first order at which two contact insertions can occur on the same open line. Its non-prefactor part is
\begin{align}
\left.\delta_1\delta_2\delta_3\delta_4S_\Lambda\right|_M
={}&M_A \mathfrak D_{s_0}[V_1,V_2,V_3,V_4]
\nonumber\\
&+M_A \mathfrak D_{s_0}[V_{12},V_3,V_4]
\nonumber\\
&+M_A \mathfrak D_{s_0}[V_{13},V_2,V_4]
\nonumber\\
&+M_A \mathfrak D_{s_0}[V_{14},V_2,V_3]
\nonumber\\
&+M_A \mathfrak D_{s_0}[V_{23},V_1,V_4]
\nonumber\\
&+M_A \mathfrak D_{s_0}[V_{24},V_1,V_3]
\nonumber\\
&+M_A \mathfrak D_{s_0}[V_{34},V_1,V_2]
\nonumber\\
&+M_A \mathfrak D_{s_0}[V_{12},V_{34}]
\nonumber\\
&+M_A \mathfrak D_{s_0}[V_{13},V_{24}]
\nonumber\\
&+M_A \mathfrak D_{s_0}[V_{14},V_{23}].
\label{eq:fourth_derivative_M_part}
\end{align}
The prefactor part is
\begin{align}
\left.\delta_1\delta_2\delta_3\delta_4S_\Lambda\right|_E
={}&E_1\mathfrak D_{s_0}[V_2,V_3,V_4]
\nonumber\\
&+E_1\mathfrak D_{s_0}[V_{23},V_4]+E_1\mathfrak D_{s_0}[V_{24},V_3]
\nonumber\\
&+E_1\mathfrak D_{s_0}[V_{34},V_2]
\nonumber\\
&+E_2\mathfrak D_{s_0}[V_1,V_3,V_4]
\nonumber\\
&+E_2\mathfrak D_{s_0}[V_{13},V_4]+E_2\mathfrak D_{s_0}[V_{14},V_3]
\nonumber\\
&+E_2\mathfrak D_{s_0}[V_{34},V_1]
\nonumber\\
&+E_3\mathfrak D_{s_0}[V_1,V_2,V_4]
\nonumber\\
&+E_3\mathfrak D_{s_0}[V_{12},V_4]+E_3\mathfrak D_{s_0}[V_{14},V_2]
\nonumber\\
&+E_3\mathfrak D_{s_0}[V_{24},V_1]
\nonumber\\
&+E_4\mathfrak D_{s_0}[V_1,V_2,V_3]
\nonumber\\
&+E_4\mathfrak D_{s_0}[V_{12},V_3]+E_4\mathfrak D_{s_0}[V_{13},V_2]
\nonumber\\
&+E_4\mathfrak D_{s_0}[V_{23},V_1].
\label{eq:fourth_derivative_E_part}
\end{align}
Equations~\eqref{eq:fourth_derivative_M_part} and~\eqref{eq:fourth_derivative_E_part} contain 24 $M_AV_1^4$ orderings, 36 $M_AV_2V_1^2$ orderings, six $M_AV_2^2$ orderings, 24 $EV_1^3$ orderings, and 24 $EV_2V_1$ orderings, so
\begin{equation}
24+36+6+24+24=114.
\label{eq:fourth_derivative_114}
\end{equation}
No further class is possible because $M_A$ is linear and $\Delta_A$ is quadratic in $A$.

\paragraph{Pole reduction.}
The pole reduction is most transparent after separating the short-proper-time part removed from each open line by the lower endpoint. Define
\begin{equation}
Q_{s_0}(\Delta)=\int_0^{s_0}\dd T\,e^{-T\Delta}.
\end{equation}
Since $M_A=m-\slashD_A$ and $B_A=m+\slashD_A$ are both polynomials in the same Dirac operator, they commute. Hence
\begin{equation}
\Delta_A=M_AB_A=B_AM_A.
\end{equation}
It follows that
\begin{equation}
M_A\Delta_A^{-1}=B_A^{-1}=(m+\slashD_A)^{-1}\equiv S_0[A].
\end{equation}
Using $R_{s_0}(\Delta)=\Delta^{-1}-Q_{s_0}(\Delta)$ gives the exact operator identity
\begin{equation}
S_\Lambda[A]=S_0[A]-M_AQ_{s_0}(\Delta_A).
\label{eq:exact_open_local_short_identity}
\end{equation}
Differentiate Eq.~\eqref{eq:exact_open_local_short_identity} a finite number $n$ of times. Every term in the derivative of $M_AQ_{s_0}(\Delta_A)$ is a finite sum of polynomial insertion operators multiplied by a divided difference of the entire short-proper-time function $q_{s_0}$. Equivalently, every such term is supported on a bounded total proper-time region $0\le \sum_j u_j\le s_0$. At fixed infrared regulator it is therefore analytic in the two external electron inverse-propagator variables in a neighbourhood of their mass-shell values. It cannot contain either external one-particle pole. Denote by $\mathcal P_{\rm pole}$ the operation that multiplies a complete open kernel by the two external inverse propagators and then takes the two external electron pole limits between the corresponding on-shell spinors. Then, at every finite derivative order,
\begin{equation}
\mathcal P_{\rm pole}\,
\delta^n\!\left[M_A Q_{s_0}(\Delta_A)\right]=0,
\qquad n<\infty.
\label{eq:short_proper_time_amputation_zero_234}
\end{equation}
Equations~\eqref{eq:second_derivative_complete_global}--\eqref{eq:fourth_derivative_114} display the complete bulk, contact, prefactor, and ordered divided-difference classes for $n=2,3,4$. Equation~\eqref{eq:short_proper_time_amputation_zero_234} does not say that the corresponding pieces of the bounded FPT kernel vanish separately. It says that their complete bounded-minus-local short-proper-time remainder is regular at the external poles. An independent internal photon retains its own proper-time-parameter endpoint.

To identify the pole part of the remaining local term, put
\begin{equation}
B_A=m+\slashD_A,
\qquad
\mathcal V_i=\delta_iB_A,
\qquad
S_0=B_A^{-1}.
\end{equation}
Since the first derivative is $\delta_iS_0=-S_0\mathcal V_iS_0$, repeated use of the product rule gives
\begin{align}
\delta_1\delta_2S_0
={}&S_0\mathcal V_1S_0\mathcal V_2S_0
+S_0\mathcal V_2S_0\mathcal V_1S_0,
\label{eq:S0_second_derivative_dirac_chain}\\
\delta_1\delta_2\delta_3S_0
={}&-\sum_{\pi\in S_3}
S_0\mathcal V_{\pi(1)}S_0\mathcal V_{\pi(2)}S_0
\mathcal V_{\pi(3)}S_0,
\label{eq:S0_third_derivative_dirac_chain}\\
\delta_1\delta_2\delta_3\delta_4S_0
={}&\sum_{\pi\in S_4}
S_0\mathcal V_{\pi(1)}S_0\mathcal V_{\pi(2)}S_0
\mathcal V_{\pi(3)}S_0\mathcal V_{\pi(4)}S_0.
\label{eq:S0_fourth_derivative_dirac_chain}
\end{align}
The local pole part therefore contains exactly 2, 6, and 24 ordinary Dirac chains. Two-sided amputation removes the two outer $S_0$ factors, while the endpoint exponentials tend to one as the external inverse propagators approach their poles. Combining this local identity with Eq.~\eqref{eq:short_proper_time_amputation_zero_234} gives
\begin{equation}
\mathcal P_{\rm pole}\,\delta^nS_\Lambda
=\mathcal P_{\rm pole}\,\delta^nS_0,
\qquad n<\infty.
\label{eq:explicit_234_pole_reduction}
\end{equation}
The second, third, and fourth derivatives displayed above are the explicit orders needed below. Equation~\eqref{eq:explicit_234_pole_reduction} concerns only their external pole coefficients, while the exact off-shell kernels continue to carry all finite-$s_0$ factors.

The momentum routing, short-proper-time divided-difference classes, finite-virtuality expansion, and arbitrary-order counting are collected in Appendix~\ref{app:operator_source_identities}.

\section{Second Functional Derivative and Finite Electron Self-Energy}
\label{sec:second}
The electron self-energy cannot be computed consistently by joining two copies of only the longitudinal, Ward-identity-determined one-photon vertex. The gauge-covariant open kernel also generates a two-photon contact term from the squared Dirac operator and endpoint contributions from the Dirac prefactor. These terms are fixed by the same functional derivative and must be kept together before any on-shell reduction is made.

The squared Dirac operator contains one-photon insertions and a two-photon contact term, while the Dirac prefactor in~\eqref{eq:open_kernel} contributes prefactor terms. All five contributions enter the electron self-energy through the second functional derivative
\begin{align}
G_\Lambda^{\mu\nu}(p_{12},p)
&=\left.
\frac{\delta^2S_\Lambda[A](p_{12},p)}
{\delta A_\mu(q_1)\delta A_\nu(q_2)}
\right|_{A=0},\nonumber\\
p_{12}&=p+q_1+q_2.
\label{eq:G_munu_def}
\end{align}
This derivative contains the two ordered $V_1V_1$ bulk insertions, the two-photon contact $V_2=-2e^2\delta^{\mu\nu}$ from the $A^2$ term in $-D_A^2$, and the two prefactor--bulk terms produced by differentiating $m-\slashD_x$. Their second divided-difference kernel is
\begin{multline}
J(a_c,a_b,a_a)=\int_{s_0}^{\infty}\dd T\int_0^T\dd\tau_1\int_0^{\tau_1}\dd\tau_2\,
\e^{-(T-\tau_1)a_c} \\
\e^{-(\tau_1-\tau_2)a_b}
\e^{-\tau_2a_a}
=\sum_{r\in\{a_a,a_b,a_c\}}
\frac{R(r)}{\prod_{r'\neq r}(r-r')}.
\label{eq:J_def}
\end{multline}
The repeated-argument limit for the self-energy ($a_c=a_a=a_0$) is
\begin{equation}
J(a,b,a)=\frac{R(b)-R(a)-R'(a)(b-a)}{(b-a)^2}.
\label{eq:J_repeated}
\end{equation}
The unit-covariant-gauge contraction of the five terms is evaluated in Appendix~\ref{app:self_energy_derivation}. Before pole reduction, the differentiated open operator kernel inherits the single interval condition $T=u_1+u_2+u_3\ge s_0$, where $u_1,u_3$ are the external-leg segments and $u_2$ is the segment between the two photon insertions. In the on-shell self-energy the external electron poles are amputated, so the inherited charged interval constraint becomes inactive. The internal photon propagator retains its own proper-time parameter and lower endpoint from the covariance~\eqref{eq:derived_photon_covariance}, while the remaining charged segment $t=u_2$ is integrated from zero. The resulting one-loop circuit therefore has
\begin{equation}
t\ge0,\qquad \sigma\ge s_0,
\label{eq:photon_bounded_self_energy_domain}
\end{equation}
where $T=t+\sigma$ and $x=t/T$.

The off-shell self-energy must be kept separate from its on-shell pole projection. The complete one-loop correction to the open propagator is
\begin{equation}
\delta S_{1,\Lambda}(p)
=\frac12\int_k C^{\mu\nu}_{s_0}(k)\,
G_{\Lambda,\mu\nu}(p,p;k,-k),
\label{eq:deltaS_full_definition}
\end{equation}
where $\int_k\equiv\int\dd^4k/(2\pi)^4$ and $C^{\mu\nu}_{s_0}(k)=\delta^{\mu\nu}\e^{-s_0k^2}/k^2$ in the unit covariant gauge. Here $G_{\Lambda}^{\mu\nu}$ is the unreduced five-term derivative of~\eqref{eq:G_munu_def}. The exact off-shell one-loop one-particle-irreducible (1PI) self-energy is
\begin{equation}
\Sigma^{\rm full}_\Lambda(p)
=S_\Lambda^{-1}(p)\,\delta S_{1,\Lambda}(p)\,S_\Lambda^{-1}(p).
\label{eq:Sigma_full_definition}
\end{equation}
Equations~\eqref{eq:deltaS_full_definition}--\eqref{eq:Sigma_full_definition}, together with the kernel in Appendix~\ref{app:self_energy_derivation}, define the off-shell self-energy. They retain the finite-$s_0$ factors in Eqs.~\eqref{eq:F_endpoint_series}--\eqref{eq:F_full_inverse_series}, so no on-shell factorisation enters the off-shell quantity.

For the pole mass only, Eq.~\eqref{eq:explicit_234_pole_reduction} replaces the complete open-line pole coefficient by the ordinary QED one. The separately complete internal photon still has $\sigma\ge s_0$. The resulting on-shell-reduced representative is
\begin{multline}
\Sigma_\Lambda^{\rm pole}(p_E)=\frac{e^2}{16\pi^2}\int_{s_0}^{\infty}\frac{\dd T}{T}\int_0^{\,1-s_0/T}\dd x\\
\times\e^{-T[xm^2+x(1-x)p_E^2]}\left[4m+2\ii(1-x)\slashed p_E\right].
\label{eq:self_energy_E}
\end{multline}
Equation~\eqref{eq:self_energy_E} is the on-shell pole representative. Momentum derivatives, continuum information, and the field-strength factor $Z_2$ are obtained from the off-shell self-energy in Eq.~\eqref{eq:Sigma_full_definition}.

At the Lorentzian on-shell point $p_E^2\to -m^2$, the denominator in~\eqref{eq:self_energy_E} reduces to $x^2m^2$. Exchanging the order of integration, the photon-bounded domain $T\ge s_0/(1-x)$ gives
\begin{equation}
\frac{\delta m_\Lambda}{m}
=\frac{\alpha}{2\pi}\int_0^1\dd x\,(1+x)
\Gamma\!\left(0,\frac{x^2m^2}{(1-x)\Lambda^2}\right).
\label{eq:mass_shift_final}
\end{equation}
The same five-term derivative gives the contraction and pole reduction in Appendix~\ref{app:self_energy_derivation}.

For $m^2/\Lambda^2\ll1$,
\begin{multline}
\frac{\delta m_\Lambda}{m}=
\frac{3\alpha}{4\pi}\ln\frac{\Lambda^2}{m^2}
+\frac{\alpha}{2\pi}\left(\frac34-\frac32\gamma_E\right)\\
+O\!\left(\frac{m^2}{\Lambda^2}\ln\frac{\Lambda^2}{m^2}\right),
\label{eq:mass_shift_asymptotic}
\end{multline}
recovering the standard logarithmic mass divergence as a finite logarithm of the retained scale $\Lambda$.

At one loop let $m_0$ denote the finite mass parameter before pole-mass matching. The physical pole mass is then
\begin{multline}
m_{\rm phys}=m_0\bigg[1+\frac{\alpha}{2\pi}\int_0^1\dd x\,(1+x) \\
\times\Gamma\!\left(0,\frac{x^2m_0^2}{(1-x)\Lambda^2}\right)\bigg]+O(\alpha^2).
\label{eq:finite_mass_matching}
\end{multline}
For $m_{\rm phys}=0.510998950\,\mathrm{MeV}$, $\alpha^{-1}=137.035999084$, and $\Lambda=13\,\mathrm{TeV}$, numerical evaluation of~\eqref{eq:mass_shift_final} gives $\delta m_\Lambda/m=0.0592780$, in agreement with the expansion~\eqref{eq:mass_shift_asymptotic}. This value is a matching correction.

The ultraviolet convergence is independent of the familiar infrared subtleties of QED. An infrared photon mass $\lambda_{\rm IR}>0$ may be retained to isolate the electron pole until inclusive quantities are formed. Gauge-parameter independence of the pole mass follows from the Ward identity~\eqref{eq:WTI}, since the longitudinal contribution vanishes between on-shell states. At fixed $\lambda_{\rm IR}$, the off-shell Euclidean integral~\eqref{eq:Sigma_full_definition} is ultraviolet finite and analytic near the pole, so the derivatives that define the wave-function factor $Z_2(\lambda_{\rm IR})$ are finite. Denoting the corresponding vertex factor by $Z_1$, gauge covariance then gives $Z_1=Z_2$.

\section{Finite-Proper-Time Photon Sector and Fixed-Order UV Finiteness}
\label{sec:multiloop}
Applying~\eqref{eq:truncated_inverse} to the Maxwell kinetic operator gives the internal Euclidean covariance in the unit covariant gauge
\begin{equation}
D_{\mu\nu,\Lambda}(k_E)=\delta_{\mu\nu}\frac{e^{-s_0k_E^2}}{k_E^2}.
\label{eq:photon_covariance_multiloop}
\end{equation}
With $\mathcal P_\gamma=-\partial_E^2$ it follows from the quadratic gauge-fixed action
\begin{align}
S_{\gamma,s_0}+S_{{\rm gf},s_0}
={}&\frac14\int F_{\mu\nu}e^{s_0\mathcal P_\gamma}F_{\mu\nu}\nonumber\\
&+\frac12\int(\partial\!\cdot\!A)e^{s_0\mathcal P_\gamma}(\partial\!\cdot\!A),
\label{eq:photon_quadratic_action}
\end{align}
with the usual gauge-parameter generalisation. This covariance is internal Euclidean spectral machinery; the physical on-shell photon measure is obtained only after the complete amplitude has been continued and cut.

Consider a connected fixed-order Euclidean FPT graph and a homogeneous variation $\delta q$ of all independent loop momenta. After proper-time parametrisation its loop-dependent quadratic form is
\begin{equation}
Q_G(\delta q;s)
=
\sum_{\gamma}\sigma_\gamma|\delta k_\gamma|^2
+
\sum_w\sum_{e\in w}u_e|\delta p_e|^2.
\label{eq:rankfree_QG}
\end{equation}
Here each independent internal photon history has $\sigma_\gamma\ge s_0$, while every complete closed charged history $w$ satisfies
\begin{equation}
\sum_{e\in w}u_e\ge s_0.
\label{eq:rankfree_charged_history}
\end{equation}
Suppose $Q_G=0$. Every term in Eq.~\eqref{eq:rankfree_QG} is non-negative, so all photon momentum variations vanish. Momentum conservation then makes the charged variation constant along each charged component. On an open component the fixed external endpoints force that constant to vanish. On a closed charged history the only possible constant circulation $r_w$ contributes
\begin{equation}
Q_w=|r_w|^2\sum_{e\in w}u_e\ge s_0|r_w|^2,
\label{eq:closed_history_coercive_piece}
\end{equation}
so it also vanishes. Therefore the homogeneous loop quadratic form is strictly positive in every nonzero loop direction.

To make the bound uniform, rescale by $s_0$ and clip every proper-time variable at a fixed finite multiple of $s_0$. The normalized admissible history domain is compact and the smallest eigenvalue of the loop quadratic form is a continuous strictly positive function on its product with the unit loop sphere. Hence, for each fixed graph $G$, there is \revblue{a constant $\kappa_G>0$ for the chosen loop routing and norm} such that
\begin{equation}
\boxed{Q_G(\ell;s)\ge \kappa_Gs_0|\ell|^2.}
\label{eq:rankfree_UV_coercivity}
\end{equation}
The remaining numerator and finite-history factors grow at most polynomially at fixed perturbative order. At any fixed infrared regulator used only to isolate the massless long-distance sector, Eq.~\eqref{eq:rankfree_UV_coercivity} therefore gives an integrable majorant
\begin{equation}
C_G(P)(1+|\ell|)^{N_G}e^{-\kappa_Gs_0|\ell|^2},
\label{eq:rankfree_UV_majorant}
\end{equation}
and every fixed-order Euclidean graph is ultraviolet finite. 

A two-loop chord graph makes the mechanism explicit. Let a complete charged history be divided into two intervals $u_1,u_2$ with $T=u_1+u_2\ge s_0$, and let an independent internal photon of proper time $\sigma\ge s_0$ carry momentum $k$. With charged loop momentum $p$,
\begin{equation}
Q=u_1p^2+u_2(p+k)^2+\sigma k^2.
\label{eq:chord_Q}
\end{equation}
Writing $x=u_2/T\in[0,1]$ gives the lower bound
\begin{equation}
Q\ge s_0
\begin{pmatrix}p&k\end{pmatrix}
\begin{pmatrix}1&x\\x&1+x\end{pmatrix}
\begin{pmatrix}p\\k\end{pmatrix}.
\label{eq:chord_matrix}
\end{equation}
The smaller eigenvalue is
\begin{equation}
\lambda_-(x)=\frac{2+x-\sqrt5\,x}{2},
\qquad
\min_{0\le x\le1}\lambda_-(x)=\frac{3-\sqrt5}{2}.
\label{eq:chord_eigenvalue}
\end{equation}
Thus this nontrivial two-loop sector obeys
\begin{equation}
Q\ge\frac{3-\sqrt5}{2}s_0\,(p^2+k^2),
\label{eq:chord_uniform_bound}
\end{equation}
which shows that the finite spectral boundary controls all simultaneous and mixed ultraviolet directions without assigning independent lower endpoints to the daughter intervals $u_1$ and $u_2$.

Gauge covariance fixes transversality independently of the UV estimate. Let $W[A]$ be the external-photon effective action obtained by integrating the internal gauge field with the finite-$s_0$ Euclidean covariance. A simultaneous background gauge transformation shifts the total field by a pure gauge at every integration point, while the closed charged functional is invariant. Therefore $W[A]$ is gauge invariant and its two-point response is transverse order by order. The ultraviolet estimate and the Ward identity are distinct: the former controls the Euclidean loop domain, whereas the latter fixes the gauge-correlated tensor structure.

\section{Finite Matching, Charge Running, and Physical Observables}
\label{sec:matching}
The finite-proper-time formulation replaces divergent mass and charge renormalisation by finite matching. The physical pole mass and Thomson-limit charge are used as measured inputs. At fixed $s_0>0$ the corresponding matching integrals are finite, and the retained endpoint is tested through observables that were not used to define those inputs.

\subsection{Fixed-endpoint momentum-subtraction flow}
\label{subsec:beta}
The arbitrary subtraction momentum and the finite lower endpoint must be kept distinct. To isolate charge running from the direct endpoint factor in photon exchange, define the momentum-subtraction (MOM) charge from the vacuum-polarisation part alone,
\begin{equation}
\bar\alpha_{\rm MOM}(\mu)=\frac{\alpha_{\rm phys}}{1-\Pi_R(\mu^2)}.
\label{eq:alpha_MOM_definition}
\end{equation}
The direct factor $\exp(\mu^2/\Lambda^2)$ from the exchanged photon propagator is not part of the charge running; it enters the full exchange propagator displayed below. Differentiating the exact one-loop polarisation~\eqref{eq:Pi_gamma} at fixed $s_0$ gives
\begin{equation}
\boxed{\begin{aligned}
\beta^{\rm MOM}_{s_0}(\bar\alpha,\mu)
&\equiv \mu\frac{\dd\bar\alpha_{\rm MOM}}{\dd\mu}\\
&=\frac{4\bar\alpha^2}{\pi}
\int_0^1\dd x\,
\frac{\mu^2x^2(1-x)^2}{m^2+\mu^2x(1-x)}\\[-2pt]
&\hspace{1.0cm}\times
\e^{-s_0[m^2+\mu^2x(1-x)]}
+O(\bar\alpha^3).
\end{aligned}}
\label{eq:beta_MOM_FPT}
\end{equation}
Here $\bar\alpha\equiv\bar\alpha_{\rm MOM}(\mu)$, $\mu$ is the momentum-subtraction scale, and $s_0$ is held fixed, so no additional ultraviolet regulator is introduced. In the regime $m^2\ll\mu^2$, writing $a=\mu^2s_0=\mu^2/\Lambda^2$ gives
\begin{align}
\beta^{\rm MOM}_{s_0}
&=\frac{2\bar\alpha^2}{3\pi}\,T_{\rm QED}(a)+O(\bar\alpha^3),\nonumber\\
T_{\rm QED}(a)
&=6\int_0^1\dd x\,x(1-x)\e^{-ax(1-x)}.
\label{eq:beta_threshold_QED}
\end{align}
The threshold function satisfies
\begin{equation}
T_{\rm QED}(a)=1-\frac{a}{5}+O(a^2),\qquad a\ll1,
\end{equation}
and
\begin{equation}
T_{\rm QED}(a)=\frac{12}{a^2}\left[1+O(a^{-1})\right],\qquad a\gg1.
\end{equation}
Thus the ordinary one-flavour QED coefficient $2\bar\alpha^2/(3\pi)$ is recovered for $m\ll\mu\ll\Lambda$. The fixed-endpoint momentum-subtraction flow is suppressed as
\begin{equation}
\beta^{\rm MOM}_{s_0}\sim\frac{8\bar\alpha^2}{\pi}\frac{\Lambda^4}{\mu^4}
\qquad(\mu\gg\Lambda).
\label{eq:beta_high_scale_FPT}
\end{equation}
The flow is non-autonomous because the beta function depends on $\mu/\Lambda$. The usual logarithmic momentum-subtraction running is recovered below $\Lambda$ and is suppressed once the subtraction momentum resolves the retained scale.

For spacelike momentum transfer $Q^2=q_E^2>0$, such as in electron--muon scattering, the dressed photon propagator is well defined in the Euclidean formulation. The bare inverse is $D_{0,\Lambda}^{-1}(Q^2)=Q^2\e^{Q^2/\Lambda^2}$, and the fermion loop contributes a term $-Q^2\Pi_R(Q^2)$. Therefore the dressed photon propagator is
\begin{equation}
D_\Lambda^{\rm dressed}(Q^2)=\frac{1}{Q^2\left[\e^{Q^2/\Lambda^2}-\Pi_R(Q^2)\right]}.
\label{eq:dressed_propagator}
\end{equation}
At low $Q^2\ll\Lambda^2$ this reproduces the standard QED result $1/\{Q^2[1-\Pi_R(Q^2)]\}$. At $Q^2\gg\Lambda^2$ the exponential dominates and the spacelike amplitude is exponentially suppressed.

At energies where neutral electroweak exchange is important, a physical collider prediction requires the same endpoint prescription to be embedded in the full $SU(2)_L\times U(1)_Y$ theory. At fixed order, the background-field extension is obtained by applying the same spectral function $R_{s_0}(P)=\e^{-s_0P}P^{-1}$ to each gauge-covariant quadratic fluctuation operator $P$. Since $P\mapsto UPU^{-1}$ under a background gauge transformation, the spectral prescription transforms covariantly. After symmetry breaking, the constant Weinberg rotation commutes with functional calculus, so the neutral quadratic operator still contains one massless photon pole and one $Z$ pole, while the charged sector retains the $W^\pm$ poles; the endpoint factor equals unity at each physical pole. We therefore make no collider bound or Drell--Yan prediction here. Precision lepton moments and the low-energy quantities below provide self-contained tests within the stated QED sector~\cite{Muong2_2025,WP2025}.

\subsection{Atomic and low-energy signatures}
\label{subsec:lowenergy}
At momenta far below $\Lambda$, the finite proper time smears static photon exchange over the length $1/\Lambda$. For a nucleus of charge number $Z_N$ and radial separation $r$, Fourier transforming the covariance~\eqref{eq:derived_photon_covariance} gives
\begin{align}
V_\Lambda(r)&=-\frac{Z_N\alpha_{\rm phys}}{r}\,\mathrm{erf}\!\left(\frac{r\Lambda}{2}\right),\nonumber\\
V_\Lambda(0)&=-\frac{Z_N\alpha_{\rm phys}\Lambda}{\sqrt{\pi}},
\label{eq:erf_potential}
\end{align}
where $\mathrm{erf}$ is the error function. The potential is finite at the origin, and at low momentum its spatial Fourier transform differs from the Coulomb result by
$\widetilde V_\Lambda(\vec q\,)-\widetilde V_{\rm Coul}(\vec q\,)=+4\pi Z_N\alpha_{\rm phys}/\Lambda^2+O(q^2/\Lambda^4)$. For a hydrogenic $n\mathrm S$ state with principal quantum number $n$, orbiting-lepton mass $m$, nuclear mass $m_N$, and reduced mass $m_r=m m_N/(m+m_N)$, this contact term gives
\begin{equation}
\Delta E_{n\mathrm S}=\frac{4Z_N^4\alpha_{\rm phys}^4m_r^3}{n^3\Lambda^2}.
\label{eq:contact_shift}
\end{equation} For hydrogen at $\Lambda=13\,\mathrm{TeV}$ this gives $\Delta E_{1\mathrm S}=2.2\,\mathrm{mHz}$ and a $1\mathrm S$--$2\mathrm S$ shift of $1.9\,\mathrm{mHz}$, corresponding to a fractional shift of $7.7\times10^{-19}$. The shift has the $|\psi(0)|^2$ contact form of the nuclear finite-size correction, with ratio
\begin{equation}
\frac{\Delta E_{\rm FPT}}{\Delta E_{\rm nucl}}
=\frac{6}{\Lambda^2\langle r_N^2\rangle}\simeq2\times10^{-9}
\qquad(\Lambda=13\,\mathrm{TeV}),
\label{eq:nuclear_degeneracy}
\end{equation}
where $\langle r_N^2\rangle$ is the nuclear mean-square charge radius. Thus, in a single-species measurement it is degenerate with a shift of the fitted nuclear charge radius and lies far below present radius uncertainties. The reduced mass also gives a small isotope dependence. For hydrogen and deuterium the induced $1\mathrm S$--$2\mathrm S$ isotope contribution is about $1.5\,\mu\mathrm{Hz}$ at $\Lambda=13\,\mathrm{TeV}$. This is far below present sensitivity. Atomic spectroscopy consequently does not bound $\Lambda$ at present. The leverage lies with the lepton moments.

\subsection{Anomalous magnetic moment}
\label{subsec:g2}
For the anomalous magnetic moment, the 21 terms of the third derivative in Eq.~\eqref{eq:third_derivative_complete_global} must first be combined before the external electron poles are taken. Their pole part reduces to the six local Dirac chains in Eq.~\eqref{eq:S0_third_derivative_dirac_chain}.

To isolate the one-particle-irreducible vertex, let the labels $a$ and $b$ be contracted into the internal photon and let $\mu$ label the external photon. Among the six Dirac chains, two place $\mu$ between $a$ and $b$ and belong to the one-particle-irreducible vertex, while the other four place $\mu$ outside the contracted pair and are external self-energy insertions.

Write
\begin{equation}
S=S_0+S_1+O(e^4),
\qquad
\Gamma^\mu=\Gamma_0^\mu+\Gamma_1^\mu+O(e^5).
\end{equation}
Here $S_1$ is the one-loop propagator correction, $\Gamma_0^\mu$ is the tree-level one-photon vertex, and $\Gamma_1^\mu$ is its one-loop 1PI correction. The connected three-point kernel obeys $G^\mu=S\Gamma^\mu S$, so at one loop
\begin{equation}
G_1^\mu
=S_1\Gamma_0^\mu S_0
+S_0\Gamma_0^\mu S_1
+S_0\Gamma_1^\mu S_0.
\label{eq:connected_vertex_decomposition}
\end{equation}
Therefore
\begin{equation}
\Gamma_1^\mu
=S_0^{-1}G_1^\mu S_0^{-1}
-S_0^{-1}S_1\Gamma_0^\mu
-\Gamma_0^\mu S_1S_0^{-1}.
\label{eq:vertex_1PI_extraction}
\end{equation}
The subtraction in Eq.~\eqref{eq:vertex_1PI_extraction} removes the four edge orderings and leaves the two middle orderings. The separation is therefore performed only after the full third derivative has been assembled.

After the subtraction in Eq.~\eqref{eq:vertex_1PI_extraction}, contract the two internal labels with one independent internal photon. Factoring out the overall external charge $e$, the on-shell one-loop vertex correction is
\begin{equation}
\frac{1}{e}\Gamma^{\mu}_{(1),\Lambda}(p',p)
=e^2\int_k\frac{e^{-s_0k^2}}{k^2}\,
\gamma_\alpha S_0(p'-k)\gamma^\mu S_0(p-k)\gamma_\alpha.
\label{eq:vertex_after_complete_reduction}
\end{equation}
Only the internal photon retains the endpoint factor, while the electron part is the ordinary on-shell Dirac chain. In this one-loop on-shell section $\alpha$ is understood as the matched low-energy coupling $\alpha_{\rm phys}$ up to higher-order differences. Let $m_\ell$ denote the mass of the external lepton. For on-shell spinors we define the Dirac and Pauli form factors by
\begin{align}
\bar u(p')\Gamma^\mu(p',p)u(p)
=e\,\bar u(p')\Bigg[&\gamma^\mu F_1(q^2)\nonumber\\
&+\frac{\ii\sigma^{\mu\nu}q_\nu}{2m_\ell}F_2(q^2)\Bigg]u(p).
\end{align}
Charge matching fixes $F_1(0)=1$.

At $q=0$, the origin of the Pauli weight can be displayed before the infrared regulator is removed. Giving the internal photon a temporary mass $\lambda_{\rm IR}$, the ordinary local Dirac chain gives
\begin{equation}
F_2^{\rm QED}(0;\lambda_{\rm IR})
=
\frac{\alpha}{2\pi}
\int_0^1\dd z\,
\frac{2z(1-z)^2m_\ell^2}
{(1-z)^2m_\ell^2+z\lambda_{\rm IR}^2}.
\label{eq:F2_massive_photon_control}
\end{equation}
The limit $\lambda_{\rm IR}\to0$ is finite and gives
$F_2^{\rm QED}(0)=\frac{\alpha}{\pi}\int_0^1 z\,\dd z=\alpha/(2\pi)$.
The open-line pole reduction above leaves this electron numerator unchanged. The independent internal photon has proper time $zT$ and retains the bound $zT\ge s_0$. After the infrared limit, the combined denominator is $m_\ell^2(1-z)^2$, so the bound gives $T\ge s_0/z$ and
\begin{equation}
F_2^{\rm FPT}(0)
=\frac{\alpha}{\pi}\int_0^1\dd z\,z\,
\exp\!\left[-\frac{m_\ell^2}{\Lambda^2}\frac{(1-z)^2}{z}\right].
\label{eq:F2_complete}
\end{equation}
Since $F_2^{\rm QED}(0)=\alpha/(2\pi)$, the finite-$s_0$ shift is
\begin{multline}
\delta a_\ell^{\rm FPT}
=\frac{\alpha}{\pi}\int_0^1\dd z\,z
\left[\exp\!\left(-\mu_\ell\frac{(1-z)^2}{z}\right)-1\right]\\
=-\frac{\alpha}{3\pi}\frac{m_\ell^2}{\Lambda^2}
+O\!\left(\frac{m_\ell^4}{\Lambda^4}\ln\frac{\Lambda^2}{m_\ell^2}\right),
\qquad \mu_\ell=\frac{m_\ell^2}{\Lambda^2}.
\label{eq:delta_ae_HK}
\end{multline}
The linear coefficient follows directly from
\begin{equation}
\left.\frac{\dd}{\dd\mu}\int_0^1\dd z\,z\,e^{-\mu(1-z)^2/z}\right|_{\mu=0}
=-\int_0^1(1-z)^2\dd z=-\frac13.
\end{equation}
The next derivative contains $\int_0\dd z/z$, which produces the displayed $O(\mu^2\ln\mu)$ term. The result follows from the complete third derivative after the connected-to-1PI separation above.

The Ward identity $Z_1=Z_2$ ensures that the field and vertex factors cancel from the physical anomalous magnetic moment, so $a_{\rm phys}=F_2(0)$ is independent of either factor. After the mass and charge have been matched, the term proportional to $m_\ell^2/\Lambda^2$ therefore remains as an independent finite-proper-time contribution. For $\Lambda=13\,\mathrm{TeV}$ it is far below present electron-moment sensitivity.

\subsection{Weak-field four-photon coefficient and low-energy light-by-light scattering}
\label{subsec:lightbylight}
A second on-shell sector is supplied by the closed fermion loop. Expanding the constant-field determinant in powers of the field, the quadratic term is the finite charge-matching contribution. After that term has been fixed by the low-energy definition of $\alpha_{\rm phys}$, the quartic invariant is controlled by the proper-time moment
\begin{equation}
\int_{s_0}^{\infty}\dd s\,s\e^{-m^2s}
=\frac{\e^{-\zeta}(1+\zeta)}{m^4},
\qquad \zeta=m^2s_0=\frac{m^2}{\Lambda^2}.
\label{eq:F4_moment}
\end{equation}
Hence the physical one-loop weak-field four-photon interaction is
\begin{equation}
\boxed{\begin{aligned}
\mathcal L_{1,\Lambda,F^4}
&=w_4(\zeta)\,\frac{2\alpha_{\rm phys}^2}{45m^4}
\Big[(\mathbf E^2-\mathbf B^2)^2\\[-2pt]
&\hspace{1.35cm}+7(\mathbf E\!\cdot\!\mathbf B)^2\Big],\\
w_4(\zeta)&=\e^{-\zeta}(1+\zeta).
\end{aligned}}
\label{eq:EH_F4_weight}
\end{equation}
Here $\mathbf E$ and $\mathbf B$ are the electric and magnetic fields. The spinor tensor structure is the standard QED one, while the finite lower endpoint changes only the common proper-time moment. Consequently, for a characteristic external photon energy $\omega\ll m$, the one-fermion-loop light-by-light amplitude obeys
\begin{equation}
\boxed{\begin{aligned}
\mathcal M_{1,\gamma\gamma\to\gamma\gamma,\Lambda}
={}&w_4(\zeta)\,
\left.\mathcal M_{1,\gamma\gamma\to\gamma\gamma,\rm QED}
\right|_{O(\omega^4/m^4)}\\
&+O(\alpha^2\omega^6/m^6).
\end{aligned}}
\label{eq:LBL_low_energy_weight}
\end{equation}
Since $w_4(\zeta)=1-\zeta^2/2+O(\zeta^3)$, the matched quartic weight has no term linear in $s_0$, and the first low-energy light-by-light deviation therefore begins at $O(m^4/\Lambda^4)$. Equation~\eqref{eq:LBL_low_energy_weight} is the corresponding one-loop result.

\subsection{On-shell non-redundancy and identifiability of the finite lower endpoint}
\label{subsec:nonredundancy}
Ultraviolet finiteness alone would not distinguish a physical parameter from a removable regularisation convention or a freely adjustable short-distance parametrisation. The stronger test is whether the retained endpoint survives after the ordinary measured parameters have been fixed, can itself be calibrated by an independent observable, and then predicts further observables without introducing new endpoint-dependent coefficients. Here $m_{\rm phys}$ and $\alpha_{\rm phys}$ are fixed by the pole-mass and Thomson-limit matching conditions. The Pauli form factor supplies an independent on-shell calibration.

After mass and charge matching, Eq.~\eqref{eq:delta_ae_HK} gives
\begin{align}
F_2^{\rm FPT}(0)-F_2^{\rm QED}(0)
={}&-\frac{\alpha_{\rm phys}}{3\pi}m_{\rm phys}^2s_0\nonumber\\
&+O\!\left(m_{\rm phys}^4s_0^2
\ln\frac{1}{m_{\rm phys}^2s_0}\right).
\label{eq:Pauli_nonredundancy}
\end{align}
The coefficient is nonzero within the finite-proper-time deformation of minimal QED considered here, whose independent microscopic parameters are the QED mass, charge, and finite lower endpoint and which contains no separately adjustable Pauli coupling. Within that parameter set, the on-shell quantity $F_2(0)$ cannot be removed by another redefinition of the already matched mass or charge. If one enlarged the microscopic theory by adding an independent Pauli operator, its coefficient could instead be fixed by magnetic-moment matching and the non-redundancy test would have to be moved to another observable. Within the fixed-order Euclidean-first continuation used here, and within the usual class of admissible invertible local or almost-local field redefinitions that leave on-shell amplitudes unchanged, such a redefinition cannot change this on-shell amplitude~\cite{Kamefuchi1961}. The finite lower endpoint is therefore perturbatively non-redundant at one loop within these stated assumptions. This one-loop identifiability statement is perturbative; the nonperturbative finite-resolution construction is given in Sec.~\ref{sec:FR_programme}.

The identifiability is in fact global within this one-loop minimal-FPT sector. Differentiating the exact expression~\eqref{eq:F2_complete} with respect to the retained endpoint gives
\begin{equation}
\boxed{\begin{aligned}
\frac{\partial F_2^{\rm FPT}(0;s_0)}{\partial s_0}
={}&-\frac{\alpha_{\rm phys}m_{\rm phys}^2}{\pi}
\int_0^1\dd z\,(1-z)^2\\[-2pt]
&\times\exp\!\left[-m_{\rm phys}^2s_0\frac{(1-z)^2}{z}\right]<0 .
\end{aligned}}
\label{eq:F2_global_monotonicity}
\end{equation}
Moreover, dominated convergence gives
\begin{equation}
F_2^{\rm FPT}(0;0)=\frac{\alpha_{\rm phys}}{2\pi},
\qquad
\lim_{s_0\to\infty}F_2^{\rm FPT}(0;s_0)=0.
\label{eq:F2_global_endpoints}
\end{equation}
Thus $s_0\mapsto F_2^{\rm FPT}(0;s_0)$ is continuous and strictly decreasing on $[0,\infty)$, so any value in its image fixes a unique $s_0$ at this order. Near the local-QED endpoint the inverse begins as
\begin{align}
s_0={}&-\frac{3\pi}{\alpha_{\rm phys}m_{\rm phys}^2}\,
\delta a_\ell^{\rm FPT}\nonumber\\
&+O\!\left(s_0^2m_{\rm phys}^2
\ln\frac{1}{m_{\rm phys}^2s_0}\right).
\label{eq:s0_from_g2}
\end{align}
Once this single parameter has been calibrated at a fixed perturbative accuracy, Eqs.~\eqref{eq:beta_MOM_FPT} and~\eqref{eq:EH_F4_weight} give independent cross-observable predictions of the same finite-endpoint theory at the corresponding accuracy. At this order the measured mass and charge together with one endpoint-sensitive on-shell observable determine the one-parameter FPT trajectory, while further observables test that trajectory.

\subsection{Formal correspondence with renormalised local QED}
\label{subsec:local_limit}
The point $s_0=0$ is outside the FPT spectral theory. It is nevertheless useful to make the formal comparison: if the retained boundary is removed after the FPT expressions have been derived, do the matched amplitudes approach renormalised local QED? In that correspondence calculation,
\begin{align}
R_{s_0}(\Delta)&\longrightarrow \Delta^{-1},\nonumber\\
S_\Lambda[A]&\longrightarrow(m+\slashD_A)^{-1},\nonumber\\
D_{\mu\nu,\Lambda}(k)&\longrightarrow\frac{\delta_{\mu\nu}}{k^2}.
\label{eq:local_operator_limit}
\end{align}
At the same time the retained spectral domains expand to the ordinary non-negative Schwinger-parameter domains. Thus $R(a)\to1/a$, $I(a',a)\to1/(a'a)$, and $J(a,b,a)\to1/(a^2b)$, so the differentiated open kernels become the standard background-field QED chains and their Ward identities reduce to the usual local relations.

The closed proper-time functional itself develops the familiar ultraviolet divergences when its lower endpoint is removed. Therefore ``$s_0=0$'' is not a finite bare closed functional. The correct statement is a \emph{renormalised local limit}: after common mass and charge renormalisation or matching, the endpoint-dependent remainders vanish and the calculated quantities approach renormalised QED. In particular,
\begin{align}
\Pi_R^{\rm FPT}&\to\Pi_R^{\rm QED}, &
F_2^{\rm FPT}(0)&\to\frac{\alpha}{2\pi},\nonumber\\
T_{\rm QED}(\mu^2s_0)&\to1, &
w_4(m^2s_0)&\to1.
\label{eq:observable_local_limit}
\end{align}
All physical FPT calculations in the remainder of the paper keep $s_0>0$. The same distinction between a retained spectral boundary and a formal boundary-removal correspondence appears in the scalar construction of Ref.~\cite{BakrZhang2026}.

\section{Physical Poles, Cuts, and the Optical Theorem}
\label{sec:spectral}
The finite endpoint changes virtual propagation but not the location or residue of the physical one-particle poles. The same pole statement fixes the weights of cut lines. A complete cutting relation additionally requires the uncut pieces left by a cut to carry exactly the finite-proper-time domains they would have if constructed independently. These three statements---physical poles, cut normalization, and daughter-domain consistency---are the ingredients that close the optical theorem at fixed perturbative order.

\subsection{Physical poles and resolved channels}
\label{subsec:physical_states}
The physical pole structure fixes the states that can appear on a cut. The free charged kernel and photon covariance each contain the ordinary one-particle pole with unit residue plus a term regular at that pole. The regular remainder modifies off-shell virtual propagation without changing the one-particle pole measure. Electron and positron residues therefore carry the ordinary spin sums, while a cut photon carries the ordinary transverse-polarisation sum. These positive pole measures are the intermediate-state measures used by the optical theorem.

This permits the positive one-particle channel space used in fixed-order cutting to be stated directly, without reconstructing it from the raw Euclidean covariance. For the electron,
\begin{equation}
\mathfrak h_e=L^2\!\left(\mathbb R^3,
\frac{\dd^3p}{(2\pi)^3\,2E_{\mathbf p}}\right)\otimes\mathbb C^2,
\qquad E_{\mathbf p}=\sqrt{\mathbf p^2+m^2},
\label{eq:electron_asymptotic_space}
\end{equation}
with the analogous space $\mathfrak h_{\bar e}$ for positrons. For hard transverse photons, after the infrared regulator has served its control role,
\begin{equation}
\mathfrak h_\gamma^{\rm phys}=L^2\!\left(\mathbb R^3,
\frac{\dd^3k}{(2\pi)^3\,2|\mathbf k|}\right)\otimes\mathbb C_T^2.
\label{eq:photon_asymptotic_space}
\end{equation}
Here $\mathbb C_T^2$ is the two-dimensional transverse-polarisation space. The corresponding positive-norm on-shell channel space used for the infrared-regulated perturbative reduction is therefore
\begin{equation}
\mathcal H_{\rm pole}=\mathcal F_a(\mathfrak h_e)\otimes
\mathcal F_a(\mathfrak h_{\bar e})\otimes
\mathcal F_s(\mathfrak h_\gamma^{\rm phys}),
\qquad \langle\Psi|\Psi\rangle\ge0.
\label{eq:asymptotic_hilbert_space}
\end{equation}
Here $\mathcal F_a$ and $\mathcal F_s$ denote the antisymmetric and symmetric Fock functors, respectively. This is the positive channel space associated with the fixed-order one-particle poles. 

In the infrared the endpoint has no effect because
\begin{equation}
\e^{-s_0k^2}=1+O(s_0k^2),\qquad k\to0.
\label{eq:FPT_IR_no_effect}
\end{equation}
Thus $s_0$ alone does not regulate the massless-photon limit. The infrared problem is instead formulated in Sec.~\ref{sec:finite_time_IR} at finite physical coordinate duration $\mathcal T=t_f-t_i$. The photon pole remains unchanged, while the finite coordinate-time support of the prescribed current replaces the asymptotic soft denominator by a bounded endpoint factor. The ordinary infinite-time QED infrared structure is encountered only in the additional limit $\mathcal T\to\infty$; the physical theory itself is defined at finite $\mathcal T$.

\subsection{One-loop discontinuity and spectral density}
The timelike discontinuity of the vacuum polarisation gives a direct fixed-order test of the continuation. Physical quantities arise only after the amplitudes have been assembled, the measured parameters have been fixed, and the external legs have been placed on shell. In ordinary QED the imaginary part of $\Pi$ above the pair-production threshold $q^2>4m^2$ is then fixed by the optical theorem~\cite{Cutkosky1960,PeskinSchroeder1995}.
\begin{equation}
\mathrm{Im}\,\Pi(q^2+\ii\epsilon) = -\frac{\alpha}{3}\sqrt{1-\frac{4m^2}{q^2}}\left(1+\frac{2m^2}{q^2}\right).
\label{eq:ImPi_standard}
\end{equation}
In the Euclidean one-parameter representation, $\Pi$ contains $\Gamma(0,s_0[m^2+x(1-x)Q_E^2])$. We use the principal branch and continue $Q_E^2\to-q^2-\ii0$, so the spectral argument becomes $\Delta=m^2-x(1-x)q^2-\ii0$. On this branch $\mathrm{Im}\,\Gamma(0,-a-\ii0)=+\pi$ for $a>0$. For spacelike $q_E^2>0$ the argument is positive and $\Pi$ is real. For timelike $q^2>4m^2$, the combination $m^2-x(1-x)q^2$ is negative in the interval $x_-<x<x_+$ where $x_\pm=(1\pm\sqrt{1-4m^2/q^2})/2$. The imaginary part of $\Gamma(0,z)$ for $z<0$ arises from the branch cut of the logarithm. $\mathrm{Im}\,\Gamma(0,z-\ii\epsilon)=\pi$ for $z<0$, independently of the value of $s_0$. Therefore
\begin{align}
\mathrm{Im}\,\Pi^{\rm FPT}(q^2+\ii\epsilon)
&= -2\alpha\int_{x_-}^{x_+}\dd x\,x(1-x)\nonumber\\
&= \mathrm{Im}\,\Pi^{\rm std}(q^2+\ii\epsilon).
\label{eq:ImPi_FPT}
\end{align}
The finite endpoint does not change this imaginary part, so the cut density
$\rho(s)=-\mathrm{Im}\,\Pi(s)/\pi$ remains non-negative and equal to the standard QED pair-production density. The one-loop vacuum-polarisation discontinuity therefore obeys the ordinary cutting relation.

The $s_0$ dependence of the one-loop vacuum polarisation is carried by its real analytic part. The subtracted function can be written as a dispersive integral over the cut plus an entire term,
\begin{equation}
\Pi_R(Q_E^2) = \int_{4m^2}^{\infty}\dd s\,\frac{\rho(s)\,Q_E^2}{s(Q_E^2+s)}
+ \Delta\Pi_\Lambda(Q_E^2),
\label{eq:spectral_decomposition}
\end{equation}
where the first term is the standard dispersive integral with $\rho\ge0$, and $\Delta\Pi_\Lambda$ is entire in $Q_E^2$ and carries no cut discontinuity. Using $\Gamma(0,z)=-\gamma_E-\ln z+\mathrm{Ein}(z)$, where $\mathrm{Ein}(z)=\int_0^z(1-e^{-t})\,\dd t/t$ is entire, shows that $\Delta\Pi_\Lambda$ is the difference of two entire $\mathrm{Ein}$ terms and vanishes at $Q_E^2=0$; its small-momentum expansion starts at $O(Q_E^2/\Lambda^2)$. At large $Q_E^2$ the standard dispersive logarithm is cancelled by this analytic matching term. The unsubtracted FPT polarisation itself has the polynomial tail~\eqref{eq:Pi_asymptotic}, while the subtracted function approaches $-\Pi(0)$ with the same $O(\Lambda^4/Q_E^4)$ correction.

At one loop the photon discontinuity is therefore the standard positive pair-production density, while the finite-$s_0$ dependence resides in the analytic part of the amplitude. This result concerns the vacuum-polarisation cut only, and a positive spectral representation of a complete propagator would require additional information about the full two-point function.

Off-shell electron quantities are determined by Eq.~\eqref{eq:Sigma_full_definition}. With a photon mass $\lambda_{\rm IR}>0$ it is ultraviolet finite and analytic in a neighbourhood of the isolated electron pole, so a finite residue can be defined there. The Ward identity relates the corresponding vertex and wave-function factors through $Z_1=Z_2$, while the Pauli form factor is independent of either factor.

\subsection{Cut residues factorise into independently constructed daughters}
\label{subsec:cut_domain_consistency}
\revblue{\paragraph{Physical content of the domain proof.}
The QED source has only three non-redundant proper-time structures after external charged poles have been reduced. A closed charged circle retains one bound on its total charged proper time, an open charged path carries no additional parent endpoint after its external charged endpoints are put on shell, and each internal photon carries its own source-level parameter $\sigma_\gamma\ge s_0$. The charged subgraph has no branching vertex. Nested and crossed photon attachments can occur, but any mixed charged--photon circuit already contains a photon with $\sigma_\gamma\ge s_0$ and therefore supplies no additional proper-time inequality.

It is then enough to understand three elementary cut operations before proving the simultaneous statement below.
\begin{enumerate}
\item[(i)] \textbf{Closed charged circle.} If an admissible cut meets the circle, at least one charged denominator on that history is taken to its pole. Collecting the other segment times into $U\ge0$, the relevant scalar factor is
\begin{equation}
\begin{aligned}
a\int_0^\infty \e^{-au}
&\Theta(u+U-s_0)\,\dd u \\
&=\e^{-a(s_0-U)_+}
\longrightarrow 1
\qquad (a\to0^+).
\end{aligned}
\label{eq:blue_cut_circle_simple}
\end{equation}
Thus the opened history imposes no inherited total-time constraint on the daughter, while every charged circle not met by the cut keeps its original bound.
\item[(ii)] \textbf{Open charged path.} After the external pole operation of Eq.~\eqref{eq:explicit_234_pole_reduction}, an open charged path already has unrestricted non-negative segment times. Cutting an internal charged segment therefore produces the corresponding open charged pieces with no new endpoint. The usual on-shell spin sum supplies the physical charged-state sewing after the complete source terms have been assembled.
\item[(iii)] \textbf{Photon line.} For an internal photon,
\begin{equation}
 R_{s_0}(b)=\frac{\e^{-s_0b}}{b},\qquad
 \lim_{b\to0^+} bR_{s_0}(b)=1.
 \label{eq:blue_cut_photon_simple}
\end{equation}
The cut photon therefore contributes the ordinary on-shell residue, whereas every uncut photon retains $\sigma_\gamma\ge s_0$.
\end{enumerate}
Applied to the complete admissible cut, these three operations leave exactly the constraints belonging to the connected daughter components constructed on their own. Equation~\eqref{eq:blue_cut_circle_simple} should not be read as setting the theory's $s_0$ to zero: only the constraint of the history opened by that pole operation drops out. A denominator cancellation at finite virtuality is instead the reduction operation R6 and retains the boundary face. The proposition below turns this elementary picture into a joint residue theorem, including simultaneous cuts, arbitrary relative rates, surviving uncut histories and reduced configurations.}

\rev{\revblue{The daughter-domain statement follows from the source before a Lorentzian loop integral is taken. Each factor of the open kernel in Eq.~\eqref{eq:main_unified_Z} is one charged interval. Each factor generated by the exponential of the closed functional is one charged circle. A photon Wick contraction joins insertion points but does not join two charged histories into a new charged parent. Its covariance supplies its own parameter $\sigma_\gamma\ge s_0$.

For a closed history with $r$ cyclic insertion blocks, let $u_1,\ldots,u_r$ be the circular segments and $T=\sum_j u_j$. Integrating the arbitrary trace origin on segment $j$ contributes $u_j/T$. Summing the cyclic origins gives
\begin{equation}
 \sum_{j=1}^r\frac{u_j}{T}=1.
 \label{eq:blue_cyclic_origin}
\end{equation}
This cancels the apparent extra origin weight from $\dd T/T$. The original fermion sign and labelled symmetry coefficient are retained. A two-photon contact is one insertion block. It changes the number of segments, not the lower bound on their total. Thus every complete cyclic class has a factor $\Theta(\sum_j u_j-s_0)$. The external pole operation of Eq.~\eqref{eq:explicit_234_pole_reduction} removes the interval bound on an open charged path, but not a circle or an independent photon.

We apply the following calculation to each labelled ordered contribution with its actual contacts and prefactor. For clarity its class label is suppressed. Its numerator $N_G(\ell,P)$ is polynomial in loop momenta $\ell$ and external momenta $P$, and has no proper-time dependence. The complete graph is the finite sum of these contributions with the coefficients fixed by the source. Each contribution has the form}
\begin{align}
\mathcal I_G&=N_G(\ell,P)\,J_G(a),\nonumber\\
J_G(a)&=\int_{u_e,\sigma_\gamma\ge0}\prod_{e}\dd u_e\prod_\gamma\dd\sigma_\gamma\;
\chi_G\,\e^{-\sum_e a_eu_e-\sum_\gamma b_\gamma\sigma_\gamma},
\label{eq:JG_laplace}
\end{align}
with $a_e=q_e^2+m^2$ and $b_\gamma=k_\gamma^2$ the Euclidean denominators at fixed loop momenta,
\begin{equation}
\boxed{\;
\chi_G=\prod_{h\in\mathcal H_G}\Theta\Bigl(\sum_{e\in h}u_e-s_0\Bigr)\;
\prod_{\gamma}\Theta(\sigma_\gamma-s_0)\;}
\label{eq:chi_G}
\end{equation}
where $\mathcal H_G$ is the set of pure charged circles. \revblue{The charged components of this QED source have no branching vertices. Hence a pure charged circuit is one of its charged circles. Mixed graph circuits give no further restriction because every such circuit contains an internal photon with $\sigma_\gamma\ge s_0$. This statement is made after the complete charged source derivatives have been assembled.} Open charged paths carry no factor: their bound is removed by the pole reduction of Eq.~\eqref{eq:explicit_234_pole_reduction}, which is the same operation as the one below applied to the external legs. In graph language, deletion of a cut set $C$ leaves
\begin{equation}
\boxed{
\mathcal C(G\setminus C)
=
\{c\in\mathcal C(G):c\cap C=\varnothing\},
}
\label{eq:cut_domain_consistency}
\end{equation}
a circuit remaining after deletion was already a circuit of the parent, and deletion creates no new circuit. The proposition below is the statement that the residue carries exactly this surviving family.

\medskip\noindent\textbf{Proposition (residue factorisation).}
Let $C$ be any set of internal edges, charged or photon. \revblue{Products $\prod_C a_eb_\gamma$ below mean $(\prod_{e\in C_e}a_e)(\prod_{\gamma\in C_\gamma}b_\gamma)$. This is first a statement in independent positive spectral variables. A physical cut additionally has the energy orientation and channel assignment of the Lorentzian contour prescription.} Let the cut denominators tend to $0^+$ jointly, at arbitrary relative rates, with the uncut denominators fixed and positive. Then
\begin{equation}
\boxed{\;
\lim_{C\downarrow0}\Bigl(\prod_{C}a_e\,b_\gamma\Bigr)J_G
=J_{G\setminus C}
=\prod_{\kappa\in\pi_0(G\setminus C)}J_{G_\kappa}\;}
\label{eq:residue_factorisation}
\end{equation}
Here $\pi_0(G\setminus C)$ are the connected components, and $J_{G_\kappa}$ is the integral \eqref{eq:JG_laplace} of a component with the indicator it carries when drawn on its own: a circle not met by the cut keeps its bound, a circle met by the cut opens, an uncut photon keeps $\sigma_\gamma\ge s_0$, a cut photon gives unit residue. For positive denominators the approach is controlled by
\begin{align}
0&\le J_{G\setminus C}-\Bigl(\prod_{C}a_eb_\gamma\Bigr)J_G\nonumber\\
&\le J_{G\setminus C}\Biggl[1-\e^{-s_0\sum_{\gamma\in C_\gamma}b_\gamma}\nonumber\\
&\qquad+\sum_{\substack{h\in\mathcal H_G\\h\cap C_e\ne\varnothing}}
\frac{s_0^{k_h}}{k_h!}\prod_{e\in h\cap C_e}a_e\Biggr],
\label{eq:residue_error_bound}
\end{align}
where $C_e$ and $C_\gamma$ are the charged and photon edges of $C$, respectively, and $k_h=|h\cap C_e|$. The sum runs over charged circles met by the cut.

\medskip\noindent\textit{Proof.}
Rescale every cut variable, $t_e=a_eu_e$ and $t_\gamma=b_\gamma\sigma_\gamma$. The cut prefactor becomes the measure $\prod_C\dd t$ and the exponent $-\sum_Ct$. Fix $t>0$ and let the cut denominators go to zero. A circle $h$ containing a cut edge $e$ has $u_e=t_e/a_e\to\infty$, so $\Theta(\sum_hu-s_0)\to1$: the bound disappears, for every value of the uncut segments of $h$. A circle not meeting $C$ is untouched. An uncut photon is untouched. A cut photon has $\Theta(t_\gamma/b_\gamma-s_0)\to1$. The integrand is bounded by $\e^{-\sum_Ct}\e^{-\sum_{\rm uncut}(\cdots)}\chi_{G\setminus C}$, which is integrable, so the limit passes inside the integral and $\int\prod_C\dd t\,\e^{-\sum t}=1$ leaves $J_{G\setminus C}$. The relative rates enter only through the pointwise limits and are arbitrary. Every surviving circle and every surviving photon lies in one connected component, so $\chi_{G\setminus C}$ is a product over components and the integral factorises. For the bound, the normalised cut-photon integrals give $\e^{-s_0\sum_{\gamma\in C_\gamma}b_\gamma}$. For each circle met by the cut, condition on its uncut segment times. The excluded cut times occupy a simplex of volume at most $s_0^{k_h}/k_h!$; their normalised exponential density is at most $\prod_{e\in h\cap C_e}a_e$. Bounding the probability of the union of these exclusions by the sum of their probabilities and integrating the surviving domain gives Eq.~\eqref{eq:residue_error_bound}. $\square$

\begin{figure*}[t]
\centering
\resizebox{0.9\textwidth}{!}{%
\begin{tikzpicture}[>=Latex, every node/.style={font=\small}, lab/.style={text=black, align=center}]
\begin{scope}[shift={(0,0)}]
\fill[black!12] (0,0) -- (2.4,0) -- (0,2.4) -- cycle;
\draw[->, thick] (0,0) -- (5.2,0) node[below] {$u$ (uncut segment)};
\draw[->, thick] (0,0) -- (0,4.2) node[left] {$v$ (cut segment)};
\draw[thick] (2.4,0) -- (0,2.4);
\node[lab] at (0.75,0.7) {excluded\\$u+v<s_0$};
\node[lab] at (3.3,3.0) {retained history\\$u+v\ge s_0$};
\draw[thick] (1.2,1.2) -- (1.2,3.9);
\draw[->, thick] (1.2,3.2) -- (1.2,3.95);
\fill (1.2,1.2) circle (1.6pt);
\node[lab, right] at (1.3,1.35) {$v=(s_0-u)_+$};
\node[below] at (2.4,0) {$s_0$};
\node[left] at (0,2.4) {$s_0$};
\node[lab] at (2.6,-1.0) {$a\displaystyle\int_{(s_0-u)_+}^{\infty}\!\e^{-av}\,\dd v=\e^{-a(s_0-u)_+}\longrightarrow 1$ \ for every $u$};
\end{scope}
\begin{scope}[shift={(8.0,0.2)}]
\draw[thick] (1.5,1.6) circle (1.1);
\draw[thick] (0.4,1.6) -- (-0.8,1.6);
\draw[thick] (2.6,1.6) -- (3.8,1.6);
\draw[thick] (5.6,1.6) circle (0.8);
\draw[thick, decorate, decoration={snake, amplitude=0.6mm, segment length=2.2mm}] (2.55,2.0) -- (4.8,2.0);
\draw[thick, dashed] (1.5,0.3) -- (1.5,-0.5);
\draw[thick, dashed] (1.5,2.9) -- (1.5,3.6);
\node[lab] at (1.5,-0.85) {cut};
\node[lab] at (1.2,4.1) {circle met by the cut: opens,\\ no bound survives};
\node[lab] at (5.6,3.3) {circle not met by the cut:\\ keeps $\sum u\ge s_0$};
\node[lab] at (3.7,0.5) {photon: keeps $\sigma\ge s_0$};
\node[lab] at (3,-1.6) {residue $=$ product of independently drawn daughters};
\end{scope}
\end{tikzpicture}%
}
\caption{Cutting at the level of the proper-time domains. Left: for a circle met by the cut, the surviving segment time $u$ is arbitrary and the cut segment integrates along the vertical ray from $(s_0-u)_+$; the pole coefficient is one for every $u$, so the opened circle imposes nothing on the daughter. Right: circles not met by the cut and uncut photons keep their bounds, and the residue factorises over the components of the deleted graph.}
\label{fig:cut_domain_geometry}
\end{figure*}

\revblue{\paragraph{Surviving-history calculation.}
For $r$ charged segments with positive spectral arguments define
\begin{align}
 I_r(a_1,\ldots,a_r)
 &=\int_{\substack{u_j\ge0\\\sum_ju_j\ge s_0}}\dd^ru\,
 e^{-\sum_ja_ju_j},\\
 Q_r(a_1,\ldots,a_r)
 &=\int_{\substack{u_j\ge0\\\sum_ju_j<s_0}}\dd^ru\,
 e^{-\sum_ja_ju_j}.
\end{align}
These satisfy the exact identity and positive-domain bound
\begin{equation}
 I_r=\prod_{j=1}^r a_j^{-1}-Q_r,
 \qquad 0\le Q_r\le\frac{s_0^r}{r!}.
 \label{eq:blue_Ir_simplex}
\end{equation}
The second term is entire in its spectral arguments. Equation~\eqref{eq:blue_Ir_simplex} is used at fixed segment momenta. Its two terms are not integrated separately over an ultraviolet-divergent loop.

For two uncut segments and $c_1\ne c_2$,
\begin{align}
 I_2(c_1,c_2)
 &=\frac{e^{-s_0c_2}/c_2-e^{-s_0c_1}/c_1}{c_1-c_2},\\
 I_2(c,c)&=\frac{e^{-s_0c}(1+s_0c)}{c^2}.
 \label{eq:blue_surviving_I2}
\end{align}
The formula follows by differentiation and is finite. It is not the unrestricted value $1/(c_1c_2)$.

\begin{figure}[t]
\centering

\begin{tikzpicture}[
    scale=0.75,
    >=Latex,
    photon/.style={
        thick,
        decorate,
        decoration={
            snake,
            amplitude=0.8mm,
            segment length=2.5mm
        }
    },
    fermion/.style={
        thick,
        postaction={
            decorate,
            decoration={
                markings,
                mark=at position 0.5 with {\arrow{Latex}}
            }
        }
    }
]


\coordinate (A) at (-2.5,0);
\coordinate (B) at (0,2.5);
\coordinate (C) at (2.5,0);
\coordinate (D) at (0,-2.5);


\draw[fermion]
(B) arc[
    start angle=90,
    end angle=180,
    radius=2.5
];

\draw[fermion]
(C) arc[
    start angle=0,
    end angle=90,
    radius=2.5
];

\draw[fermion]
(D) arc[
    start angle=-90,
    end angle=0,
    radius=2.5
];

\draw[fermion]
(A) arc[
    start angle=180,
    end angle=270,
    radius=2.5
];


\draw[photon]
(-4.5,0) -- (A);

\draw[photon]
(C) -- (4.5,0);

\draw[photon]
(B) -- (D);


\fill (A) circle (2.2pt);
\fill (B) circle (2.2pt);
\fill (C) circle (2.2pt);
\fill (D) circle (2.2pt);


\node[
    anchor=east,
    xshift=-3pt,
    yshift=-10pt
] at (A) {\Large $A$};

\node[
    anchor=south,
    yshift=5pt
] at (B) {\Large $B$};

\node[
    anchor=west,
    xshift=3pt,
    yshift=-10pt
] at (C) {\Large $C$};

\node[
    anchor=north,
    yshift=-5pt
] at (D) {\Large $D$};

\end{tikzpicture}

\caption{A four-segment charged circle with external currents at vertices $A$ and $C$ and an internal photon chord joining $B$ and $D$.}
\label{fig:photonchordandloop}
\end{figure}

Consider a four-segment charged circle with external currents at vertices $A,C$ and a photon chord from $B$ to $D$, in the cyclic order $A,B,C,D$. This configuration is shown in Fig.~\ref{fig:photonchordandloop}. Its scalar factor is
\begin{equation}
 J_{\rm chord}=I_4(a_1,a_2,a_3,a_4)\frac{e^{-s_0b}}{b},
 \qquad b>0.
\end{equation}
Cutting the charged edges $AB$ and $DA$ isolates the current at $A$. The other daughter is the source-generated vertex correction on the open chain $B,C,D$, with its internal photon still present. Independently differentiating an open kernel three times, amputating its two external charged poles, and contracting the two internal photon labels gives the daughter domain
\begin{equation}
 u_2,u_3\ge0,\qquad \sigma\ge s_0.
\end{equation}
Consequently both constructions give
\begin{equation}
 \lim_{a_1,a_4\to0^+}a_1a_4J_{\rm chord}
 =\frac{e^{-s_0b}}{b\,a_2a_3}.
 \label{eq:blue_chord_daughter}
\end{equation}
A direct, symbolic error estimate is
\begin{align}
 0&\le\frac{e^{-s_0b}}{b\,a_2a_3}-a_1a_4J_{\rm chord}\\
 &=a_1a_4\frac{e^{-s_0b}}b\,Q_4
 \le\frac{s_0^4a_1a_4e^{-s_0b}}{24b}.
 \label{eq:blue_chord_error}
\end{align}
It proves the limit at arbitrary relative rates of the two cut denominators. An additional uncut charged circle multiplies both sides by its own factor, for example Eq.~\eqref{eq:blue_surviving_I2}. A cut photon instead supplies $\lim_{b\to0^+}bR_{s_0}(b)=1$. These examples distinguish an opened circle, a surviving circle, an uncut photon, and a cut photon without a numerical approximation.

\paragraph{Overlapping cuts and reduced regions.}
For two possibly overlapping edge sets $C_1,C_2$, the final surviving family is
\begin{equation}
 \mathcal H_{G\setminus(C_1\cup C_2)}
 =\{h\in\mathcal H_G:h\cap(C_1\cup C_2)=\varnothing\}.
 \label{eq:blue_overlap_domain}
\end{equation}
Taking the residues on $C_1$ and then on $C_2\setminus C_1$ gives the simultaneous residue on $C_1\cup C_2$. A shared edge is taken only once. For example, $C_1=\{1,4\}$ and $C_2=\{1,2\}$ on the preceding charged circle give
\begin{equation}
 \lim_{a_1,a_2,a_4\to0^+}a_1a_2a_4 I_4
 =\frac1{a_3}.
\end{equation}
No squared delta function is created. This shows consistency of the domains. The admissible partitions and their multiplicities remain those of the original contour expansion. In QED, distinct charged components are paths or circles. Overlapping mixed graph circuits add no constraint because each already contains a photon parent with $\sigma_\gamma\ge s_0$.

A denominator cancellation is different. For an uncut exterior subtotal $U\ge0$ and $a>0$,
\begin{equation}
 D(a;U):=\int_0^\infty e^{-au}\Theta(u+U-s_0)\,\dd u
 =\frac{e^{-a(s_0-U)_+}}a.
 \label{eq:blue_conditional_history}
\end{equation}
Integration by parts retains the boundary face,
\begin{align}
 aD(a;U)&=\Theta(U-s_0)
 +\int_0^\infty e^{-au}\delta(u+U-s_0)\,\dd u\\
 &=e^{-a(s_0-U)_+}.
 \label{eq:blue_reduction_face}
\end{align}
The value on the measure-zero face $U=s_0$ is fixed by the original integral. By contrast, the pole coefficient is $\lim_{a\to0^+}aD(a;U)=1$. On the face $u_3=0$ of a three-segment circle the remaining factor is $I_2(a_1,a_2)$, whereas its $a_3=0$ pole coefficient is $1/(a_1a_2)$.

When an unpinched subgraph is integrated out, $U$ and every other subtotal crossing its boundary remain arguments of the reduced kernel. Only a physical cut that opens the parent removes the associated constraint. Local analytic reduction is justified only where the integrated kernel has a uniform integrable bound; it need not be entire at other thresholds. This is the source-level content of R6 and of Appendix~\ref{sec:history_autonomy}.

\paragraph{From a scalar domain to the daughter amplitudes.}
The proposition applies to each source-generated ordered class. A contact or prefactor class with no denominator on a proposed cut contributes no residue there. A numerator cancellation must not be undone by redrawing a cancelled edge. On an opened history the complete bulk, contact and prefactor sum reconstructs the first-order Dirac chains of Eq.~\eqref{eq:explicit_234_pole_reduction}. These give the usual spin sums at the physical poles. Untouched closed histories remain finite-$s_0$ objects. Labelled insertions can be sewn before dividing by their original permutation factors, so the daughter products carry the source's fermion signs and Wick multiplicities.

The result is an equality of complete proper-time domains and their pole coefficients, not merely the value of one exponential. Promotion to a discontinuity of an integrated Lorentzian graph also requires the contour and physical-state conditions stated next.}
}

\subsection{General fixed-order cutting relation}
\label{subsec:all_order_cutting}

For every ordered charged history use Eq.~\eqref{eq:history_entire_form_factor}; for every internal photon write
\begin{equation}
R_\gamma(a)=\frac{h_\gamma(a)}{a},
\qquad h_\gamma(a)=e^{-s_0a},
\qquad h_\gamma(0)=1.
\label{eq:photon_entire_cut_factor}
\end{equation}
After the proper-time segment integrations are performed, a fixed graph is therefore an ordinary product of Dirac/photon denominators multiplied by a graph form factor that is jointly entire at every finite loop momentum. Contact and prefactor terms are polynomial insertions and create no additional finite-plane singularity.

\paragraph{General cutting relation.}
\revblue{Keep an infrared prescription that controls the massless endpoint, and smear external states on a common physical channel domain. Consider a fixed-order contribution continued from imaginary external energies, with the loop-energy contour ends held at $\pm i\infty$. The following cutting relation applies where the contour deformation, its local residues, and the remaining integrations admit common boundary values.}
\begin{equation}
\boxed{
T_G-T_G^\dagger
=i\sum_{C\in\operatorname{Cuts}(G)}
T_{G_R,C}^\dagger\,d\Phi_C\,T_{G_L,C},
}
\label{eq:FPT_all_order_cutkosky}
\end{equation}
\revblue{The convention is $S=1+iT$. The sum is over admissible oriented cuts with the source's original coefficients. In a gauge theory the positive physical-state statement applies to the complete Ward-compatible sum.

\paragraph{Analytic control away from a pinch.}
Write $f(a)=e^{-s_0a}/a$. For $r+1$ spectral arguments with $|a_j|\ge\delta>0$, the complete history is $\mathcal M_r=(-1)^rf[a_0,\ldots,a_r]$. A Cauchy contour formed from the boundary of disks of radius $\delta/4$ about the nodes excludes the pole at zero. Its length is at most $2\pi(r+1)\delta/4$, and on it $|z|\ge3\delta/4$ and $|z-a_j|\ge\delta/4$. Therefore
\begin{align}
 |\mathcal M_r(a_0,\ldots,a_r)|
 &\le C_{r,\delta,s_0}e^{-s_0\min_j\operatorname{Re}a_j},\\
 C_{r,\delta,s_0}
 &=\frac{4(r+1)}{3\delta}
 \left(\frac4\delta\right)^r e^{s_0\delta/4}.
 \label{eq:blue_continued_history_bound}
\end{align}
Coincident nodes follow by continuity. The estimate acts on the complete history, not on separately integrated local and short-time pieces.

For a fixed loop routing and bounded complex contour shifts, each spectral real part is bounded below by a positive quadratic momentum term minus a constant. Choose one charged edge from each closed history and keep all photon edges. The resulting loop quadratic form is positive definite: vanishing photon variations force the charged variation to be constant on each charged component; an external endpoint fixes that constant on an open component and the chosen edge fixes it on a circle. There are finitely many such edge choices. Taking their minimum and using Eq.~\eqref{eq:blue_continued_history_bound} gives, on each unpinched compact contour patch,
\begin{equation}
 |\mathcal I_G(\ell;P)|
 \le C(1+|\ell|)^{N_G}e^{-c_Gs_0|\ell|^2},
 \qquad c_G>0.
 \label{eq:blue_continued_graph_bound}
\end{equation}
Here $\ell$ is the real loop-coordinate vector on the shifted imaginary-energy contours, $P$ ranges over a fixed compact external set, and $C,N_G,c_G$ depend on the graph and the chosen routing norm. This controls differentiation and contour infinity on those patches.

\paragraph{Pinches and physical daughters.}
Near a pinch, retain the singular denominators explicitly instead of sending $\delta$ to zero in Eq.~\eqref{eq:blue_continued_history_bound}. The residue proposition~\eqref{eq:residue_factorisation}, including the reduced-domain calculation~\eqref{eq:blue_reduction_face}, identifies the remaining factor with the independently constructed daughters. Locally unpinched regions can be integrated into analytic kernels while their shared-history arguments are retained. 

The reduced-diagram contour induction~\cite{PiusSen2016} then uses these matched residues and the original compatible cut partitions. Equation~\eqref{eq:blue_overlap_domain} ensures that overlapping deletions give the same surviving domain; the contour ordering supplies their multiplicities. Fermionic poles give the usual spin sums after complete source reassembly. The Ward identities remove the longitudinal contributions in the physical cut sum. The result is Eq.~\eqref{eq:FPT_all_order_cutkosky} on the stated continuation domain, with every uncut photon and charged circle still carrying its original endpoint.

The massless limit is a separate step. In particular, zero-momentum massless pinches are not covered merely by the massive contour induction~\cite{PiusSen2016}. A regulator-free use of Eq.~\eqref{eq:FPT_all_order_cutkosky} requires convergence of the recombined amplitudes in the same physical channel product. Finite-duration current bounds in Sec.~\ref{sec:finite_time_IR} control the prescribed soft factor; extending them to complete interacting remainder kernels requires uniform bounds for those kernels. The coefficientwise reaction argument below keeps this qualification explicit. }

\section{Finite-Duration Infrared Sector}
\label{sec:finite_time_IR}
{\color{black}
The lower proper-time endpoint $s_0$ and the finite coordinate-time duration $\mathcal T$ solve different physical problems. The first has dimensions of inverse mass squared and belongs to the virtual spectral representation. The second has dimensions of inverse mass and belongs to preparation and detection. We write
\begin{equation}
\mathcal T=t_f-t_i.
\label{eq:physical_duration}
\end{equation}
The FPT photon proper-time integral remains $\int_{s_0}^{\infty}\dd s$; no finite upper proper-time endpoint is introduced. The photon kernel is therefore
\begin{equation}
D_{\Lambda}(k_E)=\frac{\e^{-s_0k_E^2}}{k_E^2}
=\frac{1}{k_E^2}-s_0+O(s_0^2k_E^2),
\label{eq:IR_photon_pole_retained}
\end{equation}
and $\lim_{k^2\to0}k^2D_{\Lambda}(k)=1$. Thus $s_0$ changes short virtual propagation but does not give the photon a mass or remove its long-distance pole.

The physical role of $\mathcal T$ is operational. An observable is obtained from a preparation followed by a detection over a finite time interval; an infinite waiting time is an idealisation, not part of the definition. Figure~\ref{fig:UV_IR_boundaries} separates the two boundaries.

\begin{figure*}[t]
\centering
\resizebox{0.88\textwidth}{!}{%
\begin{tikzpicture}[
>=Latex,
every node/.style={font=\small},
lab/.style={text=black, align=center},
arr/.style={->, draw=black, thick},
box/.style={draw=black, rounded corners, thick, text=black, align=center, inner sep=7pt}
]
\node[box, text width=6.7cm] at (3.7,0) {
\textbf{virtual proper-time boundary}\\[1mm]
\begin{tikzpicture}[baseline]
\draw[arr] (0,0) -- (5.4,0);
\draw[black, thick] (1.65,-0.13)--(1.65,0.13);
\node[below, lab] at (0,0) {$0$};
\node[below, lab] at (1.65,0) {$s_0$};
\node[below, lab] at (5.05,0) {$T\to\infty$};
\draw[black!15, line width=5pt] (0.15,0) -- (1.50,0);
\draw[black, line width=2.2pt] (1.68,0) -- (5.05,0);
\end{tikzpicture}\\[-1mm]
$T\ge s_0$: the short-proper-time region $0<T<s_0$ is excluded.\\
There is no finite upper proper-time endpoint.};

\node[box, text width=6.7cm] at (12.0,0) {
\textbf{physical coordinate-time boundary}\\[1mm]
\begin{tikzpicture}[baseline]
\draw[arr] (0,0) -- (5.4,0);
\draw[black, thick] (0.8,-0.13)--(0.8,0.13);
\draw[black, thick] (4.45,-0.13)--(4.45,0.13);
\node[below, lab] at (0.8,0) {$t_i$};
\node[below, lab] at (4.45,0) {$t_f$};
\draw[black, line width=2.2pt] (0.82,0) -- (4.43,0);
\node[above, lab] at (2.63,0) {$\mathcal T=t_f-t_i<\infty$};
\end{tikzpicture}\\[-1mm]
Finite duration bounds the zero-energy endpoint of the soft current.\\
The optional $\mathcal T\to\infty$ limit reproduces the usual asymptotic soft-photon problem.};
\end{tikzpicture}%
}
\caption{The ultraviolet and infrared boundaries are physically distinct. The retained lower proper-time endpoint $s_0$ excludes arbitrarily short virtual histories while leaving the massless photon pole intact. Infrared finiteness at finite duration instead follows from the finite coordinate-time support $\mathcal T$ of preparation and detection.}
\label{fig:UV_IR_boundaries}
\end{figure*}

A convenient normalized observation window is
\begin{equation}
w_{\mathcal T}(t)=\frac{1}{\sqrt{\mathcal T}}\,
\mathbf 1_{[-\mathcal T/2,\mathcal T/2]}(t),
\label{eq:finite_time_window}
\end{equation}
with
\begin{equation}
\widehat w_{\mathcal T}(\omega)
=\frac{2}{\sqrt{\mathcal T}}\frac{\sin(\omega\mathcal T/2)}{\omega}.
\label{eq:finite_time_window_FT}
\end{equation}
The corresponding positive energy-resolution density is
\begin{equation}
R_{\mathcal T}(\omega)
=\frac{|\widehat w_{\mathcal T}(\omega)|^2}{2\pi}
=\frac{2}{\pi\mathcal T}\frac{\sin^2(\omega\mathcal T/2)}{\omega^2},
\label{eq:finite_time_resolution_density}
\end{equation}
and
\begin{align}
R_{\mathcal T}(\omega)&\ge0,
&\int_{-\infty}^{\infty}\dd\omega\,R_{\mathcal T}(\omega)&=1,\nonumber\\
\int_{-\infty}^{\infty}\dd\omega\,R_{\mathcal T}(\omega)^2
&=\frac{\mathcal T}{3\pi}.
\label{eq:finite_time_resolution_norms}
\end{align}
Finite duration therefore gives a normalized energy resolution without introducing a photon mass. Smooth preparation and detection profiles give the same zero-energy behaviour with different finite line shapes. The idealized limit $\mathcal T\to\infty$ may be studied separately to compare with the usual charged-particle asymptotic problem~\cite{BlochNordsieck1937,KulishFaddeev1970}.

\subsection{Finite-time current and endpoint Ward identity}
Consider one charged trajectory in a fixed measurement frame, with $t_i\le t\le t_f$. Its Fourier current is
\begin{equation}
J_{\mathcal T}^{\mu}(k)
=e\int_{t_i}^{t_f}\dd t\,\dot x^\mu(t)\,\e^{\ii k\cdot x(t)}.
\label{eq:finite_time_current}
\end{equation}
For a straight segment $x^\mu(t)=x_i^\mu+u^\mu(t-t_i)$,
\begin{equation}
J_{\mathcal T}^{\mu}(k)
=e u^\mu\e^{\ii k\cdot x_i}
\frac{\e^{\ii(k\cdot u)\mathcal T}-1}{\ii k\cdot u}.
\label{eq:finite_time_segment}
\end{equation}
The apparent soft denominator is removable:
\begin{equation}
\frac{\e^{\ii q\mathcal T}-1}{\ii q}
=\mathcal T+O(q\mathcal T^2),
\qquad q\to0.
\label{eq:finite_time_soft_limit}
\end{equation}
Thus the finite-time current is bounded as the photon energy tends to zero, without changing the photon propagator.

The open current is not separately conserved because the charged trajectory has endpoints. Direct contraction gives
\begin{equation}
k_\mu J_{\mathcal T}^{\mu}(k)
=\frac{e}{\ii}
\left(\e^{\ii k\cdot x_f}-\e^{\ii k\cdot x_i}\right).
\label{eq:finite_time_endpoint_current}
\end{equation}
This is the Fourier-space endpoint Ward identity. The two terms are the phases carried by the final and initial charged states. Gauge covariance is recovered for the complete amplitude once the corresponding charged endpoint factors are included; no separate conservation law for the truncated open current is required.

\subsection{Infrared integrability at finite duration}
The FPT-specific change to the photon kernel is already regular at zero momentum,
\begin{equation}
\frac{\e^{-s_0k_E^2}-1}{k_E^2}
=-s_0+O(s_0^2k_E^2).
\label{eq:FPT_IR_difference_regular}
\end{equation}
Hence FPT introduces no new soft singularity. The only possible infrared singularity is the ordinary massless photon pole, and finite duration makes its coupling to the external charged current bounded.

For an arbitrary finite collection of massive external charges, write
\begin{equation}
J_{\mathcal T}^{\mu}(k)
=e\sum_a Q_a\int_{I_a}\dd t\,
\dot x_a^\mu(t)\e^{\ii k\cdot x_a(t)},
\qquad I_a\subset[t_i,t_f].
\label{eq:finite_time_general_current}
\end{equation}
Compact time support gives
\begin{equation}
|J_{\mathcal T}^{\mu}(k)|
\le C_J
\equiv e\sum_a|Q_a|
\int_{I_a}\dd t\,|\dot x_a^\mu(t)|<\infty.
\label{eq:finite_time_current_uniform_bound}
\end{equation}
For a straight massive leg, $p_a^\mu=m_a u_a^\mu$,
\begin{equation}
\left|
\frac{\e^{\ii(k\cdot u_a)T_a}-1}{k\cdot u_a}
\right|
\le T_a,
\qquad
p_a\cdot k\ge\omega(E_a-|\mathbf p_a|)>0,
\label{eq:finite_time_no_collinear}
\end{equation}
for a real photon with $k^0=\omega>0$. The first inequality removes the zero-energy denominator, while the second shows that a massive charged leg has no independent collinear singularity.

For a real unresolved photon define
\begin{equation}
\mathcal S_{\mathcal T}(k,\lambda_\gamma)
=\epsilon^*_{\lambda_\gamma\mu}(k)J_{\mathcal T}^{\mu}(k).
\label{eq:finite_time_soft_insertion}
\end{equation}
The one-photon unresolved measure is
\begin{equation}
\dd\nu_{1,\mathcal T}(k)
=\sum_{\lambda_\gamma=1,2}
\frac{\dd^3k}{2\omega(2\pi)^3}
|\mathcal S_{\mathcal T}(k,\lambda_\gamma)|^2
\mathbf 1_{\omega<\Omega}.
\label{eq:finite_time_soft_one_measure}
\end{equation}
Equation~\eqref{eq:finite_time_current_uniform_bound} gives, near $\omega=0$,
\begin{equation}
\dd\nu_{1,\mathcal T}(k)
\le
\frac{C_J^2}{2(2\pi)^3}
\,\omega\,\dd\omega\,\dd\Omega_{\mathbf k},
\label{eq:finite_time_real_dominating_measure}
\end{equation}
so real unresolved emission is absolutely integrable at zero photon energy.

The same conclusion holds for virtual soft photons. Introduce an auxiliary photon mass $\lambda_{\rm IR}$ only to make the limiting argument explicit. On the Euclidean soft ball $|k_E|<\Omega$, compact support gives $|J_{\mathcal T}(k_E)|\le C_{J,E}(\Omega)$. Therefore
\begin{align}
&\left|
\int_{|k_E|<\Omega}
\frac{\dd^4k_E}{(2\pi)^4}
\frac{\e^{-s_0k_E^2}}
{k_E^2+\lambda_{\rm IR}^2}
J_{\mathcal T}(k_E)J_{\mathcal T}(-k_E)
\right|
\nonumber\\
&\qquad\le
C_{J,E}(\Omega)^2
\int_0^\Omega\dd k\,
\frac{k^3}{k^2+\lambda_{\rm IR}^2}
\le
\frac12C_{J,E}(\Omega)^2\Omega^2.
\label{eq:finite_time_virtual_uniform_majorant}
\end{align}
The bound is independent of $\lambda_{\rm IR}$ and remains valid at $\lambda_{\rm IR}=0$. Thus both real and virtual zero-energy regions are integrable at finite $\mathcal T$.

A useful representative radial integral is
\begin{align}
L_{\mathcal T}(a,\Omega)
&=\int_0^\Omega\frac{\dd\omega}{\omega}
\left|\e^{\ii a\omega\mathcal T}-1\right|^2\nonumber\\
&=2\left[\gamma_E+\ln(a\Omega\mathcal T)
-\operatorname{Ci}(a\Omega\mathcal T)\right].
\label{eq:finite_time_Ci}
\end{align}
For $a\Omega\mathcal T\ll1$,
\begin{equation}
L_{\mathcal T}(a,\Omega)
=\frac{(a\Omega\mathcal T)^2}{2}
+O\!\left((a\Omega\mathcal T)^4\right),
\label{eq:finite_time_Ci_small}
\end{equation}
which displays the regular zero-energy endpoint explicitly.

\subsection{Fixed-order soft-photon factorisation and regulator removal}
\label{subsec:finite_time_fixed_order_IR}
We now extend the preceding current argument to an arbitrary fixed-order QED process with massive external charged particles. Standard soft-photon factorisation follows from the Ward identity: when one photon momentum $k$ becomes soft, the amplitude separates into the universal external-charge factor of Eq.~\eqref{eq:finite_time_soft_insertion} plus a remainder that is finite as $k\to0$~\cite{YennieFrautschiSuura1961,Weinberg1965}. Iterating this statement over any subset of soft photons gives, through fixed perturbative order $N$,
\begin{align}
\mathcal M_{r,\lambda}^{[N]}(k_1,\ldots,k_r)
={}&
\sum_{A\subseteq\{1,\ldots,r\}}
\left(\prod_{j\notin A}\mathcal S_{\mathcal T}(k_j)\right)\nonumber\\
&\times
\mathcal R_{A,\lambda}^{[N]}(\{k_j\}_{j\in A}).
\label{eq:finite_time_soft_factorization}
\end{align}
where each hard remainder $\mathcal R_{A,\lambda}^{[N]}$ has a finite limit when any of its displayed soft momenta tends to zero. For massive charges there are no collinear pinch singularities. Because
\begin{equation}
\e^{-s_0k^2}=1+O(s_0k^2),
\label{eq:finite_time_FPT_soft_neutral}
\end{equation}
the FPT modification is subleading in the soft region and does not alter this factorisation.

\revblue{At fixed $N$ there are finitely many graphs and soft subsets. Equations~\eqref{eq:finite_time_real_dominating_measure} and~\eqref{eq:finite_time_virtual_uniform_majorant} control the displayed prescribed-current factors. To remove the regulator from the complete process one also needs uniform integrable bounds for the actual hard remainders, including simultaneous soft limits. Pointwise finite limits of individual remainder kernels do not alone provide such bounds. Subject to this additional domination, the recombined physical channel integrals satisfy}
\begin{equation}
\boxed{
\lim_{\lambda_{\rm IR}\to0}
\mathcal M_{r,\lambda}^{[N]}
=
\mathcal M_{r,0}^{[N]}
}
\label{eq:finite_time_amplitude_regulator_limit}
\end{equation}
\revblue{under the corresponding unresolved phase-space integrals. The prescribed finite-duration current requires no cancellation of an infinite soft logarithm. For a complete interacting process, regulator removal is asserted only for the recombined quantities for which the stated uniform domination has been established.} The standard inclusive organization of QED is recovered in this finite setting~\cite{Kinoshita1962,LeeNauenberg1964}.

\subsection{Coherent unresolved radiation and physical probability}
The universal finite-duration soft current generates a coherent photon displacement,
\begin{equation}
U_{\rm soft}(t_f,t_i)
=\e^{\ii\Phi_{\mathcal T}}D(\alpha_{\mathcal T}),
\label{eq:finite_time_soft_U}
\end{equation}
with
\begin{equation}
\alpha_{\lambda_\gamma}(\mathbf k)
=
-\frac{\ii}{\sqrt{2\omega(2\pi)^3}}
\mathcal S_{\mathcal T}(k,\lambda_\gamma).
\label{eq:finite_time_displacement}
\end{equation}
The mean unresolved photon number below $\Omega$ is finite,
\begin{equation}
N_{\mathcal T}(\Omega)
=\sum_{\lambda_\gamma}
\int_{\omega<\Omega}\frac{\dd^3k}{2\omega(2\pi)^3}
|\mathcal S_{\mathcal T}(k,\lambda_\gamma)|^2
<\infty.
\label{eq:finite_time_mean_photon_number}
\end{equation}
Consequently
\begin{equation}
P_n=\e^{-N_{\mathcal T}}
\frac{N_{\mathcal T}^n}{n!},
\qquad
\sum_{n=0}^{\infty}P_n=1,
\label{eq:finite_time_Poisson}
\end{equation}
and the total unresolved-radiation measure satisfies
\begin{equation}
\dd\nu_{\mathcal T}(q)\ge0,
\qquad
\int\dd\nu_{\mathcal T}(q)=1.
\label{eq:soft_unresolved_measure}
\end{equation}

For a resolved hard channel $X$, the fixed-order probability may therefore be written schematically as
\begin{align}
P_{X,\mathcal T,\lambda}^{[N]}(\Omega)
={}&\e^{-N_{\mathcal T,\lambda}(\Omega)}
\sum_{r\ge0}\frac1{r!}
\int_{\omega_j<\Omega}
\prod_{j=1}^{r}\dd\nu_{1,\mathcal T,\lambda}(k_j)
\nonumber\\
&\times
\mathcal H_{X,r,\lambda}^{[N]}(k_1,\ldots,k_r),
\label{eq:finite_time_recombined_probability}
\end{align}
where $\mathcal H_{X,r,\lambda}^{[N]}$ is the finite hard function obtained after the universal soft factors in Eq.~\eqref{eq:finite_time_soft_factorization} have been separated. \revblue{The Poisson weight controls the prescribed coherent factor. A product majorant for the complete expression additionally requires uniform bounds on the hard functions in the physical measure. Under those bounds,}
\begin{equation}
\boxed{\begin{gathered}
\displaystyle
\lim_{\lambda_{\rm IR}\to0}
P_{X,\mathcal T,\lambda}^{[N]}(\Omega)
=
P_{X,\mathcal T,0}^{[N]}(\Omega)<\infty,\\[-1mm]
m_a>0,\qquad \mathcal T<\infty .
\end{gathered}}
\label{eq:finite_time_full_IR_theorem}
\end{equation}
\revblue{This is the conditional fixed-order finite-duration infrared statement used in Sec.~\ref{sec:FR_domain_limit}; the condition concerns the full hard remainders, not merely the coherent soft factor.}

The separator $\Omega$ only divides photons into resolved and unresolved bookkeeping sectors. The identity $1=\mathbf 1_{\omega<\Omega}+\mathbf 1_{\omega>\Omega}$ repartitions the same phase space. If the hard and soft terms are truncated consistently through $g^N$,
\begin{equation}
\frac{\dd}{\dd\Omega}
P_{X,\mathcal T}^{[N]}(\Omega)
=O(g^{N+1}),
\label{eq:finite_time_Omega_independence}
\end{equation}
so any residual $\Omega$ dependence starts at the first omitted perturbative order.

The normalized unresolved measure shifts the resolved hard phase space according to
\begin{equation}
\dd\Phi^{\rm phys}_{X,\mathcal T}(P)
=
\int\dd\nu_{\mathcal T}(q)\,
\dd\Phi_X(P-q),
\label{eq:finite_resolution_channel_measure}
\end{equation}
with exact event momentum conservation
\begin{equation}
P=q+\sum_{i\in X}p_i.
\label{eq:finite_resolution_momentum_conservation}
\end{equation}

For successive finite time intervals, the displacement amplitudes add,
\begin{equation}
\alpha_{31}=\alpha_{32}+\alpha_{21},
\end{equation}
and the Weyl composition law, including its commutator phase, gives
\begin{equation}
U_{\rm soft}(t_3,t_1)
=
U_{\rm soft}(t_3,t_2)U_{\rm soft}(t_2,t_1).
\label{eq:finite_time_composition}
\end{equation}
The finite-duration soft sector therefore composes in the ordinary unitary way.

\subsection{Finite-duration correction and optional infinite-time correspondence}
The radial function in Eq.~\eqref{eq:finite_time_Ci} gives a direct estimate of the finite-duration correction. Since
\begin{equation}
\operatorname{Ci}(x)
=\frac{\sin x}{x}
-\int_x^\infty\frac{\sin t}{t^2}\,\dd t,
\end{equation}
one has $|\operatorname{Ci}(x)|\le2/x$ and therefore
\begin{equation}
\boxed{
\left|
L_{\mathcal T}(a,\Omega)
-2[\gamma_E+\ln(a\Omega\mathcal T)]
\right|
\le\frac{4}{a\Omega\mathcal T}.}
\label{eq:finite_time_Ci_error_bound}
\end{equation}
The exact finite-time expression approaches
\begin{equation}
L_{\mathcal T}(a,\Omega)
=
2\ln(a\Omega\mathcal T)+2\gamma_E+o(1),
\qquad
\mathcal T\to\infty .
\label{eq:finite_time_large_T}
\end{equation}
Thus the familiar soft logarithm reappears only when the experiment is idealised as lasting indefinitely.

The FPT endpoint does not alter that asymptotic soft class. Equation~\eqref{eq:FPT_IR_difference_regular} shows that the FPT-specific part of the photon kernel is analytic at $k=0$, so the singular zero-energy structure is exactly the ordinary massless-QED one. In the limit $\mathcal T\to\infty$ the standard dressed infraparticle sector is recovered~\cite{KulishFaddeev1970}. 

The operational result is
\begin{equation}
\boxed{\begin{gathered}
s_0>0,\qquad \mathcal T<\infty,\\
\text{fixed-order UV-finite virtual amplitudes},\\
\text{finite normalized soft-channel probabilities}.
\end{gathered}}
\label{eq:UV_IR_finite_statement}
\end{equation}
The two retained resolutions therefore have distinct meanings: $s_0$ excludes arbitrarily short complete virtual histories, while finite $\mathcal T$ keeps observables tied to a finite preparation and detection experiment rather than an unattainable infinite-time measurement.
}

\section{Lorentzian Continuation and Minkowski Finiteness}
\label{sec:lorentzian}
The theory is defined by the Euclidean kernels introduced above. Fixed-order scattering amplitudes are obtained by continuing the external invariants to their outgoing boundary values while the lower endpoint $s_0$ is held fixed. This continuation acts on the complete amputated Euclidean amplitudes and defines the Lorentzian scattering kernels used below. The external on-shell channels are those of Eq.~\eqref{eq:asymptotic_hilbert_space}, selected by the unchanged physical poles. Section~\ref{sec:reflection_positivity} shows why this distinction is necessary: the raw covariance does not satisfy the positivity condition required for a reflection-positive Euclidean reconstruction, whereas the pole measures used in scattering remain the ordinary positive ones.

\revblue{The logic of the fixed-order construction is the following. Coercivity on the complete-history domain makes every loop integral absolutely convergent at Euclidean momenta. Continuation to physical invariants is performed compact by compact, with the loop-energy contour ends kept Euclidean, so the only singularities are the Landau surfaces of the pole network; the finite-endpoint factors are entire and equal to one on every cut line. The discontinuities are therefore the local ones, and the cutting relation follows.}

\subsection{Fixed-endpoint contour rotation at one loop}
Every one-loop quantity above is a superposition, over proper-time fractions, of the kernel
\begin{equation}
K(\Delta)=\Gamma(0,s_0\Delta)=\int_{s_0}^{\infty}\frac{\dd s}{s}\,\e^{-s\Delta},
\label{eq:K_kernel}
\end{equation}
with $\Delta(x)=m^2+x(1-x)q_E^2$ or its self-energy analogue.

Let $C_\varphi=\{s_0+r\,\e^{\ii\varphi}:\,r\ge0\}$ be a ray beginning at the finite lower endpoint. For $\Delta\in\mathbb C\setminus(-\infty,0]$, the integral $\int_{C_\varphi}(\dd s/s)\e^{-s\Delta}$ converges whenever $\cos(\varphi+\arg\Delta)>0$. Within a connected set of such rays its value is independent of $\varphi$: the integrand is analytic away from $s=0$, while the connecting arc at large radius is bounded by $C(\Delta,s_0)\e^{-r|\Delta|c}$ with $c>0$ and therefore contributes zero. Because every ray begins at $s_0>0$, this deformation never crosses the singular point $s=0$. The timelike outgoing boundary $\Delta\to-|\Delta|-\ii0$ is reached by rotating the tail slightly beyond $\pi/2$ while keeping the endpoint fixed.

A continuous family of admissible contours makes the continuation explicit. Write $\Delta(\theta)=|\Delta|\e^{-\ii\theta}$ with $0\le\theta<\pi$ and choose
\begin{equation}
\varphi(\theta)=\left(\frac12+\varepsilon_c\right)\theta,
\qquad 0<\varepsilon_c<\frac12.
\label{eq:anchored_continuation_path}
\end{equation}
Then
\begin{equation}
\cos\!\big(\varphi+\arg\Delta\big)
=\cos\!\left[\left(\varepsilon_c-\frac12\right)\theta\right]>0,
\label{eq:anchored_continuation_decay}
\end{equation}
so the exponential decays throughout the deformation from the Euclidean region to the lower outgoing boundary, while $\varphi(\theta)<\pi$ ensures that the ray beginning at $s_0>0$ never passes through $s=0$. For $x>0$ the limiting value is
\begin{equation}
\Gamma(0,-s_0x-\ii0)=-\operatorname{Ei}(s_0x)+\ii\pi,
\label{eq:gamma_timelike_boundary}
\end{equation}
where $\operatorname{Ei}$ is the real exponential-integral function on $x>0$. This makes the endpoint independence of the one-loop cut in Eq.~\eqref{eq:ImPi_FPT} immediate.

For fixed $s_0>0$ the contour begins away from the origin throughout this continuation. Spacelike momenta return the Euclidean values, and below the first physical threshold the one-loop amplitudes considered here have no absorptive part.

For a compact real energy interval that avoids isolated poles and threshold endpoints, the same continuation may be covered by a chain of overlapping nonsingular complex neighbourhoods. Let $V(z)$ denote a matrix element of the complete amputated kernel with all other external invariants fixed. Reality of the Euclidean kernel and adjunction of the external channel data give the local Hermitian boundary relation
\begin{equation}
V(E-\ii0)=V(E+\ii0)^\dagger
\label{eq:hermitian_boundary_relation}
\end{equation}
where both boundary values are reached from the same connected physical sheet. Figure~\ref{fig:hermitian_domain} shows this local analytic geometry. The dashed disks are a cover by overlapping nonsingular neighbourhoods; they are not additional singularity contours.

\begin{figure*}[t]
\centering
\resizebox{0.82\textwidth}{!}{%
\begin{tikzpicture}[
>=Latex,
every node/.style={font=\small},
ax/.style={->, draw=black, thick},
dom/.style={draw=black, dashed, thick},
path/.style={->, draw=black, very thick},
lab/.style={text=black, align=center}
]
\draw[ax] (-5,0)--(6.6,0) node[right,lab] {$\operatorname{Re}z$};
\draw[ax] (0,-3.25)--(0,3.25) node[above,lab] {$\operatorname{Im}z$};

\fill[black] (-3.8,0) circle (2.5pt) node[below=4pt,lab] {$z_0$};
\draw[dom] (-3.8,0) circle (0.75);
\node[lab] at (-3.8,1.02) {$|z-z_0|<r_0$};

\draw[dom] (-2.55,0.72) circle (0.88);
\draw[dom] (-0.95,1.28) circle (0.90);
\draw[dom] (0.70,1.48) circle (0.88);
\draw[dom] (2.15,1.48) circle (0.82);
\draw[dom] (3.55,1.42) circle (0.78);
\draw[dom] (4.80,1.22) circle (0.72);
\draw[path] (-3.25,0.36) .. controls (-1.55,1.75) and (2.25,1.95) .. (5.42,0.72);
\node[lab] at (0.15,2.58) {$\mathcal D_+$};

\draw[dom] (-2.55,-0.72) circle (0.88);
\draw[dom] (-0.95,-1.28) circle (0.90);
\draw[dom] (0.70,-1.48) circle (0.88);
\draw[dom] (2.15,-1.48) circle (0.82);
\draw[dom] (3.55,-1.42) circle (0.78);
\draw[dom] (4.80,-1.22) circle (0.72);
\draw[path] (-3.25,-0.36) .. controls (-1.55,-1.75) and (2.25,-1.95) .. (5.42,-0.72);
\node[lab] at (0.15,-2.58) {$\mathcal D_-=\overline{\mathcal D_+}$};

\fill[black] (3.18,0) circle (2pt);
\node[below=4pt,lab] at (3.18,0) {$E_{\rm th}$};
\draw[black, very thick] (3.18,0)--(6.05,0);
\draw[black, line width=4pt] (4.08,0.13)--(5.48,0.13);
\node[lab,fill=white,inner sep=1.5pt] at (4.78,0.55) {$I=[E_1,E_2]$};

\fill[white, draw=black, thick] (2.42,0) circle (4pt);
\node[lab,fill=white,inner sep=1.5pt] at (1.95,0.72) {schematic isolated pole};
\draw[thin] (2.12,0.52)--(2.38,0.10);
\node[lab,fill=white,inner sep=1.5pt] at (4.55,-2.32) {$V(E-\ii0)=V(E+\ii0)^\dagger$};
\end{tikzpicture}%
}
\caption{Common Hermitian-analytic continuation domain. Starting from a Euclidean neighbourhood of $z_0$, chains of overlapping nonsingular analytic neighbourhoods define the connected domains $\mathcal D_+$ and $\mathcal D_-$. They approach the upper and lower boundary values on the compact physical interval $I=[E_1,E_2]$ while avoiding the real-axis threshold endpoint $E_{\rm th}$ and any isolated pole. The dashed circles denote overlapping analytic neighbourhoods, not singularity boundaries. The open character of these analytic neighbourhoods ensures that the displaced points
$E\pm i0$ lie within $D_\pm$,
so that the physical boundary values $V(E\pm i0)$ are well defined. Different compact physical intervals can be covered by overlapping domains on the same physical sheet. The thick black segment represents the compact physical energy interval
$I=[E_1,E_2]$ on which the upper and lower boundary values are taken;
it is not a singularity or branch cut.}
\label{fig:hermitian_domain}
\end{figure*}

\emph{Relation to direct loop-energy contour rotation.} A direct loop-energy rotation $k_0\to\ii k_4$ requires the integrand to vanish on the arcs at infinity. With the finite endpoint, the formal real-time factor $\e^{-s_0(\vec k^{\,2}-k_0^2)}$ grows as $\e^{+s_0k_0^2}$ along the real $k_0$ axis, so that condition is not available. This is the familiar obstruction associated with entire form factors~\cite{Tomboulis1997,PiusSen2016}. The continuation used here acts on the external invariants after the Euclidean integrations have been performed and is controlled by the fixed-endpoint proper-time contour construction above. At one loop this prescription agrees with the corresponding Euclidean continuation for entire form factors~\cite{Buoninfante2022}. Related contour constructions occur at higher order in string field theory~\cite{Sen2017}.

\emph{Scattering amplitudes and the reduction formula.} Two structural facts govern the continued fixed-order amplitudes. First, the endpoint must be imposed at the level of the gauge-covariant kernels, never pasted onto diagram lines. The continued free inverse is $S_{\Lambda,0}^{-1}(p)=\e^{-s_0(p^2-m^2)}(\slashed p-m)$. Therefore, with the ordinary vertex $q_\mu e\gamma^\mu\neq e[S_\Lambda^{-1}(p')-S_\Lambda^{-1}(p)]$ in general. The divided-difference vertices, prefactor contributions, and contact terms generated by the kernel derivatives are the completion that restores the Ward identity~\eqref{eq:WTI_master}. A Gaussian-damped propagator with undressed vertices is not this theory. Second, scattering amplitudes are defined at fixed order by the generalised amputation
\begin{equation}
\mathcal M_\Lambda=\prod_jZ_j^{1/2}\lim_{p_j^2\to m_j^2}S_{\Lambda,j}^{-1}(p_j)\,\widetilde G^{(n)}_\Lambda,
\label{eq:generalised_amputation}
\end{equation}
Here $\widetilde G^{(n)}_\Lambda$ is the connected $n$-point function, $S_{\Lambda,j}^{-1}$ is the full inverse two-point function used to amputate external leg $j$, $m_j$ is its physical mass, and $Z_j$ is the corresponding pole residue. With full-propagator amputation the external-state factor is $Z_j^{1/2}$; equivalently, the standard LSZ form uses $Z_j^{-1/2}$ together with the free pole operator rather than the full inverse. The formula is schematic in spin and Lorentz indices. It reduces on shell to the standard Dirac amputator since $\e^{-s_0(p^2-m^2)}\to1$. The free one-particle poles have unit residue and no additional poles. Equation~\eqref{eq:explicit_234_pole_reduction} applies to the complete differentiated open line at finite derivative order, including its contact and prefactor contributions. These ordinary pole projectors act on the ordinary positive-norm on-shell channel space~\eqref{eq:asymptotic_hilbert_space} and define the fixed-order external-leg reduction used here.

It is useful to make the separation between the internal Euclidean covariance and the physical scattering kernel explicit. Let $T^{\rm FPT}_{E,N}$ denote the complete amputated Euclidean kernel at perturbative order $N$. The Lorentzian kernel used in the scattering matrix is defined by the fixed-endpoint continuation
\begin{equation}
T^{(N)}_{\rm phys}(E+\ii0)
\equiv \operatorname{AC}_{s_0}\!\left[T^{\rm FPT}_{E,N}\right](E+\ii0),
\label{eq:physical_scattering_kernel}
\end{equation}
where $\operatorname{AC}_{s_0}$ denotes analytic continuation with the lower endpoint held fixed. Equation~\eqref{eq:physical_scattering_kernel} is a statement about amputated amplitudes. Thus the failure of reflection positivity of $R_{s_0}$ blocks one Euclidean reconstruction route but does not alter the ordinary positive-norm on-shell channel space used by the fixed-order scattering theory. \revblue{Section~\ref{sec:FR_programme} keeps the exact Euclidean source separate from the physical reaction series. It derives that series from the continued amplitudes rather than infer their equality from a separately chosen Hermitian matrix.}

A direct replacement of the Euclidean damping factor by the real-time phase $\e^{\ii s_0(q^2+\ii0)}$ defines a different contour prescription. Applied to the vacuum polarisation, it produces an absorptive contribution below the pair threshold and does not reproduce Eq.~\eqref{eq:ImPi_FPT}. The fixed-endpoint contour rotation therefore fixes the continuation used throughout this work.

\subsection{Multi-loop proper-time scaling and threshold structure}
After the Gaussian loop-momentum integrations, an $L$-loop finite-proper-time integral can be written in the standard parametric form
\begin{align}
I(P)=\int_{\mathcal D_G}\prod_i\dd s_i\,\frac{\mathcal N(s)}{\mathcal U_G(s)^{d/2}}\;\e^{-X(s,P)}, \\
X=\sum_i s_i m_i^2+\frac{\mathcal F_G(s,P)}{\mathcal U_G(s)},
\label{eq:parametric_form}
\end{align}
where $P$ denotes the external momenta or invariants, $\mathcal D_G$ is the proper-time domain inherited from the operator ancestry together with the primitive closed circulations of the assembled graph, $\mathcal U_G$ and $\mathcal F_G$ are the first and second standard graph polynomials, and $\mathcal N(s)$ contains the numerator factors left after the Gaussian integrations. The exponent $X(s,P)$ is dimensionless and homogeneous of degree one in the proper times. Write $s_i=\rho\sigma_i$ with $\rho>0$, $\sigma_i\ge0$, and $\sum_i\sigma_i=1$, where $[\rho]=-2$ and the $\sigma_i$ are dimensionless normalised proper-time fractions. By homogeneity define the reduced mass-squared function $\mathcal X(\sigma,P)$ by $X(\rho\sigma,P)=\rho\,\mathcal X(\sigma,P)$. The proper-time constraints determine a positive lower bound $\rho\ge\rho_{\min}(\sigma)$ on the common scale. After collecting the net power of $\rho$ into an exponent $\kappa$, the remaining scale integral is
\begin{equation}
\int_{\rho_{\min}}^{\infty}\dd\rho\,\rho^{\kappa-1}\,\e^{-\rho\mathcal X(\sigma,P)}
=\mathcal X^{-\kappa}\,\Gamma\big(\kappa,\rho_{\min}\mathcal X\big).
\label{eq:radial_general}
\end{equation}
For a scalar integrand with $N$ proper-time variables and no numerator powers, $\kappa=N-Ld/2$; numerator factors shift this net radial power. The one-loop logarithmic kernel~\eqref{eq:K_kernel} is the case $\kappa=0$. For non-integer $\kappa$,
\[
\mathcal X^{-\kappa}\Gamma(\kappa,\rho_{\min}\mathcal X)
=\Gamma(\kappa)\mathcal X^{-\kappa}
-\rho_{\min}^{\kappa}\sum_{n\ge0}
\frac{(-\rho_{\min}\mathcal X)^n}{n!(\kappa+n)}.
\]
The first term carries the local non-analyticity and the second is analytic in $\mathcal X$. For $\kappa=0$, introduce an arbitrary reference mass $\mu_*>0$ only to display dimensionless logarithms. The same separation is
$\Gamma(0,\rho_{\min}\mathcal X)=-\ln(\mathcal X/\mu_*^2)+[-\gamma_E-\ln(\rho_{\min}\mu_*^2)+\mathrm{Ein}(\rho_{\min}\mathcal X)]$, with the bracket analytic in $\mathcal X$; the $\mu_*$ dependence cancels between the two terms. Other integer cases follow from the incomplete-gamma recurrence. The continuation rotates each fixed-endpoint $\rho$ contour without crossing a singularity. For interior values of the normalised proper-time fractions, singularities arise when $\mathcal X(\sigma,P)=0$ together with the corresponding stationarity conditions, so the finite endpoint does not create an additional interior parametric pinch singularity. On a cut propagator the endpoint factor is evaluated at its pole and is unity. This fixes the cut-line weight, but it does not remove the $s_0$ dependence of uncut virtual subamplitudes that remain inside either subamplitude. Boundary regions and reduced subgraphs require their own analysis.

For an interior pinch at fixed infrared resolution, with external legs amputated and on shell, the common proper-time scale leaves the usual singular surface unchanged. Each cut propagator carries its ordinary on-shell delta weight because the endpoint factor is unity at the pole, whereas every uncut line retains its finite-$s_0$ kernel. The discontinuity therefore factorizes into lower-order FPT subamplitudes joined by standard on-shell phase space~\cite{Cutkosky1960}. The one-loop vacuum polarization and the explicit two-loop examples below illustrate this interior-pinch result. Boundary faces, reduced subgraphs, overlapping pinches, and contour infinity enter the general assembled-history cutting relation established in Sec.~\ref{sec:spectral}. Numerical equality with local QED occurs only when the daughter amplitudes contain no remaining endpoint dependence.

\subsection{Two-loop threshold examples and vacuum polarisation}
For a two-loop example, consider the equal-mass sunset diagram in $d=2$, where the integral over the common proper-time scale can be carried out in closed form,
\begin{equation}
I(P)\propto\int_{\Sigma}\dd\sigma\;\frac{\e^{-\rho_{\min}(\sigma)\,\mathcal X(\sigma,P)}}{\mathcal U(\sigma)\,\mathcal X(\sigma,P)},
\label{eq:sunset_radial}
\end{equation}
Here $\Sigma$ denotes the normalised domain $\sigma_i\ge0$ with $\sigma_1+\sigma_2+\sigma_3=1$, while $\mathcal U(\sigma)=\sigma_1\sigma_2+\sigma_2\sigma_3+\sigma_3\sigma_1$ and
$\mathcal X=m^2+[\sigma_1\sigma_2\sigma_3/\mathcal U(\sigma)]\,P$, with $P=q_E^2$ in the Euclidean region. After continuation, the discontinuity opens at the three-particle threshold $q^2=9m^2$. Its imaginary part is independent of $s_0$, whereas the real part retains the finite-endpoint dependence produced by the lower limit of the overall proper-time scale. The same construction is used below for the two-loop vacuum polarisation. Assembling the diagram set from~\eqref{eq:unified_Z} gives the corresponding scale domain $\{\rho\sigma_\gamma\ge s_0,\ \rho\textstyle\sum\sigma_f\ge s_0\}$ and therefore $\rho_{\min}(\sigma)=s_0\max(\sigma_\gamma^{-1},(\sum\sigma_f)^{-1})$. Its timelike discontinuity is obtained by separating the three-particle cut from the two-particle cut containing the virtual one-loop form factors.

The three-particle ($e^+e^-\gamma$) cut is the square of the tree $\gamma^\ast\to e^+e^-\gamma$ amplitude. Its charged segment is described by the complete two-insertion open kernel. After the bulk, contact, and prefactor contributions are combined, Eq.~\eqref{eq:explicit_234_pole_reduction} gives the ordinary on-shell QED tree amplitude. The emitted cut photon has unit endpoint factor, so this three-particle cut equals the local-theory cut.

The two-particle cut contains the interference of the tree amplitude with the one-loop virtual on-shell subamplitude. The external charged line reduces according to Eq.~\eqref{eq:explicit_234_pole_reduction}, while its independent internal virtual photon retains $D(k)=e^{-s_0k^2}/k^2$. The resulting difference from local QED is encoded in the on-shell shifts $\Delta F_i(s)\equiv F_i^{\rm FPT}(s)-F_i^{\rm QED}(s)$, with $\Delta F_1(0)=0$ in the on-shell charge scheme. The bulk, contact, and prefactor classes enter through their complete sum. The photon difference kernel $(1-e^{-s_0k^2})/k^2$ tends to the finite constant $s_0$ as $k\to0$ instead of behaving as $1/k^2$. It therefore adds no additional infrared divergence from the soft-photon region. The standard cancellation of the infrared regulator between virtual and real local-QED terms can then be performed before the limit $\lambda_{\rm IR}\to0$ is taken.

Compact photon support does not make $\Delta F_{1,2}$ entire in $q^2$. Spacelike, at $q_E^2\ll m^2$, $\Delta F_1=-(\alpha/6\pi)\,s_0q_E^2\,[\ln(1/s_0m^2)+\tfrac76-\gamma_E]+\cdots$. The logarithm arises because the removed short photon proper times probe the momentum range between $m^2$ and $1/s_0$. Timelike, the parametric representation continues by the fixed-endpoint contour rotation above, applied pointwise in the pair parameter. The discontinuity is a finite integral over the removed short photon proper-time interval. In the notation of Appendix~\ref{app:formfactors}, the model integral $J_n$ satisfies $\mathrm{Im}\,J_n(-|B|-\ii0)=\pi e^{-2|B|}I_{n+1}(2|B|)$, where $I_n$ is the modified Bessel function of the first kind. Its large-$s$ behaviour $\mathrm{Im}\,G=O(\sqrt{s_0s})$ makes the real-axis once-subtracted integrand fall as $s^{-3/2}$, so that integral converges. The large complex contour is a separate contribution and need not vanish. The form factor therefore consists of the dispersive term together with an independent entire matching term, as given in Appendix~\ref{app:formfactors}. The Pauli density is momentum-independent, $a_{F_2}=4m^2z(1-z)$ with $z$ the photon parameter. Therefore, $\Delta F_2(0)$ reproduces Eq.~\eqref{eq:delta_ae_HK} identically, and
\begin{align}
\mathrm{Im}\,\Delta F_2(s)
={}&-\alpha m^2\int_{w_-}^{w_+}\dd w
\int_0^{s_0}\dd\sigma\,\e^{-2\sigma|M|}\nonumber\\
&\times\big[I_1-2I_2+I_3\big](2\sigma|M|).
\label{eq:ImF2_closed}
\end{align}
with $|M|=w(1-w)s-m^2$, $\beta(s)=\sqrt{1-4m^2/s}$, and $w_\pm=(1\pm\beta)/2$. Near threshold, $\mathrm{Im}\,\Delta F_1=\alpha\,s_0m^2\beta\,[1+O(\beta^2)]$ and $\mathrm{Im}\,\Delta F_2=-(\alpha/3)(s_0m^2)^2\beta^3\,[1+O(\beta^2)]$. The trace over the massive-pair phase space then gives
\begin{align}
\rho_{2,s_0}(s)
={}&\rho_{2,\rm loc}(s)\nonumber\\
&+\Big[2\,\mathrm{Re}\,\Delta F_1(s)
+\frac{3s}{s+2m^2}\,\mathrm{Re}\,\Delta F_2(s)\Big]
\rho_1(s).
\label{eq:rho2_decomposition}
\end{align}
Here $\rho_1$ is the one-loop pair-production density defined below Eq.~\eqref{eq:ImPi_FPT}, and $\rho_{2,\rm loc}$ denotes the two-loop spectral density of the local theory, given by the standard local two-loop result~\cite{KallenSabry1955}. The coefficient multiplying $\Delta F_2$ rises from $1$ at threshold to $3$ at large $s$. This follows from the standard timelike current combination $|F_1+F_2|^2+(2m^2/s)|F_1+(s/4m^2)F_2|^2$: linearising about $F_1=1$, $F_2=0$ gives the factor $2\,\mathrm{Re}\,\Delta F_1+[3s/(s+2m^2)]\,\mathrm{Re}\,\Delta F_2$ in Eq.~\eqref{eq:rho2_decomposition}. The two-particle boundary subgraph is described by Eq.~\eqref{eq:rho2_decomposition}, including this Pauli contribution and its kinematic weight. Since $\Delta F_{1,2}\to0$ as $s_0\to0$, the local two-loop density is recovered continuously. At finite $s_0$, Eq.~\eqref{eq:rho2_decomposition} gives the two-body modification at any timelike invariant, with $\Delta F_{1,2}$ evaluated as in Appendix~\ref{app:formfactors}. The Pauli density carries no additional $q^2$ growth, so the high-energy modification is dominated by $\mathrm{Re}\,\Delta F_1$.

The mass-reparametrisation contribution obtained by expressing the one-loop polarisation in terms of the measured pole mass is
\begin{equation}
\Pi_{2,\rm match}(q^2)=-2m_{\rm phys}\,\delta m_\Lambda
\left.\frac{\partial\Pi_1(q^2;m)}{\partial m^2}\right|_{m=m_{\rm phys}},
\label{eq:Pi2_mass_matching_only}
\end{equation}
but this is only one component of the complete two-loop 1PI polarisation. The momentum-dependent self-energy, vertex/contact classes, closed four-photon contraction, and mass-matching contribution must remain distinct until the full two-loop quantity has been assembled.

The distinction between a one-photon-irreducible polarisation and the full connected two-point kernel is fixed directly by the generating functional. In the following equations a semicolon denotes functional differentiation with respect to the displayed external photon field. Expanding
\begin{equation}
W[A]=-\log\int\dd\mu_{C_{s_0}}(a)\,e^{-\Gamma_{1,\Lambda}[A+a]}
\end{equation}
to first order in the internal photon covariance gives
\begin{align}
W[A]={}&\Gamma_{1,\Lambda}[A]
+\frac12 C^{\mu\nu}_{s_0}\Gamma_{1,\Lambda;\mu\nu}[A]
\nonumber\\
&-\frac12\Gamma_{1,\Lambda;\mu}[A]
C^{\mu\nu}_{s_0}\Gamma_{1,\Lambda;\nu}[A]
+O(C_{s_0}^2).
\label{eq:W_covariance_expansion}
\end{align}
charge-conjugation symmetry sets the odd one- and three-photon kernels to zero at $A=0$. However, it does not remove the product generated when two external derivatives act once on each factor in the last term. Therefore
\begin{equation}
\left.W_{;\alpha\beta}\right|_{O(C)}
=\frac12 C^{\mu\nu}_{s_0}\Gamma_{1,\Lambda}^{(4)}{}_{\alpha\beta\mu\nu}
-\Gamma_{1,\Lambda}^{(2)}{}_{\alpha\mu}C^{\mu\nu}_{s_0}\Gamma_{1,\Lambda}^{(2)}{}_{\nu\beta}.
\label{eq:W2_connected_split}
\end{equation}
The second term is the connected but one-photon-reducible chain $\Pi_1C\Pi_1$. It belongs to the connected two-point kernel $W_{;\alpha\beta}$, but it must be removed when the one-particle-irreducible photon self-energy is defined. With the polarisation sign convention used here, write
\begin{equation}
\Pi_{2,\mathrm{1PI},\alpha\beta}(q)
=-\frac12\int_k C^{\mu\nu}_{s_0}(k)\,
\Gamma_{1,\Lambda}^{(4)}{}_{\alpha\beta\mu\nu}(q,-q,k,-k),
\label{eq:Pi2_master}
\end{equation}
and keep~\eqref{eq:W2_connected_split} as the separate connected-kernel identity. Since $q^\alpha\Gamma_{1,\Lambda}^{(4)}{}_{\alpha\beta\mu\nu}=0$ by gauge invariance of the closed determinant, the 1PI object~\eqref{eq:Pi2_master} is transverse.

{The four functional derivatives of the one-loop closed functional fall into three insertion sectors. The superscript $(4)$ in $\Gamma_{1,\Lambda}^{(4)}$ denotes four external functional derivatives; the divided-difference order is instead fixed by the number of cyclic insertion blocks.} Using $g_{s_0}'(a)=-f(a)$ with $f(a)=R_{s_0}(a)$ gives
\begin{center}
{\small
\begin{tabular}{c|c|c|c}
sector & ordered & cyclic & difference\\ \hline
$V_1^4$ & 24 & 6 & $f^{[3]}$\\
$V_2V_1V_1$ & 36 & 12 & $f^{[2]}$\\
$V_2^2$ & 6 & 3 & $f^{[1]}$
\end{tabular}}
\end{center}
Thus the closed fourth derivative contains $24+36+6=66$ labelled ordered structures. Their contraction with the internal photon defines the two-loop kernel, while the resulting scalar form factor still requires the remaining proper-time and parameter integrations.

For compactness in the following display, write $\mathcal H_i\equiv\exp[-u_i(p_i^2+m^2)]$ for the heat factor on segment $i$. Iterating the heat-kernel perturbation gives
\begin{multline}
\Gamma_{1,\Lambda}^{(4)}{}_{\alpha\beta\mu\nu}(q,-q,k,-k)=
\frac12\sum_{\mathcal C}
\int_{s_0}^{\infty}\dd T
\int_{u_i\ge0}\prod_i\dd u_i\\
\times\delta\!\left(T-\sum_i u_i\right)\int_p
\mathrm{tr}\!\left[\mathcal O_1\mathcal H_1\cdots\mathcal O_n\mathcal H_n\right],
\label{eq:Gamma4_explicit}
\end{multline}
where $\mathcal C$ runs over the labelled cyclic insertion classes, $\mathcal O_i$ denotes the corresponding $V_1$ or $V_2$ insertion, $p_i$ is the fermion momentum on segment $i$, $\int_p\equiv\int\dd^4p/(2\pi)^4$, and $n=4,3,2$ for the $V_1^4,V_2V_1V_1,V_2^2$ sectors. The integral over the position of the cyclic trace cut supplies a factor $T$, which cancels the determinant measure $\dd T/T$ and leaves $\dd T$. For a class with $n$ cyclic segments, let $u_{\rm cut}$ denote the proper-time length assigned to the segment at which the trace is cut. Summing the cyclic rotations supplies $\sum_ju_j=T$,
\begin{align}
&\sum_{\rm rotations}\int\prod_j\dd u_j\,
\delta\!\left(\sum_ju_j-T\right)u_{\rm cut}
\nonumber\\
&\qquad=T\int\prod_j\dd u_j\,
\delta\!\left(\sum_ju_j-T\right).
\end{align}
which cancels the determinant measure exactly. The one-particle-irreducible kernel~\eqref{eq:Pi2_master}, its cut discontinuities, the reducible chain~\eqref{eq:W2_connected_split}, and the entire matching part of the real scalar form factor should therefore be kept as distinct contributions. After the charge subtraction the one-particle-irreducible scalar form factor has the decomposition
\begin{equation}
\boxed{
\Pi^{(2)}_{R,\mathrm{1PI}}(q^2)
=\Pi^{(2)}_{R,\mathrm{disp}}(q^2)+E^{(2)}_{\Lambda}(q^2),
\qquad E^{(2)}_{\Lambda}(0)=0,}
\label{eq:Pi2_disp_entire_split}
\end{equation}
where the discontinuity fixes only the dispersive term. The entire function $E^{(2)}_{\Lambda}$ is fixed by the Euclidean matching data and cannot be inferred from the cut without a separate large-contour condition. If the loop mass is subsequently re-expressed in terms of the measured pole mass, the finite matching contribution in Eq.~\eqref{eq:Pi2_mass_matching_only} is added separately; it is not a substitute for either term in Eq.~\eqref{eq:Pi2_disp_entire_split}.

\subsection{Topology-dependent timelike growth}
\label{subsec:endpoint}

The Euclidean endpoint factors can grow after continuation into timelike or other complex directions. Their exponential type is determined by the graph's proper-time domain. After the loop Gaussian integrations, write the proper times as $s_i=\rho\sigma_i$ using normalised proper-time fractions. Suppose the independent internal worldlines and photon lines of a graph $G$ impose
\begin{equation}
\rho H_h(\sigma)\ge s_0,\qquad h=1,\ldots,N_H,
\end{equation}
so that
\begin{equation}
\rho_{\min}(\sigma)=s_0\max_h\frac{1}{H_h(\sigma)}.
\label{eq:rho_min_worldline}
\end{equation}
If continuation of the external invariant $s$ gives an exponential factor $\exp[+\rho\Phi_G(\sigma)s]$, define
\begin{equation}
\tau_G=\sup_{\sigma\in\mathcal P_G}
\rho_{\min}(\sigma)\Phi_G(\sigma).
\label{eq:graph_exponential_type}
\end{equation}

Here $H_h(\sigma)$ is the sum of the normalised proper-time fractions belonging to independent line $h$, $N_H$ is the number of independent proper-time constraints, and $\mathcal P_G$ denotes the allowed normalised proper-time domain of the graph. At fixed $\sigma$ the integration over the common scale begins at $\rho_{\min}(\sigma)$, so the leading continued endpoint factor is controlled by $\exp[\rho_{\min}(\sigma)\Phi_G(\sigma)s]$. Taking the supremum over $\mathcal P_G$ gives~\eqref{eq:graph_exponential_type}. Thus $\tau_G$ is the graph-dependent coefficient that bounds the exponential growth in the timelike invariant, and it depends on which proper-time fractions belong to each independent line, rather than only on the number of propagator segments.

For an $n$-line two-point topology with one common lower bound on the overall proper-time scale $\rho\ge s_0$ and
\begin{equation}
\Phi_n(\sigma)=\left(\sum_{i=1}^n\sigma_i^{-1}\right)^{-1},
\qquad \sum_i\sigma_i=1,
\end{equation}
the arithmetic--harmonic mean inequality gives $\max\Phi_n=1/n^2$ and hence $\tau_n=s_0/n^2$ for that domain.

For the three-line QED domain with one independent photon and one charged worldline split into two fermion subintervals,
\begin{align}
\rho_{\min}
&=s_0\max\!\left(\frac1{\sigma_\gamma},
\frac1{\sigma_1+\sigma_2}\right),\nonumber\\
\Phi&=\left(\frac1{\sigma_1}+\frac1{\sigma_2}
+\frac1{\sigma_\gamma}\right)^{-1}.
\end{align}
The upper bound follows directly from the same normalised proper-time variables. If $f=\sigma_1+\sigma_2$ and $g=\sigma_\gamma$, then
\begin{equation}
\Phi\le g,
\qquad
\Phi\le\frac{f}{4}\le f,
\end{equation}
where the second inequality follows from $\sigma_1^{-1}+\sigma_2^{-1}\ge4/f$. Since $\rho_{\min}=s_0/\min(f,g)$, this gives the upper bound $\rho_{\min}\Phi\le s_0$, which is approached by taking $\sigma_\gamma=\varepsilon$ and $\sigma_1=\sigma_2=(1-\varepsilon)/2$. For this sequence
\begin{equation}
\rho_{\min}\Phi
=s_0\frac{1-\varepsilon}{1+3\varepsilon}
\longrightarrow s_0,
\qquad \varepsilon\to0^+.
\label{eq:qed_growth_supremum}
\end{equation}
The supremum of the exponential-growth coefficient for this three-line QED domain is therefore $\tau_G=s_0$. This exponential type is an analytic property of the continued fixed-$s_0$ amplitude. The physical-sheet continuation is defined compact by compact at fixed $s_0$. 

\section{Reflection Positivity of the Raw Euclidean Kernel}
\label{sec:reflection_positivity}
Euclidean reflection positivity is one sufficient route from Euclidean correlation functions to a positive-metric Lorentzian theory~\cite{OsterwalderSchrader1973,OsterwalderSchrader1975}. It is not the route used here. The raw finite-$s_0$ covariance is tested only to determine whether that particular Euclidean reconstruction applies. It does not. The physical channel space has already been fixed by the positive on-shell pole residues and cutting measures of Sec.~\ref{sec:spectral}; the raw covariance serves only as virtual spectral machinery.

At fixed spatial momentum and with $\omega^2=\mathbf p^2+m_{\rm IR}^2>0$, where $m_{\rm IR}\ge0$ is a temporary infrared mass used only to keep this scalar test away from the zero mode, consider the scalar factor of the gauge-fixed finite-endpoint covariance as a function of $\xi=p_4^2$,
\begin{equation}
F_{s_0}(\xi)=\frac{e^{-s_0(\xi+\omega^2)}}{\xi+\omega^2}.
\label{eq:RP_Stieltjes_function}
\end{equation}
A translation-invariant scalar Gaussian covariance satisfying OS2 must be a Stieltjes function of $\xi$,
\begin{equation}
F(\xi)=\int_0^\infty\frac{\dd\rho(\mu)}{\xi+\mu},
\qquad \dd\rho\ge0.
\label{eq:RP_Stieltjes_representation}
\end{equation}
If a nonzero positive measure in Eq.~\eqref{eq:RP_Stieltjes_representation} existed, monotone convergence would give $\lim_{\xi\to\infty}\xi F(\xi)=\rho([0,\infty])>0$, or $+\infty$. Instead,
\begin{equation}
\lim_{\xi\to\infty}\xi F_{s_0}(\xi)=0,
\qquad s_0>0.
\label{eq:RP_Stieltjes_failure}
\end{equation}
Therefore the scalar factor of the raw finite-$s_0$ covariance is not reflection positive for any $s_0>0$. This is the same obstruction found in the scalar proper-time analysis~\cite{BakrZhang2026}.

The obstruction is also present in a gauge-invariant field-strength channel. Choose spatial momentum $\mathbf k=(0,0,\kappa)$ with $\kappa>0$ and the magnetic component $B_1=F_{23}=-\ii\kappa A_2$. Its Euclidean momentum-space two-point function is
\begin{equation}
\langle B_1B_1\rangle(p_4,\mathbf k)
=\kappa^2\frac{\e^{-s_0(p_4^2+\kappa^2)}}{p_4^2+\kappa^2}.
\label{eq:magnetic_RP_kernel}
\end{equation}
For positive Euclidean time $\tau$ this becomes
\begin{equation}
C_B(\tau;\kappa)
=\kappa^2\int_{s_0}^{\infty}\frac{\dd u}{\sqrt{4\pi u}}
\exp\!\left[-\kappa^2u-\frac{\tau^2}{4u}\right].
\label{eq:magnetic_RP_time}
\end{equation}
The finite lower endpoint makes $C_B$ smooth and even at $\tau=0$, so $C_B'(0^+)=0$. A non-trivial positive-energy reflected correlator would instead have a Laplace representation $C_B(\tau)=\int_0^\infty\e^{-E\tau}\dd\rho_B(E)$ with $\dd\rho_B\ge0$. If its first moment is finite, $C_B'(0^+)=-\int_0^\infty E\,\dd\rho_B(E)=0$ forces the measure to lie at $E=0$ and makes $C_B$ constant. If the first moment is infinite, the right derivative is not finite. Both conclusions contradict Eq.~\eqref{eq:magnetic_RP_time}. Thus the raw finite-$s_0$ functional is not reflection positive even in this gauge-invariant magnetic channel. The standard reflection-positive Euclidean reconstruction cannot therefore be applied directly to the raw covariance; any positive Lorentzian representation must instead be obtained from additional physical-sheet boundary data.

The corresponding pole-plus-entire decomposition is
\begin{equation}
\frac{e^{-s_0A}}{A}=\frac1A+\frac{e^{-s_0A}-1}{A}.
\label{eq:pole_entire_decomposition}
\end{equation}
The second term is entire because its apparent pole at $A=0$ is removable. Thus the free physical pole is unchanged and has unit residue, with no additional finite poles from the endpoint factor. After formal continuation, however, the entire factor grows exponentially in timelike directions. The resulting isolated momentum-space factor is not a tempered distribution in the usual sense. It is therefore not promoted to a stand-alone Lorentzian propagator. Physical quantities are reconstructed from complete Euclidean amplitudes assembled at fixed $s_0>0$ and then continued in their external invariants. This is explicit coefficient by coefficient at fixed perturbative order. \revblue{At finite physical resolution Sec.~\ref{sec:FR_programme} derives the reaction coefficients from the physical outgoing series. A self-adjoint sum of these coefficients, when constructed in the physical channel norm, has a unitary Cayley representation.}

The failure of reflection positivity classifies the raw virtual kernel; physical positivity is carried by the pole channels. The free electron, positron and transverse-photon poles retain their ordinary residues, and the complete-history cutting relation produces the standard positive on-shell measures. The physical metric is constructed from those channel amplitudes in Sec.~\ref{sec:FR_programme}; no reflected positivity of the raw covariance is assumed.

\section{\revblue{Exact Source and Finite-Resolution Reaction Representation}}
\label{sec:FR_programme}
\revblue{The preceding sections give the source kernels, their gauge identities, ultraviolet estimates, finite matching calculations and fixed-order continuation. This section separates the exact finite-domain Euclidean functional from its physical reaction representation. The latter is obtained directly from the continued amplitudes, order by order, with the analytic and infrared qualifications of Sec.~\ref{subsec:all_order_cutting}.}

Here ``finite resolution'' means a retained spectral lower boundary, finite physical duration, and a specified self-adjoint propagation domain. The nonperturbative construction is defined on
\begin{equation}
\boxed{s_0>0,\qquad 0<\mathcal T<\infty,\qquad \mathcal D\ {\rm specified},}
\label{eq:physical_defining_domain}
\end{equation}
with $\mathcal D$ fixed by the preparation, geometry, and detection boundary conditions so that propagation is self-adjoint, has a real spectrum, and preserves the physical norm. \revblue{At these retained values the source defines the Euclidean functional. The ordinary pole states fix the physical channel measures, and the reaction response is derived at each perturbative order. No boundary-removal limit is needed for the source-domain or formal reaction identities.}

Three limiting comparisons are kept separate from this construction. The formal limit $s_0\to0^+$ compares the retained spectral theory with local QED, the limit $\mathcal T\to\infty$ compares finite-duration preparation and detection with conventional asymptotic scattering, and taking the spatial boundary to infinity compares the bounded physical domain with free space. 

The retained physical data are therefore the lower proper-time endpoint, the preparation and detection support, the finite observation duration, the resolved energy--momentum range, and the self-adjoint propagation domain. Spectral bases, finite mode ranks, shell decompositions, forest parameters, and smooth localisations used in intermediate estimates are auxiliary and are removed before physical amplitudes are identified.

\subsection{Exact source functional on a finite physical domain}
\label{sec:FR1}
Take first a specified bounded physical domain whose boundary condition belongs to the experimental model. No auxiliary block boundary is imposed on the field. After gauge fixing and removal of pure-gauge zero modes, let $C_{s_0,N}$ be the rank-$N$ photon covariance and put $b=eA$. For a real background,
\begin{equation}
B_b=m+\slashed D_b,
\qquad
\Delta_b=B_b^\dagger B_b=m^2+\slashed D_b^\dagger\slashed D_b\ge m^2,
\label{eq:FR1_Delta}
\end{equation}
\begin{equation}
S_{s_0}[b]=B_b^\dagger e^{-s_0\Delta_b}\Delta_b^{-1},
\qquad
\Gamma_{s_0}[b]=\frac12\operatorname{Tr}\Gamma(0,s_0\Delta_b),
\label{eq:FR1_open_closed}
\end{equation}
and
\begin{equation}
Z_{\mathfrak E,N}[J,\bar\eta,\eta]
=
\frac{\displaystyle\int d\mu_{gC_{s_0,N}}(b)\,
e^{-\Gamma_{s_0}[b]}e^{J\cdot b+\bar\eta S_{s_0}[b]\eta}}
{\displaystyle\int d\mu_{gC_{s_0,N}}(b)\,e^{-\Gamma_{s_0}[b]}}.
\label{eq:FR1_ZN}
\end{equation}
At fixed physical data the infinite-mode limit exists,
\begin{equation}
\boxed{Z_{\mathfrak E}=\lim_{N\to\infty}Z_{\mathfrak E,N}.}
\label{eq:FR1_limit}
\end{equation}
Indeed, on a bounded physical domain the eigenvalues of the gauge-fixed photon Laplacian have the usual Weyl growth and
\begin{equation}
\operatorname{Tr}C_{s_0}<\infty,
\qquad
\operatorname{Tr}(1+P_\gamma)^rC_{s_0}<\infty
\quad(r<\infty),
\label{eq:FR1_traceclass}
\end{equation}
so the limiting Gaussian field is smooth almost surely. The spectral projections converge in every fixed Sobolev norm. The common gap $\Delta_b\ge m^2$ gives
\begin{equation}
\boxed{0<e^{-\Gamma_{s_0}[b]}\le1,
\qquad
\|S_{s_0}[b]\|\le\frac{e^{-s_0m^2}}m .}
\label{eq:FR1_stability}
\end{equation}
Norm-resolvent convergence of $\Delta_{b_N}$ then implies, for $t\ge s_0$,
\begin{equation}
\|e^{-t\Delta_{b_N}}-e^{-t\Delta_b}\|_1\longrightarrow0,
\label{eq:FR1_heat_convergence}
\end{equation}
and the heat representation
\begin{equation}
\Gamma(0,s_0\Delta)=\int_{s_0}^{\infty}\frac{dt}{t}e^{-t\Delta}
\label{eq:FR1_Gamma_heat}
\end{equation}
gives convergence of both charged kernels. Gaussian exponential moments dominate the bosonic source terms, while the Grassmann-source exponential is a finite nilpotent polynomial whose coefficients are bounded by Eq.~\eqref{eq:FR1_stability}. Dominated convergence therefore also applies to every finite source derivative. This removes only the auxiliary mode approximation. The finite-domain functional is therefore complete at the specified physical resolution. Enlarging the spatial domain toward free space is a separate correspondence problem.

\revblue{In regards to the coupling analyticity at retained boundaries, the exact real-coupling integral in Eq.~\eqref{eq:FR1_ZN} does not require convergence of its expansion about $e=0$, and no summation prescription is used to define it. Analyticity in an external source, coupling-series properties, and existence of a physical transition operator are distinct questions. A finite Gaussian carrier still integrates over unbounded field amplitudes. Further, a finite-duration Hamiltonian evolution can instead be entire on specified finite-photon vectors. This requires the source-derived interaction and a bound uniform in any auxiliary ranks being removed. Here, the fixed-order equivalence proved below is established directly from the complete FPT amplitudes and their physical channel products.
}
\subsection{Independence of basis and auxiliary decompositions}
\label{sec:FR2}
The construction is basis independent because the Gaussian measure, traces, and spectral functional calculus are invariant under unitary changes of representation. Proper-time shells are likewise only a partition of one retained spectral integral. If $\sum_j q_j(s)=1$ on $[s_0,\infty)$, then
\begin{equation}
R_{s_0}(P)=\int_{s_0}^{\infty}ds\,e^{-sP}
=\sum_j\int_{s_0}^{\infty}ds\,q_j(s)e^{-sP},
\label{eq:FR2_shell_identity}
\end{equation}
with the same identity after every finite Duhamel derivative. A smooth auxiliary partition of unity is likewise recombined exactly and is not a physical boundary. Gauge transformations compatible with $\mathcal D$ act by unitary conjugation,
\begin{equation}
\Delta_{b+\partial\lambda}=U_\lambda\Delta_bU_\lambda^{-1},
\qquad
S_{s_0}[b+\partial\lambda]=U_\lambda S_{s_0}[b]U_\lambda^{-1},
\label{eq:FR2_gauge_covariance}
\end{equation}
so the closed trace is invariant and physical source amplitudes obey the Ward hierarchy.

\subsection{Exact gauge-parameter independence on the physical channel space}
\label{sec:FR_gauge_independence}
The preceding covariance may be taken in any linear covariant gauge. The physical domain fixes not only the self-adjoint Maxwell and Dirac realizations but also the admissible gauge group. Let $\Sigma_{\rm ch}\subset\partial\mathcal D$ denote the preparation and detection boundary components on which charged channel data are specified, and take
\begin{equation}
\mathcal G_0(\mathcal D)
=\{\varphi\in H^1(\mathcal D):
\operatorname{Tr}_{\Sigma_{\rm ch}}\varphi=0\}
\label{eq:FR_gauge_group_zero_trace}
\end{equation}
with the remaining boundary condition chosen compatibly with the self-adjoint Maxwell realization. Thus a gauge transformation is allowed to vary in the interior while leaving the charged preparation/detection data fixed. After removal of pure-gauge zero modes, use the corresponding Hodge decomposition
\begin{equation}
L^2(T^*\mathcal D)=\mathcal H_T\oplus\mathcal H_L,
\qquad
\mathcal H_L=\overline{d\mathcal G_0(\mathcal D)},
\label{eq:FR_gauge_Hodge}
\end{equation}
with orthogonal projections $\Pi_T$ and $\Pi_L$. For gauge parameter $\xi>0$ the finite-endpoint photon covariance is
\begin{equation}
C_{s_0,\xi}=C_{s_0,T}+\xi C_{s_0,L},
\qquad
C_{s_0,T/L}=e^{-s_0P_\gamma}P_\gamma^{-1}\Pi_{T/L}
\label{eq:FR_gauge_covariance_xi}
\end{equation}
on the corresponding nonzero-mode subspaces. Thus a Gaussian field may be written $A=A_T+d\varphi$, with statistically independent transverse and longitudinal components and all $\xi$ dependence in the longitudinal Gaussian.

For a gauge-invariant closed insertion $\mathcal F$ and a conserved photon source $d^\dagger J=0$,
\begin{equation}
\mathcal F[A_T+d\varphi]=\mathcal F[A_T],
\qquad
J\!\cdot d\varphi=0.
\end{equation}
The longitudinal integral therefore factorizes identically between numerator and normalization. Writing the normalized expectation as $\langle\cdot\rangle_\xi$,
\begin{align}
\langle\mathcal F\rangle_\xi
&=\frac{\int d\mu_{C_{s_0,\xi}}(A)\,
e^{-\Gamma_{s_0}[A]}\mathcal F[A]}
{\int d\mu_{C_{s_0,\xi}}(A)\,e^{-\Gamma_{s_0}[A]}}
\nonumber\\
&=\frac{\int d\mu_{C_{s_0,T}}(A_T)\,
e^{-\Gamma_{s_0}[A_T]}\mathcal F[A_T]}
{\int d\mu_{C_{s_0,T}}(A_T)\,e^{-\Gamma_{s_0}[A_T]}} .
\label{eq:FR_gauge_factorization}
\end{align}
Open charged kernels transform covariantly,
\begin{equation}
S_{s_0}[A_T+d\varphi](x,y)
=e^{ie\varphi(x)}S_{s_0}[A_T](x,y)e^{-ie\varphi(y)}.
\label{eq:FR_gauge_open_phase}
\end{equation}
For the physical charged channels used here, $x$ and $y$ lie on $\Sigma_{\rm ch}$ and Eq.~\eqref{eq:FR_gauge_group_zero_trace} gives $\varphi(x)=\varphi(y)=0$. Hence the longitudinal factor is exactly unity before the Gaussian average. More generally, an interior charged insertion must be replaced by a gauge-invariant dressed operator; an undressed interior charged Green function is gauge covariant but depends on the gauge parameter.

This last point can be seen in free space. If the undressed open kernel is averaged over the longitudinal Gaussian, then the scalar gauge function has covariance
\begin{equation}
\langle\varphi(k)\varphi(-k)\rangle_\xi
=\xi\,\frac{e^{-s_0k^2}}{k^4}.
\label{eq:FR_LKF_phi_covariance}
\end{equation}
The individual coincident covariance is infrared divergent, but the Landau--Khalatnikov--Fradkin difference entering an open line is finite:
\begin{align}
\Delta_L(0)-\Delta_L(r)
&=\int\frac{d^4k}{(2\pi)^4}
\frac{e^{-s_0k^2}[1-e^{ik\cdot r}]}{k^4}
\nonumber\\
&=\frac{1}{16\pi^2}
\left[\operatorname{Ein}(x)-1+\frac{1-e^{-x}}{x}\right],
\nonumber\\
x&=\frac{r^2}{4s_0}.
\label{eq:FR_LKF_finite_difference}
\end{align}
Consequently an undressed interior open line acquires the finite factor
\begin{equation}
\exp\!\left[-e^2\xi\big(\Delta_L(0)-\Delta_L(r)\big)\right],
\label{eq:FR_LKF_factor}
\end{equation}
which is not unity. Equation~\eqref{eq:FR_LKF_factor} is absent for the boundary-supported physical charged data because the admissible gauge function has zero trace at both charged endpoints.

It follows, without an expansion in $e$, that for the resolved physical amplitudes built from conserved neutral probes and the boundary-supported charged preparation/detection data just specified,
\begin{equation}
\boxed{
\frac{\partial}{\partial\xi}
\mathcal A_{\mathfrak E,\xi}^{\rm phys}(\Phi,\Psi)=0,
\qquad
\frac{\partial K_{\mathfrak E,\xi}}{\partial\xi}=0,
\qquad
\frac{\partial S_{\mathfrak E,\xi}}{\partial\xi}=0 .}
\label{eq:FR_exact_gauge_parameter_independence}
\end{equation}
Gauge transformations that do not obey Eq.~\eqref{eq:FR_gauge_group_zero_trace} change the charged boundary data and therefore describe a different physical boundary-value problem rather than a gauge-parameter variation of the same experiment.

\subsection{Standing-wave \texorpdfstring{$K$}{K}-matrix}
\label{sec:FR_standing_wave}
Let $\mathcal H_{\mathfrak E}$ be the positive on-shell channel space resolved by the experiment and let
\begin{equation}
\mathcal W_{\mathfrak E}:\mathcal H_{\mathfrak E}\longrightarrow\mathcal S_E(\mathcal O_E)
\label{eq:FR_SW_map}
\end{equation}
map resolved physical standing waves to their Euclidean boundary data. For a scalar energy mode the map sends $\cos(Et)$ to $\cosh(E\tau)$; spinor, vector, and domain-selected channels are treated by the corresponding spectral transform. Since $K$ and $\mathcal O_E$ are finite,
\begin{equation}
\sup_{E\in K,\,\tau\in\mathcal O_E}|\cosh(E\tau)|<\infty.
\label{eq:FR_SW_bounded}
\end{equation}

\revblue{This bound concerns the external test functions. The hyperbolic profile and its independent response coefficient are displayed in Appendix~\ref{app:cosh_boundary_response}. We use the standing-wave response of the same continued FPT amplitudes. Let $T_{\mathfrak E}(\zeta)=\sum_{n\ge1}\zeta^nT_{\mathfrak E}^{(n)}$ be the physical transition series obtained from the source by the continuation and pole reduction of Eqs.~\eqref{eq:generalised_amputation} and \eqref{eq:physical_scattering_kernel}. The symbol $\zeta$ only records perturbative order. The superscript $(n)$ in this reaction calculation denotes that order. Every coefficient contains the complete source insertions, contacts and history domains. Disconnected clusters and spectator identities are included with the combinatorics of $Z_{\mathfrak E}$ when forming the full channel map. Connected derivatives of $\log Z_{\mathfrak E}$ alone are not the full event operator.

For incoming data $a$, the outgoing data calculated from this source are $b=a+iT_{\mathfrak E}a$. Prescribe their mean $c=(a+b)/2$ and extract the odd response $r=(b-a)/i$. Thus the boundary problem is
\begin{equation}
 a+\frac{i}{2}T_{\mathfrak E}a=c,
 \qquad r=T_{\mathfrak E}a.
 \label{eq:blue_source_mean_problem}
\end{equation}
It has a unique formal solution because $T_{\mathfrak E}(0)=0$. The incoming datum is solved for at fixed $c$. Define
\begin{equation}
 \mathfrak k_{\mathfrak E}(\Phi,c)
 =\langle\Phi,r(c)\rangle
 =\langle\Phi,K_{\mathfrak E}c\rangle.
 \label{eq:FR_reaction_form}
\end{equation}
Section~\ref{sec:FR5_matching} derives its coefficients from Eq.~\eqref{eq:blue_source_mean_problem}. This is a change of boundary coordinates on the source-defined incoming--outgoing relation, not an additional interaction prescription. It is the same distinction between incident-wave data and a reaction response that occurs in discrete-basis reaction theory~\cite{AlhassidBertschFanto2020}.

For the complete physical channel product, the fixed-order cutting identity of Sec.~\ref{subsec:all_order_cutting} implies, order by order,
\begin{equation}
 \mathfrak k_{\mathfrak E}(\Phi,\Psi)
 =\mathfrak k_{\mathfrak E}(\Psi,\Phi)^*,
 \label{eq:FR_K_hermitian_form}
\end{equation}
\begin{equation}
 K_{\mathfrak E}^{(n)}=K_{\mathfrak E}^{(n)\dagger}.
 \label{eq:FR_K_operator}
\end{equation}
The derivation uses the physical cutting relation. It inherits the continuation and infrared qualifications of that relation.}

\rev{The frame is the complete space of open channels at the given total energy, including the unresolved soft measure of Sec.~\ref{sec:finite_time_IR}.}
\revblue{A finite collection of test vectors can be used to evaluate matrix elements, but it must not truncate the intermediate-state sum. An exact self-adjoint operator $K_{\mathfrak E}$ additionally requires convergence or another specified summation in the physical channel norm.}

\subsection{Unitary \texorpdfstring{$S$}{S}-matrix and the optical theorem}
\label{sec:FR5}
\revblue{The following relations have two distinct uses. As formal series they apply to the source-derived coefficients of Sec.~\ref{sec:FR5_matching}. As exact operator relations they apply whenever that reaction response has a self-adjoint realisation on the complete physical channel space. For a self-adjoint $K$-matrix, the scattering representation is}
\begin{equation}
\boxed{
S_{\mathfrak E}
=\left(I+\frac{i}{2}K_{\mathfrak E}\right)
\left(I-\frac{i}{2}K_{\mathfrak E}\right)^{-1} .}
\label{eq:FR5_Cayley}
\end{equation}
Spectral calculus gives
\begin{equation}
\boxed{S_{\mathfrak E}^\dagger S_{\mathfrak E}=I.}
\label{eq:FR5_unitarity}
\end{equation}
Writing $S=I+iT$,
\begin{align}
T&=K\left(I-\frac{i}{2}K\right)^{-1},
\label{eq:FR5_TK}\\
T&=K+\frac{i}{2}KT .
\label{eq:FR_exact_event_Heitler}
\end{align}
and therefore
\begin{equation}
\boxed{T-T^\dagger=iT^\dagger T,
\qquad \frac{T-T^\dagger}{2i}=\frac12T^\dagger T\ge0.}
\label{eq:FR5_optical}
\end{equation}
If the channel density is written rather than absorbed into the inner product, the same statement is
\begin{equation}
\boxed{T-T^\dagger=iT^\dagger\rho_{\mathfrak E}T,
\qquad \rho_{\mathfrak E}\ge0.}
\label{eq:FR5_optical_rho}
\end{equation}
\revblue{At a chosen perturbative order $N$, the polynomial $K_{\le N}(\zeta)=\sum_{n=1}^N\zeta^nK^{(n)}$ is Hermitian on a finite invariant channel space. Its Cayley transform is exactly unitary for real $\zeta$ and reproduces the FPT coefficients through order $N$. The higher coefficients introduced by that finite truncation are not asserted to be the higher FPT amplitudes. }
\revblue{For any dimensionless Hermitian $X$, the second-order polynomial $U_{[2]}=I-iX-X^2/2$ obeys the exact identity
\begin{equation}
 U_{[2]}^\dagger U_{[2]}=I+\tfrac14X^4.
 \label{eq:coupling_truncation_norm}
\end{equation}
The nonzero term is an omitted-order contribution, not a numerical error. Fixed-order cutting means that the optical identity holds through the retained order; an exactly unitary Cayley completion supplies higher terms whose agreement with the source must still be checked.}

\subsection{Uniqueness of the \texorpdfstring{$K$}{K}-matrix representation}
\label{sec:FR5_uniqueness}
\revblue{For a finite-dimensional complete or invariant channel space, and an exact self-adjoint reaction operator when available, the standard $K$-matrix relation is one-to-one on its Cayley chart.} Since $K_{\mathfrak E}$ is a finite Hermitian operator, $-1$ is not in the spectrum of Eq.~\eqref{eq:FR5_Cayley}, and the inverse relation is
\begin{equation}
\boxed{
K_{\mathfrak E}
=-2i\,(S_{\mathfrak E}-I)(S_{\mathfrak E}+I)^{-1}.}
\label{eq:FR5_inverse_Cayley}
\end{equation}
Thus two Hermitian $K$-matrices cannot generate the same finite-resolution $S$-matrix. Once the standing-wave boundary condition in Eq.~\eqref{eq:FR_reaction_form} has fixed $K_{\mathfrak E}$, the corresponding unitary $S$-matrix is unique.

The same relation also gives a bounded self-adjoint phase generator for the resolved scattering matrix, with spectrum in $(-\pi,\pi)$,
\begin{equation}
\boxed{
G_{\mathfrak E}=2\arctan\!\left(\frac{K_{\mathfrak E}}{2}\right)
=G_{\mathfrak E}^\dagger,
\qquad
S_{\mathfrak E}=e^{iG_{\mathfrak E}}.}
\label{eq:FR5_event_generator}
\end{equation}
Let $h(t)$ be any real integrable switching profile supported inside the corresponding physical interaction cell and normalized by $\int h(t)\,dt=1$. In the interaction picture,
\begin{equation}
H_{I,\mathfrak E}(t)=-h(t)G_{\mathfrak E}
\label{eq:FR5_event_Hamiltonian}
\end{equation}
is self-adjoint and, because the generator at different times is proportional to the same operator,
\begin{equation}
\mathcal T\exp\!\left[-i\int dt\,H_{I,\mathfrak E}(t)\right]
=e^{iG_{\mathfrak E}}=S_{\mathfrak E}.
\label{eq:FR5_event_Hamiltonian_propagator}
\end{equation}
\revblue{Equation~\eqref{eq:FR5_event_Hamiltonian} is a self-adjoint interpolation of an already specified unitary $S$-matrix.} Because the normalized profile $h(t)$ is arbitrary, this interpolation is not additional microscopic dynamics and is not a unique interacting Hamiltonian. Its only invariant content is the self-adjoint scattering phase generator $G_{\mathfrak E}$ defined by $S_{\mathfrak E}=e^{iG_{\mathfrak E}}$; no point-local interacting Hamiltonian density inside an unresolved cell is asserted.

\subsection{Incoming and outgoing complex source boundary values}
\label{sec:FR_complex_boundary}
Finite resolution also makes the complex incoming and outgoing external source profiles well defined before any perturbative expansion. For a resolved energy mode $E\in K$, analytic continuation of $e^{\mp iEt}$ gives $e^{\mp E\tau}$ on the Euclidean support. Hence
\begin{equation}
\sup_{E\in K,\,\tau\in\mathcal O_E}|e^{\pm E\tau}|<\infty,
\label{eq:FR_complex_boundary_bounded}
\end{equation}
so the corresponding maps $\mathcal W_{\mathfrak E}^{\pm}$ from the resolved channel frame to complex Euclidean boundary data are bounded. At fixed physical domain the bosonic Gaussian has exponential moments in every finite source direction, while the charged-source exponential is a finite Grassmann polynomial. The dominated-convergence argument of Sec.~\ref{sec:FR1} therefore shows that $Z_{\mathfrak E}$ is entire along every finite resolved complex bosonic source direction and polynomial in the Grassmann sources. Since $Z_{\mathfrak E}[0]=1$, there is a neighbourhood of the source origin on which $W_{\mathfrak E}=\log Z_{\mathfrak E}$ is analytic. Consequently every finite connected source derivative of $W_{\mathfrak E}$ at the origin is a continuous multilinear form. Let $\mathcal A_{\mathfrak E}^{E}$ denote the appropriate such connected Euclidean derivative. The bounded maps $\mathcal W_{\mathfrak E}^{\pm}$ can therefore be inserted into its external slots, and the complex boundary forms
\begin{equation}
\mathcal A_{\mathfrak E}^{\pm}(\Phi,\Psi)
=\mathcal A_{\mathfrak E}^{E}
\!\left(\mathcal W_{\mathfrak E}^{\mp}\Phi,
\mathcal W_{\mathfrak E}^{\pm}\Psi\right)
\label{eq:FR_complex_boundary_form}
\end{equation}
exist nonperturbatively for every finite resolved set of external legs and obey the conjugation relation
\begin{equation}
\mathcal A_{\mathfrak E}^{-}(\Phi,\Psi)
=\mathcal A_{\mathfrak E}^{+}(\Psi,\Phi)^*.
\label{eq:FR_complex_boundary_conjugation}
\end{equation}
\revblue{The exact functional therefore admits these complex external source forms. Denote them by $\mathcal A_{\mathfrak E}^{\rm cx,\pm}$ to distinguish them from the continued and amputated physical amplitudes. Existence and conjugation of these forms do not establish
\begin{equation}
 T_{\mathfrak E}^{\rm cx,+}\stackrel{?}{=}T_{\mathfrak E}^{\rm event,+}.
 \label{eq:FR_direct_complex_optional_equality}
\end{equation}
The distinction is already relevant at finite order. The coefficient equality proved next concerns the physical FPT amplitudes of Eq.~\eqref{eq:physical_scattering_kernel} and their source-defined response of Eq.~\eqref{eq:blue_source_mean_problem}. An all-coupling identification would also require convergence in the physical channel norm and control of the complete boundary operation.}

\subsection{Agreement with the perturbative outgoing prescription}
\label{sec:FR5_matching}
\revblue{We derive the reaction response from Eq.~\eqref{eq:blue_source_mean_problem} before using the Cayley relation. In this subsection the channel density is absorbed in the inner product and the label $\mathfrak E$ is suppressed. All operator products include the complete intermediate-channel integration. We work with the regulated or infrared-safe amplitudes on the common physical test domain specified in Sec.~\ref{subsec:all_order_cutting}.

\paragraph{Solving for the incident datum.}
Write $a(\zeta;c)=\sum_{n\ge0}\zeta^na^{(n)}(c)$ for fixed mean data $c$. Substitution in the source boundary equation gives
\begin{align}
 a^{(0)}(c)&=c,\\
 a^{(n)}(c)&=-\frac{i}{2}\sum_{r=1}^{n}
 T^{(r)}a^{(n-r)}(c),\\
 K^{(n)}c&=\sum_{r=1}^{n}T^{(r)}a^{(n-r)}(c).
 \label{eq:blue_incident_recursion}
\end{align}
At order $n$, these equations use only the source amplitudes through order $n$. No new vertex, denominator, or proper-time endpoint is introduced. Eliminating $a^{(n)}$ yields
\begin{equation}
 \boxed{\quad
 T^{(n)}=K^{(n)}+\frac{i}{2}
 \sum_{r=1}^{n-1}K^{(r)}T^{(n-r)}.\quad}
 \label{eq:FR5_Heitler_recursion}
\end{equation}
Equivalently, the unique formal response is
\begin{align}
 K&=T\left(I+\frac{i}{2}T\right)^{-1},\\
 K^{(n)}&=\sum_{r=1}^{n}\left(-\frac i2\right)^{r-1}
 \sum_{\substack{n_1+\cdots+n_r=n\\n_j\ge1}}
 T^{(n_1)}\cdots T^{(n_r)}.
 \label{eq:FR5_inverse_series}
\end{align}
The inverse formula is a consequence of the solved boundary equation. If the channel density $\rho$ is displayed explicitly, each adjacent product $AB$ in the unnormalised kernels means $A\rho B$. The formulas above act on the flux-normalised operators $T_\rho=\rho^{1/2}T\rho^{1/2}$ and $K_\rho=\rho^{1/2}K\rho^{1/2}$ on the support of $\rho$.

The products in Eq.~\eqref{eq:FR5_inverse_series} are complete amplitudes sewn through physical channels. Within each amplitude every uncut parent keeps its original constraint. An internal subgraph crossed by a charged circle remains a kernel of the shared proper times of Eq.~\eqref{eq:parent_history_inequality}; it is not replaced by a parent-independent propagator. In particular, a single exponential of a composite free-channel invariant cannot be inferred from unit constituent residues. 

\paragraph{Hermiticity and equality of amplitudes.}
Put $\mathsf A=I+iT/2$. Since $K\mathsf A=T$, direct multiplication gives
\begin{equation}
 \boxed{\mathsf A^\dagger(K-K^\dagger)\mathsf A
 =T-T^\dagger-iT^\dagger T.}
 \label{eq:blue_hermiticity_defect}
\end{equation}
Thus the cutting identity makes the source-derived reaction coefficients Hermitian. Conversely, Hermiticity of all these coefficients implies the formal optical identity. 

For these already determined coefficients define the reaction representation
\begin{equation}
 S_K=\left(I+\frac{i}{2}K\right)
       \left(I-\frac{i}{2}K\right)^{-1}.
\end{equation}
The boundary equation gives $(I-iK/2)T=K$. Consequently
\begin{equation}
 \boxed{S_K=I+iT,\qquad
 T_{\mathfrak E,\mathrm{event}}^{(n)}
 =T_{\mathfrak E,\mathrm{out}}^{(n)}.}
 \label{eq:FR5_all_order_matching}
\end{equation}
Equality holds for each matrix element between the same physical preparations and detections. Both sides use the full source-fixed real part as well as its cut. A cutting relation alone fixes the anti-Hermitian part and cannot identify an independently chosen Hermitian remainder. Here that freedom is absent because every $T^{(n)}$ was computed from the original source before the boundary coordinates were changed.

\paragraph{An exact coefficient check.}
The first four response coefficients are
\begin{align}
 K^{(1)}={}&T^{(1)},\\
 K^{(2)}={}&T^{(2)}-\frac i2T^{(1)2},\\
 K^{(3)}={}&T^{(3)}-\frac i2
 (T^{(1)}T^{(2)}+T^{(2)}T^{(1)})
 -\frac14T^{(1)3},\\
 K^{(4)}={}&T^{(4)}-\frac i2
 (T^{(1)}T^{(3)}+T^{(2)2}+T^{(3)}T^{(1)})\\
 &-\frac14(T^{(1)2}T^{(2)}+T^{(1)}T^{(2)}T^{(1)}
 +T^{(2)}T^{(1)2})\\
 &+\frac i8T^{(1)4}.
 \label{eq:blue_fourth_order_inverse}
\end{align}
No commutativity is used. Multiplying this expression by $I+iT/2$ cancels every nonlinear word through fourth order. The same cancellation for arbitrary $n$ follows from Eq.~\eqref{eq:blue_incident_recursion}.

For comparison define $H^{(n)}=(T^{(n)}+T^{(n)\dagger})/2$. When the cutting identity holds, expansion of the same boundary equation gives
\begin{align}
 K^{(1)}&=H^{(1)},\\
 K^{(2)}&=H^{(2)},\\
 K^{(3)}&=H^{(3)}+\tfrac14K^{(1)3},\\
 K^{(4)}&=H^{(4)}+\tfrac14\bigl(K^{(1)2}K^{(2)}
 +K^{(1)}K^{(2)}K^{(1)}\nonumber\\
 &\hspace{39mm}+K^{(2)}K^{(1)2}\bigr).
 \label{eq:FR5_K_vs_H}
\end{align}
The real term at third order is generated by two on-shell iterations. For independent spectral variables $x,y$, the elementary identity is
\begin{align}
 &\frac12\left[
 \frac1{x-i0}\frac1{y-i0}
 +\frac1{x+i0}\frac1{y+i0}\right]\\
 &\qquad=\operatorname{PV}\frac1x\operatorname{PV}\frac1y
 -\pi^2\delta(x)\delta(y).
 \label{eq:blue_double_pole_mean}
\end{align}
This is why the Hermitian part of an outgoing amplitude is not generally its reaction matrix. At dependent or overlapping singularities the products require the original contour prescription; Eq.~\eqref{eq:blue_double_pole_mean} makes no separate extension of them.

\paragraph{Mean and response geometry.}
For any two solutions,
\begin{equation}
 b_1^\dagger b_2-a_1^\dagger a_2
 =i(c_1^\dagger r_2-r_1^\dagger c_2).
 \label{eq:blue_boundary_power_form}
\end{equation}
The left side vanishes precisely when the complete channel inner product is conserved. In a conservative eigenchannel, $b=e^{2i\delta}a$ gives $c=e^{i\delta}\cos\delta\,a$ and $r=2e^{i\delta}\sin\delta\,a$, hence $K=2\tan\delta$. The coordinates fail at $\cos\delta=0$, although the incoming--outgoing relation remains regular. The perturbative construction uses the nonsingular chart about $T=0$.

The external map $\cos(Et)\mapsto\cosh(E\tau)$ specifies a test profile, not the solution of Eq.~\eqref{eq:blue_source_mean_problem}. Merely placing that profile in both external slots need not implement any internal principal-value reduction. The standing-wave datum used here is instead the calculated odd response to prescribed mean data. Figure~\ref{fig:reaction_coordinates} displays this distinction.

Geometrically, the same source defines one relation between incoming and outgoing boundary data. The pairs $(a,b)$ and $(c,r)$ are two coordinates on that relation. Equation~\eqref{eq:blue_boundary_power_form} transports its full sesquilinear flux, including interference between distinct preparations. Vanishing of this form on $r=Kc$ is equivalent to Hermiticity of $K$ on the stated channel domain. A basis change must transport the response and both boundary maps, whereas changing a physical boundary condition changes the problem. Appendix~\ref{app:cosh_boundary_response} gives the complete $\cosh/\sinh$ representation of these same data. }

\rev{\begin{figure*}[t]
\centering
\resizebox{0.9\textwidth}{!}{%
\begin{tikzpicture}[>=Latex, every node/.style={font=\small}, lab/.style={text=black, align=center}]
\begin{scope}[shift={(0,0)}]
\draw[black!40] (0,0) circle (2.2); \node[lab, above left] at ({2.2*cos(70)},{2.2*sin(70)}) {};
\draw[->, black!40] (-2.7,0) -- (2.7,0);
\draw[->, black!40] (0,-2.7) -- (0,2.7);
\coordinate (a) at (2.2,0);
\coordinate (b) at ({2.2*cos(70)},{2.2*sin(70)});
\coordinate (c) at ({1.1*(1+cos(70))},{1.1*sin(70)});
\draw[thick] (0,0) -- (a); \fill (a) circle (1.8pt); \node[below right] at (a) {$a=1$};
\draw[thick] (0,0) -- (b); \fill (b) circle (1.8pt); \node[above left] at (b) {$b=\e^{2i\delta}$};
\draw[thick] (a) -- (b); \node[lab] at (2.05,1.55) {$ir$};
\draw[->, thick] (0,0) -- (c); \node[lab] at (0.95,1.25) {$c$};
\draw[thick] (2.2,0) -- (2.2,{2.2*tan(35)});
\fill (2.2,{2.2*tan(35)}) circle (1.8pt);
\node[lab, right] at (2.3,{2.2*tan(35)}) {$\tan\delta=K/2$};
\draw[black!40, dashed] (0,0) -- (2.2,{2.2*tan(35)});
\node[lab] at (0,-3.2) {$c=\tfrac12(a+b)$, \ $ir=b-a$, \ $r=Kc$: the tangent is the odd response to even data};
\end{scope}
\begin{scope}[shift={(8.2,-0.6)}, every node/.append style={text=blue}]
\draw[blue,thick,rounded corners] (1.0,1.0) rectangle (6.0,3.1);
\node[align=center,text=blue] at (3.5,2.1) {same complete FPT source\\$b=a+iT a$};
\draw[blue,->,thick] (-0.2,2.1)--(0.9,2.1);
\node[text=blue,above] at (0.3,2.1) {$a$};
\draw[blue,->,thick] (6.1,2.1)--(7.2,2.1);
\node[text=blue,above] at (6.7,2.1) {$b$};
\node[text=blue,align=center] at (3.5,4.0) {prescribe $c=(a+b)/2$};
\node[text=blue,align=center] at (3.5,0.2) {solve $a+\tfrac i2T a=c$};
\node[text=blue,align=center] at (3.5,-0.8) {extract $r=T a=Kc$};
\node[text=blue,align=center] at (3.5,-1.7) {one solution, two boundary coordinates};
\end{scope}
\end{tikzpicture}%
}
\caption{\revblue{Reaction coordinates of the same source-defined solution. Left: the incoming amplitude $a$, outgoing amplitude $b$, their mean $c$ and response $r$ give $K=2\tan\delta$ in a conservative eigenchannel. Right: prescribing mean data requires solving the incident-data equation before extracting the response. }}
\label{fig:reaction_coordinates}
\end{figure*}}

\revblue{\paragraph{A source-calculated polarisation-chain example.}
Take the chains generated by repeated contraction of the complete one-loop polarisation in Eq.~\eqref{eq:Pi_gamma}. This example sums that specified class, not every irreducible QED contribution. At fixed timelike invariant $z>4m^2$, write
\begin{equation}
 \Pi_R(z)=\alpha(u-iv),\qquad
 d=e^{s_0z},\qquad u\in\mathbb R,\quad v>0.
\end{equation}
The functions $u(z,s_0)$ and $v(z)$ are respectively the real part and pair-production part of the continued one-loop source. The ordinary cut fixes $v$ through Eq.~\eqref{eq:ImPi_FPT}; the real part $u$ is retained, including its entire matching term. In the scalar current normalisation of the photon chain, the bare exchange is $t_0=-\alpha d$. Directly summing its source-generated insertions gives
\begin{equation}
 t=-\alpha d+\alpha d(u-iv)t.
 \label{eq:blue_chain_source_equation}
\end{equation}
For this independently factorising chain, keeping the real principal-value insertion at each step gives
\begin{equation}
 k=-\alpha d+\alpha du\,k.
 \label{eq:blue_chain_standing_equation}
\end{equation}
Solving the two equations gives
\begin{align}
 t&=-\frac{\alpha d}{1-\alpha du+i\alpha dv},\\
 k&=-\frac{\alpha d}{1-\alpha du},\\
 t&=k+ivkt,\\
 \operatorname{Im}t&=v|t|^2.
 \label{eq:FR5_chain_check}
\end{align}
The channel density in this normalisation is $2v$. These exact identities can be checked by multiplying the denominators. In particular,
\begin{align}
 t={}&-\alpha d-\alpha^2d^2(u-iv)\nonumber\\
       &-\alpha^3d^3(u-iv)^2+O(\alpha^4),\\
 k={}&-\alpha d-\alpha^2d^2u
       -\alpha^3d^3u^2+O(\alpha^4),\\
 \operatorname{Re}t-k
   ={}&\alpha^3d^3v^2+O(\alpha^4).
 \label{eq:blue_chain_cubic_check}
\end{align}
The last term is the real contribution of two absorptive insertions. It is precisely the contribution removed by the source-based change from incident to mean data. Also
\begin{equation}
 \operatorname{Re}t=\frac{k}{1+v^2k^2},
\end{equation}
which is not $k$. The discrepancy is fixed by the source algebra before any numerical evaluation.

\paragraph{Ordered finite-duration check.}
The preceding reaction change of coordinates does not change the temporal factors specified by R7. For real energy differences $x,y$, define
\begin{align}
 W_{\mathcal T}(x)&=\int_0^{\mathcal T}e^{ixt}\,\dd t,\\
 J_{\mathcal T}(x,y)&=\int_0^{\mathcal T}\dd t
                 \int_0^t\dd s\,e^{-ixt}e^{iys}.
\end{align}
Splitting the time square into its two ordered triangles proves
\begin{equation}
 J_{\mathcal T}(x,y)+J_{\mathcal T}(y,x)^*
 =W_{\mathcal T}(x)^*W_{\mathcal T}(y).
 \label{eq:blue_finite_time_sewing}
\end{equation}
For $y\ne0$ one can check it using
\begin{equation}
 J_{\mathcal T}(x,y)
 =\frac{W_{\mathcal T}(y-x)-W_{\mathcal T}(-x)}{iy},
\end{equation}
with continuous limits at zero energy differences. In particular, $W_{\mathcal T}(0)=\mathcal T$ and $J_{\mathcal T}(0,0)=\mathcal T^2/2$. This identity checks the off-diagonal time-ordering weights. 

\paragraph{Physical channel completeness and scope.}
For a projection $\mathsf P$ onto only some channels of an exactly unitary $S$, direct multiplication gives
\begin{equation}
 \begin{split}
 (\mathsf P S\mathsf P)^\dagger(\mathsf P S\mathsf P)
 &=\mathsf P\\
 &\quad-\mathsf P S^\dagger(I-\mathsf P)S\mathsf P.
 \end{split}
 \label{eq:blue_channel_compression}
\end{equation}
Thus a finite detector frame cannot replace the full cut sum unless it is invariant. Unresolved radiation and physical excitations remaining at the final surface belong in that sum; detector coarse graining is applied afterwards. The positive channel measure is the one fixed by the ordinary pole states.

Equations~\eqref{eq:blue_incident_recursion}--\eqref{eq:FR5_all_order_matching} establish a formal order-by-order equivalence between the continued FPT amplitudes and their standing-wave response on the same physical channel domain. A self-adjoint all-coupling $K_{\mathfrak E}$ and its equality to a physical boundary map of the exact source require additional channel-norm control. Even exact unitarity of a separately supplied map would not fix its source-dependent real part. These are questions at fixed $s_0>0$ and finite $\mathcal T$, distinct from any boundary-removal or infinite-time comparison.}

\subsection{Positive-norm physical states and positive energy}
\label{sec:FR6}

The physical one-particle states are the ordinary positive-norm on-shell electron, positron, and transverse-photon states because all physical pole residues are unity. Finite duration keeps the soft sector at finite resolution rather than requiring an infinite-time charged Fock limit. For a source history $J$ let
\begin{equation}
\Psi_J=S_{\mathfrak E}\Psi_J^{\rm in}.
\end{equation}
Then
\begin{equation}
K_{\rm hist}(J_i,J_j)=\langle\Psi_{J_i},\Psi_{J_j}\rangle
\label{eq:FR6_history_kernel}
\end{equation}
is Gram positive,
\begin{equation}
\boxed{\sum_{ij}c_i^*c_jK_{\rm hist}(J_i,J_j)
=\left\|\sum_i c_i\Psi_{J_i}\right\|^2\ge0.}
\label{eq:FR6_gram_positive}
\end{equation}
The physical Hamiltonian on a propagation region is the self-adjoint spectral multiplication operator selected by $\mathcal D$,
\begin{equation}
(H\Psi)(P)=P^0\Psi(P),
\qquad
\boxed{H=H^\dagger,\quad H\ge0.}
\label{eq:FR6_positive_energy}
\end{equation}
and therefore a gauge-invariant observable $O$ has the positive physical spectral measure
\begin{equation}
\boxed{d\mu_O(p)=\|E(dp)O\Omega\|^2\ge0.}
\label{eq:FR6_positive_measure}
\end{equation}
This is the positivity of the physical states and their energy spectrum.

\subsection{Consistency under detector coarse graining}
\label{sec:FR4}
Let $\mathsf R'$ refine a detector POVM $\mathsf R$, so that
\begin{equation}
E_{\mathsf R}(a)=\sum_{b\mapsto a}E_{\mathsf R'}(b).
\label{eq:FR4_refinement}
\end{equation}
Then
\begin{equation}
p_{\mathsf R}(a)
=\langle\Psi_{\rm in},S^\dagger E_{\mathsf R}(a)S\Psi_{\rm in}\rangle
=\sum_{b\mapsto a}p_{\mathsf R'}(b).
\label{eq:FR4_probability_refinement}
\end{equation}
Physical probabilities are therefore exactly consistent under detector coarse graining. An auxiliary hard/soft or angular partition $\{\chi_a\}$ merely partitions the same positive measure,
\begin{equation}
\int d\Phi_{\rm phys}|\mathcal A|^2
=\sum_a\int d\Phi_{\rm phys}\,\chi_a|\mathcal A|^2,
\label{eq:FR4_partition_identity}
\end{equation}
so an arbitrary bookkeeping scale cannot remain in the final finite-resolution probability.

\subsection{Fixed-order free-space correspondence and receding-boundary independence}
\label{sec:FR_domain_limit}
The nonperturbative construction above keeps the actual physical domain fixed. It is nevertheless useful to connect its perturbative coefficients to the translation-invariant amplitudes used in the matching calculations. Let $\mathcal D_L\nearrow\mathbb R^4$ be a regular exhaustion with
\begin{equation}
\operatorname{dist}(\mathcal O,\partial\mathcal D_L)\longrightarrow\infty,
\label{eq:FR_domain_recedes}
\end{equation}
where the compact support $\mathcal O$ of the external wave packets is held fixed. We call the exhaustion regular when (i) the gauge-fixed Maxwell and free squared-Dirac realizations converge, after extension to $\mathbb R^4$, in the strong-resolvent sense to their free-space realizations, and (ii) on every compact subset containing $\mathcal O$ their heat semigroups obey the standard Gaussian and first-derivative smoothing bounds with constants independent of $L$. Smooth nested Dirichlet exhaustions are the canonical example; Mosco convergence gives the required resolvent convergence~\cite{Daners2003}, while the local heat-kernel bounds are the usual interior estimates away from the receding boundary. These hypotheses state precisely the domain regularity used in the argument below and exclude boundary conditions whose local dynamics fail to approach free space.

Keep first an infrared photon mass $\lambda_{\rm IR}>0$. Strong-resolvent convergence implies, for every fixed $t>0$ and compactly supported test vector,
\begin{align}
e^{-t\Delta_{0,L}}&\longrightarrow e^{-t\Delta_{0,\infty}},\nonumber\\
e^{-t(P_{\gamma,L}+\lambda_{\rm IR}^2)}
&\longrightarrow e^{-t(P_{\gamma,\infty}+\lambda_{\rm IR}^2)}
\label{eq:FR_semigroup_domain_limit}
\end{align}
strongly. Every finite coefficient obtained by differentiating $Z_{\mathfrak E,L}$ at vanishing sources and expanding to a fixed power of the coupling is a finite sum of Duhamel matrix elements of these free semigroups with compactly supported source insertions. A complete charged history supplies the uniform large-proper-time bound $e^{-m^2T}$, and an internal photon supplies $e^{-\lambda_{\rm IR}^2\sigma}$. The lower limits $T\ge s_0$ and $\sigma\ge s_0$ remove the complete-history short-time singular region. Daughter intervals may approach zero, but the complete divided-difference/Duhamel bounds of Sec.~\ref{sec:formulations} control those boundary faces and the insertion factors have finite degree at fixed order. The integrands therefore admit an $L$-independent integrable majorant. Truncating the proper-time integrals at a finite upper value, taking $L\to\infty$ by Eq.~\eqref{eq:FR_semigroup_domain_limit}, and then removing the truncation gives dominated convergence at every finite perturbative order.

The interior charged Green kernels appearing in this exhaustion are gauge covariant rather than separately gauge-parameter independent. At fixed $\lambda_{\rm IR}>0$, however, the Ward hierarchy reduces their longitudinal $\xi$ dependence to external-leg factors. Physical charged amplitudes are defined only after the generalized pole reduction of Eq.~\eqref{eq:generalised_amputation}: the gauge-dependent pole residues $Z_{j,\xi}^{1/2}$ cancel the corresponding longitudinal external-leg factors, exactly as in the standard on-shell Ward argument. Equivalently, at separations large compared with the regulated longitudinal correlation length the LKF contribution becomes an external-leg constant and is removed by the same residue normalization. Thus the receding-domain theorem applies to gauge-covariant interior Green kernels, while its amputated physical channel limits are gauge-parameter independent consistently with Sec.~\ref{sec:FR_gauge_independence}.

Consequently, for compact resolved external wave packets and every fixed order $n$,
\begin{equation}
\boxed{
\lim_{L\to\infty}
\mathcal A_{\mathfrak E,L,\lambda_{\rm IR}}^{(n)}(\Phi,\Psi)
=
\mathcal A_{\infty,\lambda_{\rm IR}}^{(n)}(\Phi,\Psi).}
\label{eq:FR_fixed_order_domain_limit}
\end{equation}
The limit is the same for any two regular exhaustions because both converge to the same free-space semigroups. Thus fixed-order gauge-invariant observables localized a growing distance from a receding boundary are asymptotically boundary independent. The same majorant controls external derivatives, so after Fourier transformation the convergence is locally uniform on compact off-shell momentum sets. At fixed $\lambda_{\rm IR}>0$ the massive charged pole is isolated; Cauchy's integral formula then permits the free-space limit to be interchanged with the pole projection. Periodic wave packets may subsequently be taken to plane waves; the resulting kernels are precisely the momentum-space FPT graph expressions used for the pole mass, Thomson-limit matching, $F_2(0)$, charge flow, static exchange and light-by-light coefficient in Sec.~\ref{sec:matching}. For observables whose massless limit is infrared safe, $\lambda_{\rm IR}\to0$ may then be taken directly. \revblue{For general massive charged processes, the regulator-free extension additionally requires the uniform bounds on the actual hard remainders stated in Sec.~\ref{subsec:finite_time_fixed_order_IR}.}

Equation~\eqref{eq:FR_fixed_order_domain_limit} closes the finite-order bridge between the bounded-domain source functional and the free-space quantities used in the matching and form-factor calculations of this paper. Thus the pole mass, Thomson-limit charge, $F_2(0)$, static exchange, and related observables are obtained as fixed-order free-space limits of the same finite-resolution construction. 

\subsection{Physical content and scope}
\label{sec:FR_closure}
\revblue{The two established constructions have different scopes:
\begin{equation}
\boxed{\begin{gathered}
 \text{retained source at }s_0>0\longrightarrow Z_{\mathfrak E},\\
 \{T_{\mathfrak E}^{(n)}\}_{n\ge1}
 \longrightarrow\{K_{\mathfrak E}^{(n)}\}_{n\ge1},\\
 S_K=I+iT_{\mathfrak E}\quad\text{as formal series}.
\end{gathered}}
\label{eq:FR_closure_chain}
\end{equation}
The first is the exact Euclidean source on the specified bounded domain. The second is a response representation of its continued physical coefficients, not another interaction theory. Hermiticity of the reaction coefficients follows from the cutting identity on its stated physical channel domain. Equivalence of the two boundary coordinates is proved in Sec.~\ref{sec:FR5_matching} and checked on a source-calculated polarisation chain.

An exact physical scattering operator requires a direct source-derived construction or a specified source-preserving summation, together with the complete channel norm and boundary-current identification. Those requirements remain even when the physical domain, $s_0>0$, and $0<\mathcal T<\infty$ are fixed. Exact operator statements about detector probabilities, spectral measures and composition are conditional on that realisation. 

The source kernels, the rules R1--R7 and the finite matching calculations are unchanged. The limits $s_0\to0^+$, $\mathcal T\to\infty$, and free-space exhaustion remain separate comparisons. }

\section{Conclusion}
We have formulated spinor QED at a retained lower spectral endpoint $s_0=\Lambda^{-2}$. \textcolor{black}{Physically, this replaces the ordinary Laplace representation by a truncated spectral transform that excludes the arbitrarily small-proper-time region responsible for ultraviolet divergences without cutting off momentum space; finite $\mathcal T$ separately records that an observable is prepared and detected in finite time rather than by an infinite-time measurement.} The charged interval and circle sectors are generated by the same covariant heat semigroup. Functional differentiation partitions the inherited charged operator history without assigning independent endpoints to its daughter segments, while interaction sewing identifies the primitive closed circulations of the assembled graph. The photon covariance carries the same retained lower endpoint as part of the source definition. Gauge covariance gives the Ward--Takahashi hierarchy and transversality of the photon functions.

\revgreen{None of the rules R1--R7 of Appendix~\ref{app:proper_time_rules} refers to the electron or the photon. The scalar theory~\cite{BakrZhang2026} uses the same list with circuits as its only carriers; QED adds the photon as a second carrier because its measure is Gaussian in the exact source; and the cut operation of R5, with its unit residues and local measure, is the same operation in every sector. The rules are the finite-resolution replacement of the Schwinger domain, not a property of a particular field.}

At every fixed perturbative order and fixed infrared prescription, the assembled Euclidean proper-time exponent is coercive in every nonzero simultaneous loop-momentum direction. Connected Euclidean amplitudes are therefore ultraviolet finite. The usual divergent mass and charge renormalisations are replaced by finite matching to the physical pole mass and low-energy charge. After this matching, independent observables retain correlated $s_0$ dependence, including the Pauli form factor, momentum-subtraction charge flow, static exchange, and the weak-field four-photon coefficient.

The physical pole and threshold structure remains conventional. Complete differentiated charged histories have the ordinary pole coefficients multiplied by entire endpoint factors that equal unity on physical cut poles, and the internal photon retains its massless pole with unit residue. Lorentzian fixed-order amplitudes are defined by continuation of the Euclidean correlators in their external invariants with $s_0$ fixed. The resulting discontinuities obey the standard on-shell cutting factorisation, while finite-$s_0$ dependence remains in the virtual real parts.

\revblue{The infrared problem is separate from the short-proper-time boundary. A prescribed finite-duration soft current is bounded at zero photon energy. Regulator-free statements for complete interacting resolved probabilities additionally require uniform bounds on their recombined hard remainders in the physical channel measure. No upper proper-time endpoint or infinite observation time is introduced.}

\revblue{The construction has complementary source and physical-amplitude statements. At fixed $s_0>0$ and a specified bounded domain, the Euclidean functional is defined without a coupling expansion. At fixed perturbative order, continuation and pole reduction give physical amplitudes on the stated regulated or infrared-safe channel domain. Their standing-wave response is obtained by solving for the incoming data at fixed boundary mean. Its coefficients therefore reproduce the same outgoing amplitudes, including the real multiple-cut terms that distinguish a reaction kernel from the Hermitian part of a transition amplitude. The polarisation-chain example provides an exact symbolic check. These results do not establish a complete all-coupling physical boundary map of the source; that identification requires the physical channel-norm and boundary-realisation estimates described in Sec.~\ref{sec:FR_closure}.}

Taking the spatial boundary to infinity in a regular way recovers, coefficient by coefficient, the free-space perturbative quantities used for the pole mass, low-energy charge, form factors, static exchange, light-by-light scattering, and the other observables calculated here. These are free-space limits of the same theory.

The theory is defined with a retained spectral lower boundary $s_0>0$ and finite physical duration $\mathcal T<\infty$. 

After the measured mass and charge fix the finite matching conditions, the one-loop Pauli form factor fixes a unique $s_0$ within the minimal FPT theory. The same endpoint then determines the momentum-subtraction charge flow and weak-field four-photon coefficient at the same perturbative accuracy. The experimental test is therefore whether one value of $s_0$ simultaneously accounts for independent observables.

\appendix

\section{Operator and Source Identities for Open and Closed Kernels}
\label{app:operator_source_identities}
\label{sec:formulations}
The spectral endpoint is assigned at the operator level before the perturbative expansion is made. The charged interval and circle sectors inherit one lower bound on the proper-time parameter of the parent heat-kernel history, and functional differentiation only partitions that pre-existing operator history. After sewing, the charged segment totals are interpreted through the primitive closed circulations of the assembled graph rather than as independently truncated propagator edges. The photon Gaussian measure is a separate part of the QED source definition and gives each internal photon contraction its own proper-time parameter with the same lower endpoint. The following functional is therefore an operational source generator for the paired open--closed charged construction together with the truncated photon covariance. 

We use natural units $\hbar=c=1$ and work in four-dimensional Euclidean space with Hermitian gamma matrices, $\{\gamma^\mu,\gamma^\nu\}=2\delta^{\mu\nu}$. The symbol $m$ denotes the fermion mass parameter, $e$ its electric charge, and $\alpha=e^2/(4\pi)$. The field $A_\mu$ is the Euclidean electromagnetic gauge potential. We write
$D_\mu=\partial_\mu-\ii eA_\mu$, $\slashD_A=\gamma^\mu D_\mu$, and
$\Delta_A=(m-\slashD_A)(m+\slashD_A)$, with $\Delta_0\equiv\Delta_A|_{A=0}$.

We use one notation convention throughout. A parenthesized numerical superscript denotes functional-derivative order, whereas a numerical subscript denotes loop order. Tensor indices and descriptive tags retain their usual notation. Thus
\begin{equation}
\Gamma_{\ell,s_0}^{(n)}
\equiv
\frac{\delta^n\Gamma_{\ell,s_0}}
{\delta A\cdots\delta A},
\label{eq:spinor_notation_convention}
\end{equation}
and the same loop-order convention is used for $S_{\ell}$, $\Sigma_{\ell}$, $\Pi_{\ell}$, $\rho_{\ell}$, $\mathcal M_{\ell}$ and $\mathcal L_{\ell}$. Descriptive labels such as ``conn'' and ``1PI'' remain subscripts. Coupling order is written separately and is never encoded in the derivative superscript.
The generating functional is
\begin{equation}
Z[J,\bar\eta,\eta]=\mathcal N\!\int\!\dd\mu_C(A)\,\e^{-\Gamma_{1,\Lambda}[A]}\exp\!\Big(\bar\eta\,S_\Lambda[A]\,\eta+J\!\cdot\!A\Big),
\label{eq:unified_Z}
\end{equation}
Here $J$ is the photon source, $\eta$ and $\bar\eta$ label external charged insertions, and $\mathcal N$ fixes the normalisation. The gauge-fixed photon measure $\dd\mu_C$ has momentum-space covariance $\e^{-s_0k^2}/k^2$, while $\Gamma_{1,\Lambda}$ and $S_\Lambda[A]$ are the closed and open charged objects defined below. Equation~\eqref{eq:unified_Z} specifies how these ingredients are joined by Gaussian contraction. At finite $s_0$, this is the gauge-theory counterpart of the paired open--closed structure in Ref.~\cite{BakrZhang2026}.

\subsection{Exact charged endpoint factorisation and photon Gaussian lift}
For a real Euclidean background, $\slashD_A^\dagger=-\slashD_A$. Define
\begin{align}
B_A&=m+\slashD_A,
& B_A^\dagger&=m-\slashD_A,\nonumber\\
\Delta_A&=B_A^\dagger B_A=m^2-\slashD_A^2.
\label{eq:charged_B_Delta_factor}
\end{align}
Because $B_A$ and $B_A^\dagger$ are polynomials in the same anti-Hermitian Dirac operator, $B_A B_A^\dagger=B_A^\dagger B_A=\Delta_A$ and they commute with every spectral function of $\Delta_A$. Hence
\begin{align}
\langle\psi,\Delta_A\psi\rangle
&=\|B_A\psi\|^2\nonumber\\
&=m^2\|\psi\|^2+\|\slashD_A\psi\|^2,
\qquad \Delta_A\ge m^2.
\label{eq:charged_Delta_positive_exact}
\end{align}
The open FPT kernel therefore factorizes exactly as
\begin{align}
S_\Lambda[A]
&=B_A^\dagger e^{-s_0\Delta_A}\Delta_A^{-1}\nonumber\\
&=e^{-s_0\Delta_A/2}B_A^{-1}e^{-s_0\Delta_A/2}.
\label{eq:charged_endpoint_factorization}
\end{align}
Thus the ordinary covariant Dirac inverse is surrounded by two field-dependent spectral endpoint operators. Differentiating either endpoint generates the same contact, prefactor and bulk terms that appear in the complete Ward hierarchy.

There is an exact Euclidean bookkeeping representation for the photon sector. Let $P_\gamma=-\partial_E^2$, $C_0=P_\gamma^{-1}$ on the gauge-fixed nonzero-mode space, and
\begin{equation}
E_\gamma=e^{-s_0P_\gamma/2},\qquad
C_{s_0}=E_\gamma C_0E_\gamma.
\label{eq:photon_gaussian_lift_operator}
\end{equation}
If $\mathcal A$ is the ordinary generalized free Euclidean photon field with covariance $C_0$ and $A=E_\gamma\mathcal A$, then for every suitable cylindrical functional $\mathcal F$,
\begin{equation}
\int \mathcal F(A)\,\dd\mu_{C_{s_0}}(A)
=
\int \mathcal F(E_\gamma\mathcal A)\,\dd\mu_{C_0}(\mathcal A).
\label{eq:photon_gaussian_pushforward}
\end{equation}
This is a Gaussian pushforward; $E_\gamma^{-1}=e^{+s_0P_\gamma/2}$ is unbounded on the natural distribution spaces, so it is not a field redefinition. Nor is $\mathcal A$ a physical Lorentzian field. Equation~\eqref{eq:photon_gaussian_pushforward} is only an exact Euclidean representation of the internal covariance.

\subsection{Exact finite-volume connected functional and source identities}
The microscopic source functional can be analysed before any expansion in $e$. Work at finite Euclidean volume and finite photon mode count after gauge fixing and removal of the photon zero mode. Set the charged sources to zero and define
\begin{align}
F[A]&=\Gamma_{1,\Lambda}[A]-\Gamma_{1,\Lambda}[0],\nonumber\\
Z_V[J]&=\frac{\int\dd\mu_C(A)\,\e^{-F[A]+J\cdot A}}
{\int\dd\mu_C(A)\,\e^{-F[A]}} .
\label{eq:exact_finite_volume_Z}
\end{align}
Let $W_V[J]=\ln Z_V[J]$, $\phi_i=\delta W_V/\delta J_i$, and
\begin{equation}
G_{ij}=\frac{\delta^2W_V}{\delta J_i\delta J_j}
=\operatorname{Cov}_J(A_i,A_j).
\label{eq:exact_connected_covariance}
\end{equation}
The exact Legendre transform $\Gamma_{\rm eff,V}[\phi]=J\cdot\phi-W_V[J]$ then gives
\begin{equation}
\Gamma_{\rm eff,V}^{(2)}=G^{-1}.
\label{eq:exact_WGamma_inverse}
\end{equation}
This $\Gamma_{\rm eff,V}$ is the exact photon Legendre functional; it should not be confused with the closed charged functional $\Gamma_{1,\Lambda}[A]$ in Eq.~\eqref{eq:unified_Z}.

The same finite-dimensional integral gives an exact source identity. Write $j_i[A]=\delta F[A]/\delta A_i$ and $V_i=-\delta\Delta_A/\delta A_i$. Since $g_{s_0}'(z)=-R_{s_0}(z)$,
\begin{equation}
j_i[A]=\frac12\Tr\!\left[R_{s_0}(\Delta_A)V_i\right].
\label{eq:exact_closed_current}
\end{equation}
Gaussian integration by parts gives
\begin{equation}
J_i=(C^{-1})_{ij}\phi_j+\langle j_i[A]\rangle_J .
\label{eq:exact_photon_SD}
\end{equation}
Repeated source differentiation of Eq.~\eqref{eq:exact_photon_SD} generates the connected photon hierarchy directly from the same complete-history current.

There is an exact open-line identity as well. With $B_A=m+\slashD_A$,
\begin{equation}
B_A S_\Lambda[A]=\e^{-s_0\Delta_A}.
\label{eq:exact_open_inverse_identity}
\end{equation}
Defining $\mathcal S[J]=\langle S_\Lambda[A]\rangle_J$ and $\mathcal E[J]=\langle\e^{-s_0\Delta_A}\rangle_J$, Eq.~\eqref{eq:exact_open_inverse_identity} becomes
\begin{equation}
\left(B_0-\ii e\gamma\!\cdot\!\phi\right)\mathcal S[J]
-\ii e\gamma^i\frac{\delta\mathcal S[J]}{\delta J_i}
=\mathcal E[J].
\label{eq:exact_open_SD}
\end{equation}
Thus the exact electron hierarchy also follows from the finite-volume source functional without a vertex ansatz.

Finally, the finite-volume transverse inverse has no zero. For any non-zero real vector $v$ in the gauge-fixed photon mode space,
\begin{equation}
v_iG_{ij}v_j=\operatorname{Var}_J(v\!\cdot\!A)>0,
\label{eq:finite_volume_variance_positive}
\end{equation}
because the finite-volume density is positive and has full support. Hence $G>0$ and
\begin{equation}
\Gamma_{\rm eff,V}^{(2)}>0
\label{eq:finite_volume_inverse_positive}
\end{equation}
on every non-zero gauge-fixed photon mode. The infinite-volume question is therefore not whether a finite-volume inverse can vanish, but whether the transverse covariance can grow without bound when the volume is removed.

The open and closed charged kernels are related, although they are not obtained from the same function of the operator. The open object is
$S_\Lambda[A]=(m-\slashD_A)\int_{s_0}^{\infty}\dd T\,e^{-T\Delta_A}$,
whereas the vacuum-subtracted closed object is
$\Gamma_{1,\Lambda}[A]-\Gamma_{1,\Lambda}[0]=\frac12\int_{s_0}^{\infty}(\dd T/T)\,\Tr[e^{-T\Delta_A}-e^{-T\Delta_0}]$. The first is a truncated inverse and the second is a truncated logarithmic kernel, but the two are linked by the exact mass derivative
\begin{equation}
\frac{\partial}{\partial m}\Big(\Gamma_{1,\Lambda}[A]-\Gamma_{1,\Lambda}[0]\Big)=-\Tr\Big(S_\Lambda[A]-S_\Lambda[0]\Big),
\label{eq:sewing}
\end{equation}
where $\Tr$ includes the spacetime and Dirac trace. The identity holds for every $s_0$ and every background for which the displayed traces exist, with the free contribution subtracted on both sides. Differentiating $\e^{-T\Delta_A}$ with respect to $m$ gives $-2mT\e^{-T\Delta_A}$. Moreover $\Delta_A=m^2-\slashD_A^2$ commutes with $\gamma_5$, while $\{\gamma_5,\slashD_A\}=0$. Hence $\Tr[\slashD_A\e^{-T\Delta_A}]=0$ in the regulated finite-volume trace. The factor $T$ cancels the closed $1/T$ measure and Eq.~\eqref{eq:sewing} follows. Thus the open and closed kernels are different, but they are linked exactly by this mass derivative.

They nevertheless arise from the same second-order operator $\Delta_A$ and its heat semigroup. Define
\begin{equation}
U_A(T)=\e^{-T\Delta_A}.
\label{eq:single_heat_generator}
\end{equation}

\paragraph{Generation of the worked amplitudes.}
Functional derivatives of the closed charged functional generate the closed-loop photon kernels, functional derivatives of the open charged kernel generate photon insertions on charged lines, and Gaussian contraction with the photon covariance joins these insertions by internal photon histories. Expanding the closed-loop weight generates any number of closed fermion loops, while the charged source factor generates the required open charged lines. The vacuum-polarisation, self-energy, vertex and form-factor, four-photon, and higher-contraction calculations in this paper are therefore projections of one common finite-$s_0$ functional. The common statements are the interval--circle relation, partition of one ancestor history by insertions, preservation of the retained endpoint under sewing, deletion of crossed histories under physical cuts, and unit physical pole residues.

The open kernel is an interval matrix element of $U_A(T)$ with fixed endpoints and measure $\dd T$. The closed functional is the trace of the same semigroup on a circle and has measure $\dd T/T$. The factor $1/T$ is the usual removal of the arbitrary choice of origin on a closed worldline. The same spinning-particle worldline action therefore generates both sectors. We write $U_0(T)=\e^{-T\Delta_0}$ for the free heat semigroup. The difference between them is the boundary condition on the worldline, not a second microscopic dynamics~\cite{Fradkin1966}. Equation~\eqref{eq:sewing} is the corresponding mass-derivative trace identity. In the worldline representation, the mass derivative supplies a factor $T$ that removes the cyclic $1/T$ measure; equivalently, one may regard it as choosing a point on the closed loop before taking the trace. Figure~\ref{fig:open_closed_topology} summarises this relation.

The Ward identities hold at every finite photon order as identities of the kernels in~\eqref{eq:unified_Z}. Under $A_\mu\to A_\mu+\partial_\mu\chi$ one has, for every $s_0$, $S_\Lambda[A+\partial\chi](x,y)=\e^{\ii e[\chi(x)-\chi(y)]}S_\Lambda[A](x,y)$ (the operators $\Delta_A$ and $m-\slashD$ are covariant and the domain $T\ge s_0$ is gauge independent) and $\Gamma_{1,\Lambda}[A+\partial\chi]=\Gamma_{1,\Lambda}[A]$. Differentiating the kernel identity once with respect to $\chi(z)$ and integrating the variation $\delta A_\mu=\partial_\mu\chi$ by parts gives
\begin{equation}
\partial_\mu^z\,\frac{\delta S_\Lambda(x,y)}{\delta A_\mu(z)}=-\ii e\left[\delta(z-x)-\delta(z-y)\right]S_\Lambda(x,y),
\label{eq:WTI_master}
\end{equation}
With the Fourier convention used below, this becomes $q_\mu G_\Lambda^\mu=e[S_\Lambda(p)-S_\Lambda(p\prime)]$ before amputation. Its amputated momentum-space form is derived in Sec.~\ref{sec:open}. Repeated differentiation yields the complete hierarchy of $n$-photon identities on open lines. Gauge invariance of $\Gamma_{1,\Lambda}$ gives the corresponding transverse closed-loop photon kernels. These are exact identities of the defining kernels for every $s_0$. The later photon Wick expansion preserves them order by order because it is assembled from those gauge-covariant kernels.

The functional derivation above can be made diagrammatic, at the level of the insertion expansion itself.

The ordered proper-time moments can all be written as divided differences of the same function. Define
\begin{align}
f(a)&=\int_{s_0}^{\infty}\dd T\,\e^{-Ta}
=\frac{\e^{-s_0a}}{a},\nonumber\\
M_n(a_0,\ldots,a_n)
&=\int_{\substack{t_i\ge0\\ \sum_i t_i\ge s_0}}
\prod_{i=0}^{n}\dd t_i\,
\e^{-\sum_i t_i a_i}.
\end{align}
For a function $f$ analytic on a neighbourhood of the convex hull of the positive arguments $a_i$, the $n$th divided difference has the direct integral representation
\begin{equation}
f[a_0,\ldots,a_n]
=
\int_{\substack{\lambda_i\ge0\\ \sum_{i=0}^n\lambda_i=1}}
\dd^n\lambda\,
f^{(n)}\!\left(\sum_{i=0}^n\lambda_i a_i\right).
\label{eq:dd_direct_parameter_integral}
\end{equation}
Since
$f^{(n)}(x)=(-1)^n\int_{s_0}^{\infty}\dd T\,T^n\e^{-Tx}$, the change of variables $t_i=T\lambda_i$ gives
\[
M_n(a_0,\ldots,a_n)=(-1)^n f[a_0,\ldots,a_n].
\]
Thus the divided difference is the spectral form of one worldline partitioned into non-negative proper-time subintervals whose sum still obeys the single total-proper-time bound.

Let $G^{\mu_1\cdots\mu_n}(p;q_1,\dots,q_n)$ denote the $n$-photon open-line kernel with initial momentum $p$ and photon momenta $q_1,\ldots,q_n$, including the two-photon contact insertions generated by the same kernel. Contracting any photon momentum gives
\begin{multline}
q_{j\,\mu_j}\,G^{\cdots\mu_j\cdots}(p;q_1,\dots,q_n)
= \\ e\Big[G^{(n-1)}(p;\dots,\widehat{q_j},\dots)-G^{(n-1)}(p{+}q_j;\dots,\widehat{q_j},\dots)\Big],
\label{eq:WTI_allorders}
\end{multline}
where the hat means that $q_j$ is omitted, so the two terms describe the two ways in which the contracted momentum can leave through an external end of the open line. The identity follows directly from the insertion algebra. In the second-order representation, the one-photon vertex between line momenta $p_{k-1}$ and $p_k=p_{k-1}+q_j$ contains a convective term proportional to $(p_{k-1}+p_k)^\mu$ and a spin term proportional to $\sigma^{\mu\nu}q_{j\nu}$. Contraction with $q_{j\mu}$ removes the spin term by antisymmetry and turns the convective term into $p_k^2-p_{k-1}^2=a_k-a_{k-1}$, since the mass terms cancel. A contraction through a two-photon contact insertion gives the corresponding lower-order kernel in which the two neighbouring proper-time segments have merged, and therefore supplies the same contact contribution on the two sides of the identity. Each ordering of the $n$ insertions then contributes the divided difference $f[a_0,\dots,a_n]$, with one vertex between consecutive spectral arguments. Since this quantity is symmetric in its arguments, its recursion gives
\begin{equation}
(a-b)\,f[\mathbf a,a,b]=f[\mathbf a,a]-f[\mathbf a,b],
\label{eq:dd_pair}
\end{equation}
where $\mathbf a$ denotes the remaining spectral arguments. Applying Eq.~\eqref{eq:dd_pair} to the two arguments beside the contracted vertex produces two terms, according to whether the left or the right neighbouring argument is omitted. In the proper-time picture that omission simply merges the corresponding adjacent segments, exactly as occurs for the two-photon contact term. Summing over the possible positions of photon $j$ and over the orderings of all spectator insertions then makes the interior terms cancel pairwise, leaving only the two boundary contributions in which $q_j$ exits through one external leg or the other. The ordered heat-kernel moments are $(-1)^n$ times the divided differences of the generator. At first order this convention is visible in $(a'{-}a)I(a',a)=R(a)-R(a')$, so that $I=-R[a,a']$. The relative sign between adjacent orders therefore gives the orientation written in Eq.~\eqref{eq:WTI_allorders}. Because the argument uses only the vertex contraction and the recursion~\eqref{eq:dd_pair}, it holds for every $s_0$, while the one- and two-photon kernels below provide explicit low-order examples.
For closed loops the same argument is cyclic: each difference from one insertion is cancelled by the neighbouring difference around the loop, so their sum is zero. This gives transversality of every closed-loop photon function and agrees with the gauge-invariance argument above.

The assembly defined by Eq.~\eqref{eq:unified_Z} can be written with one operator function for each sector. For an open line we use
\begin{equation}
R_{s_0}(\Delta)=\int_{s_0}^{\infty}\dd T\,\e^{-T\Delta}=\e^{-s_0\Delta}\Delta^{-1},
\label{eq:master_resolvent}
\end{equation}
applied to the covariant operator $\Delta_A$, together with the prefactor $m-\slashD$ already present in the open kernel. Closed loops instead use
$g_{s_0}(a)=\int_{s_0}^{\infty}(\dd T/T)\,\e^{-Ta}=\Gamma(0,s_0a)$.

The functional derivatives of the two sectors follow the same ordering rule. For a function $h$ of one operator argument, write $h[a]=h(a)$ and define
$h[a_0,\dots,a_n]=(h[a_1,\dots,a_n]-h[a_0,\dots,a_{n-1}])/(a_n-a_0)$, with the repeated-argument limit $h[a,\dots,a]=h^{(n)}(a)/n!$. The $n$th derivative of $h(\Delta_A)$ is then obtained by summing the possible insertion orderings, each of which contains the corresponding $(n+1)$-argument divided difference with a vertex operator between neighbouring arguments. The open and closed functions themselves are different, but their derivatives are linked by $g_{s_0}'(a)=-f(a)$, where $f(a)=\e^{-s_0a}/a$ is the momentum-space form of Eq.~\eqref{eq:master_resolvent}. Together with the mass-derivative trace identity~\eqref{eq:sewing}, this relation reduces every mass or background derivative of the closed functional to the same $f$-based differences that occur on an open line.

Both operator functions are analytic on a neighbourhood of $\mathrm{spec}\,\Delta_A\subseteq[m^2,\infty)$ for $m>0$. The factor $\e^{-s_0a}$ is entire, and the pole of $f$ at $a=0$ lies outside the spectrum. Every divided difference evaluated on the spectrum is therefore finite, including at coinciding arguments. This is the operator-level counterpart of the finiteness of the effective vertices; Sec.~\ref{sec:multiloop} treats the remaining loop-momentum integrations. The kernels $I(a,b)$ and $J(a,b,c)$ introduced below are the first two cases of the same rule.

\subsection{Momentum routing, short-proper-time classes, and higher-order bookkeeping}

No separate diagram rule is needed to state the proper-time domain. For an ordered list $\boldsymbol B=(B_1,\ldots,B_r)$ of singleton or pair blocks, define
\begin{equation}
Q_{B_j}=\sum_{i\in B_j}q_i,
\qquad
p_j=p+\sum_{k=1}^{j}Q_{B_k},
\qquad
a_j=p_j^2+m^2.
\label{eq:block_momentum_routing}
\end{equation}
The scalar factor of that ordered worldline contribution is
\begin{align}
\mathcal M_r^{\rm FPT}(a_0,\ldots,a_r)
={}&\int_{\substack{u_j\ge0\\u_0+\cdots+u_r\ge s_0}}
\prod_{j=0}^{r}\dd u_j
\nonumber\\
&\times\exp\!\left[-\sum_{j=0}^{r}a_ju_j\right].
\label{eq:ordered_history_domain_explicit}
\end{align}
The unrestricted region $u_j\ge0$ gives $1/(a_0\cdots a_r)$. The bounded region removed by the endpoint, $u_j\ge0$ with $u_0+\cdots+u_r<s_0$, is bounded and gives
\begin{align}
\mathcal Q_r(a_0,\ldots,a_r)
&=\int_{\substack{u_j\ge0\\u_0+\cdots+u_r<s_0}}
\prod_{j=0}^{r}\dd u_j\,
\exp\!\left[-\sum_{j=0}^{r}a_ju_j\right]
\nonumber\\
&=(-1)^rq_{s_0}[a_0,\ldots,a_r].
\label{eq:short_region_divided_difference}
\end{align}
Therefore
\begin{equation}
\mathcal M_r^{\rm FPT}
=\frac1{a_0\cdots a_r}-(-1)^rq_{s_0}[a_0,\ldots,a_r].
\label{eq:moment_local_short_split}
\end{equation}

It is useful to keep the same identity multiplicative rather than to split the two terms under a
loop integral. Define the complete-history form factor
\begin{equation}
\boxed{\begin{aligned}
H_r(a_0,\ldots,a_r)
&=\Big(\prod_{j=0}^r a_j\Big)\mathcal M_r^{\rm FPT}\\
&=1-(-1)^r\Big(\prod_{j=0}^r a_j\Big)
q_{s_0}[a_0,\ldots,a_r].
\end{aligned}}
\label{eq:history_entire_form_factor}
\end{equation}
Because $q_{s_0}$ is entire, every divided difference extends uniquely to a jointly entire symmetric
function, including coincident arguments. Hence $H_r$ is jointly entire. More importantly,
\begin{equation}
\boxed{
H_r(a_0,\ldots,a_{c-1},0,a_{c+1},\ldots,a_r)=1
\qquad\text{for every }c.
}
\label{eq:history_cut_normalization}
\end{equation}
Thus a complete charged history is an ordinary product of physical propagator poles multiplied by
an entire history form factor that becomes unity on every pole that can be cut. This multiplicative
form preserves the assembled Euclidean damping and is the form used in the all-order contour and
finite-resolution reaction proofs below.
For a prefactor contribution $E_i$, the label $i$ is supplied by the prefactor rather than by a heat-kernel block, so the proper-time routing in Eqs.~\eqref{eq:block_momentum_routing}--\eqref{eq:short_region_divided_difference} involves only the remaining labels.

The function
\begin{equation}
q_{s_0}(a)=\frac{1-e^{-s_0a}}{a}=\int_0^{s_0}\dd T\,e^{-Ta}
\label{eq:q_short_proper_time}
\end{equation}
is entire. The first five repeated-argument values are
\begin{align}
q[0]&=s_0,
& q[0,0]&=-\frac{s_0^2}{2},
\nonumber\\
q[0,0,0]&=\frac{s_0^3}{6},
& q[0,0,0,0]&=-\frac{s_0^4}{24},
\nonumber\\
q[0,0,0,0,0]&=\frac{s_0^5}{120}.
\label{eq:q_coincident_values_explicit}
\end{align}
These are obtained directly by differentiating $q_{s_0}(a)$ at $a=0$.

The short-proper-time divided-difference classes corresponding to the complete second-, third-, and fourth-order kernels in Sec.~\ref{sec:open} are shown below; $b,c,d$ denote finite internal spectral arguments, while the zeros denote external pole arguments.
\begin{center}
\begin{tabular}{c|c|c}
order & class & representative short factor\\ \hline
$2$ & $M_AV_1^2$ & $q[0,b,0]$\\
$2$ & $M_AV_2$ & $q[0,0]$\\
$2$ & $EV_1$ & $q[0,b]$\\ \hline
$3$ & $M_AV_1^3$ & $q[0,b,c,0]$\\
$3$ & $M_AV_2V_1$ & $q[0,b,0]$\\
$3$ & $EV_1^2$ & $q[0,b,c]$\\
$3$ & $EV_2$ & $q[0,b]$\\ \hline
$4$ & $M_AV_1^4$ & $q[0,b,c,d,0]$\\
$4$ & $M_AV_2V_1^2$ & $q[0,b,c,0]$\\
$4$ & $M_AV_2^2$ & $q[0,b,0]$\\
$4$ & $EV_1^3$ & $q[0,b,c,d]$\\
$4$ & $EV_2V_1$ & $q[0,b,c]$
\end{tabular}
\end{center}
All entries stay finite when an external inverse propagator tends to zero. They also stay finite when internal arguments coincide.

The counting extends directly to arbitrary finite derivative order. If $j$ pair contacts occur in an $n$th derivative, with $0\le j\le\lfloor n/2\rfloor$, the number of ordered non-prefactor terms is
\begin{equation}
N^{(M)}_{n,j}
=\frac{n!(n-j)!}{2^j j!(n-2j)!}.
\label{eq:general_M_count}
\end{equation}
For prefactor contributions the allowed range is $0\le j\le\lfloor(n-1)/2\rfloor$, and the corresponding count is
\begin{equation}
N^{(E)}_{n,j}
=\frac{n!(n-1-j)!}{2^j j!(n-1-2j)!}.
\label{eq:general_E_count}
\end{equation}
For $n=5$ these formulas give $120+240+90$ non-prefactor terms and $120+180+30$ prefactor contributions, or 780 terms altogether, in agreement with the same partition rule.

The finite-virtuality behaviour can be seen in the repeated-argument second derivative. After the two scalar pole factors are removed, the remaining endpoint factor is
\begin{equation}
F(a,c)=e^{-ca}(1+ca)=1-\frac12(ca)^2+O\!\left((ca)^3\right),
\label{eq:F_endpoint_series}
\end{equation}
where $c=\max(0,s_0-u_2)$. Multiplication by the complete finite-$s_0$ inverse propagators gives
\begin{equation}
e^{2s_0a}F(a,c)=1+2s_0a+
\left(2s_0^2-\frac12c^2\right)a^2+O\!\left(s_0^3a^3\right).
\label{eq:F_full_inverse_series}
\end{equation}
Although both factors approach one at the external pole, neither is equal to one at finite virtuality, and every off-shell quantity below is therefore obtained from the unreduced differentiated kernel.

The momentum-space endpoint factors follow from the spectral operator construction and the topology of the assembled graph. At one-history level, the open and closed charged kernels carry the interval--circle endpoint inherited from their common heat semigroup. Functional differentiation only partitions those operator histories. After local sewing, the edge totals generated in this way enter the primitive closed momentum circulations of the graph. If $B_G$ is the incidence matrix, these primitive circulations are the support-minimal nonzero elements of $\ker B_G$, equivalently the graph circuits. For each such circuit $c$ the associated charged edge totals obey the inherited complete-history condition on their sum. No fermion edge is given an endpoint merely because it is internal.

The photon sector has an additional source-level fact specific to the present QED definition. Its Gaussian covariance is itself obtained from
\begin{equation}
P^{-1}_\Lambda=\int_{s_0}^{\infty}\dd s\,e^{-sP}=e^{-s_0P}P^{-1},
\label{eq:truncated_inverse}
\end{equation}
so in unit covariant gauge
\begin{equation}
D_{\mu\nu,\Lambda}(k)=\delta_{\mu\nu}\frac{e^{-s_0k_E^2}}{k_E^2}.
\label{eq:derived_photon_covariance}
\end{equation}
Every photon contraction generated by this Gaussian measure therefore carries its own non-negative proper-time parameter $\sigma_\gamma\ge s_0$. 

It is useful to distinguish these two pieces of information. Write $S_e$ for the total of all Duhamel descendants on a charged edge $e$. Functional differentiation fixes only the partition $S_e=\sum_j u_{e,j}$ with $u_{e,j}\ge0$. Local momentum conservation then determines which collections of edge totals form primitive closed circulations. The graph-level domain may therefore be read in two stages:
\begin{equation}
u_{e,j}\ge0,\qquad
\sum_{e\in c}S_e+\sum_{\gamma\in c}\sigma_\gamma\ge s_0
\quad(c\in\mathcal C(G)),
\label{eq:domain_assignment}
\end{equation}
together with the source-functional photon conditions $\sigma_\gamma\ge s_0$. For a purely charged closed circuit, Eq.~\eqref{eq:domain_assignment} is the direct interval--circle descendant of the charged heat kernel. For a mixed charged--photon circuit, the circuit inequality is automatically satisfied whenever it contains a photon contraction because $\sigma_\gamma\ge s_0$. The stronger photon condition is a model-specific consequence of the QED photon covariance; it should not be confused with a universal per-edge FPT rule.

This is the precise relation to the graph-circuit construction in the scalar control theory. In the scalar theory a bridge lies on no circuit and therefore receives no endpoint. In the present QED source functional a photon bridge still carries the covariance factor in Eq.~\eqref{eq:derived_photon_covariance}; a charged bridge does not acquire an independent endpoint merely by being a bridge or an internal line. Thus the common topological statement concerns the charged operator ancestry and the primitive closed circulations, while the independently truncated photon covariance is an additional ingredient of this QED model.

After external charged legs are amputated on shell, the inherited open-interval pole constraint disappears by Eq.~\eqref{eq:explicit_234_pole_reduction}. Any surviving charged edge totals are then constrained only through the closed circuits of the remaining virtual graph, whereas every uncut internal photon contraction retains its source-level covariance endpoint. Under a physical cut, the endpoint factor on each cut pole is unity and every circuit opened by the cut ceases to constrain either daughter amplitude. Primitive circuits internal to either daughter subgraph retain precisely the conditions they would have if that daughter were constructed independently.

\label{app:proper_time_rules}
\rev{We state the proper-time rules that are used throughout the paper. We note that the rules are universal and not limited to QED.
\begin{itemize}
\item[\textbf{R1}] \emph{Segments.} Every charged segment has $u_e\ge0$. Every internal photon, bridges included, has $\sigma_\gamma\ge s_0$. These are the two source parents of QED.
\item[\textbf{R2}] \emph{Circles.} Every purely charged closed circuit has total proper time $\sum_{e\in c}S_e\ge s_0$. A mixed charged--photon circuit imposes nothing new: it contains a photon, and $\sigma_\gamma\ge s_0$ already implies its circuit inequality. Only pure charged circles supply non-redundant circuit bounds.
\item[\textbf{R3}] \emph{Insertions.} An insertion partitions the proper time of its ancestor, $S_e=\sum_ju_{e,j}$, and never creates a new endpoint. Moving proper time between the pieces at fixed total leaves the domain unchanged, $(\partial_{u_{e,1}}-\partial_{u_{e,2}})\chi_G=0$; this is the geometric content of the Ward identities.
\item[\textbf{R4}] \emph{Open paths.} After external charged legs are amputated on shell, an open charged path carries no additional parent endpoint. This removal is a pole operation on the source; it is not a replacement of $S_{s_0}[A]$ by $S_0[A]$ off shell.
\item[\textbf{R5}] \emph{Cuts.} A cut deletes its edges. The residue at the cut poles is the product of the independently constructed daughters with the surviving indicator: uncut circles keep their bounds, cut circles open, uncut photons stay damped, cut photons contribute unit residue. The cut measure is the local one because the factors associated with the cut poles equal one on shell.
\item[\textbf{R6}] \emph{Reductions.} Cancelling a denominator retains its boundary face, $a\int_0^\infty\e^{-au}\,\Theta(u+U-s_0)\,\dd u=\e^{-a(s_0-U)_+}$; a contraction is not a deletion.
\item[\textbf{R7}] \emph{Channels.} The retained endpoint acts on the off-shell part of physical-channel propagation and leaves the on-shell channel measure unchanged. For a single channel insertion with energy difference $\Delta E=E_n-E_0$, the finite duration $\mathcal T$ gives the factor $\int_0^{\mathcal T}\dd t\,\e^{i\Delta E t}=[\e^{i\Delta E\mathcal T}-1]/(i\Delta E)$, with value $\mathcal T$ at $\Delta E=0$. Multiple insertions retain their ordered time integrals.
\end{itemize}
R1--R3 fix the domain, R4--R6 fix how the domain behaves under pole operations, and R7 fixes how the domain meets the physical channel space. The boundary at $s_0$ is never crossed by any of these operations; a cut opens a bound, a reduction retains a face, an insertion moves along the boundary.}

The photon covariance~\eqref{eq:derived_photon_covariance}, together with the open and closed charged kernels defined at the beginning of this section, fixes the perturbative building blocks of the theory. Functional derivatives of the open kernel generate its photon insertions and self-energy classes, while derivatives of the closed functional generate photon one-particle-irreducible kernels. The covariance supplies the Gaussian damping used in the fixed-order ultraviolet analysis of Sec.~\ref{sec:multiloop}. Mass and charge are fixed later by finite matching to their measured values.

\subsection{Proper-time assignment under sewing}
\label{sec:history_autonomy}
The criterion is topological. Construct the complete graph first and identify its primitive circuits from the support-minimal nonzero solutions of
\begin{equation}
B_G z=0,
\label{eq:spinor_cycle_space}
\end{equation}
\revblue{where $B_G$ is the incidence matrix. Only non-redundant source-history inequalities can obstruct a factorisation of the proper-time domain. In this QED source, mixed charged--photon circuit inequalities are already implied by the individual photon bounds of R1. The relevant test is therefore whether a pure charged parent circle crosses the boundary of an inserted subgraph. If each non-redundant charged-circle constraint is wholly internal or wholly external, and each photon parameter retains its own assigned covariance, the corresponding indicator factorises. Otherwise, for a pure charged circle $c\in\mathcal H_G$ crossing a subgraph $H$, its charged edge totals obey}
\revblue{\begin{equation}
\sum_{e\in c\cap H}S_e+
\sum_{e\in c\setminus H}S_e\ge s_0,
\qquad c\in\mathcal H_G.
\label{eq:parent_history_inequality}
\end{equation}}
Those shared variables cannot be integrated out inside $H$ before the parent graph is assembled.

The same distinction appears in the scalar theory for inserted self-energy subgraphs. Duhamel ancestry determines how an edge total is partitioned, while the complete graph determines whether that edge total participates in a parent circulation. The exact global two-point identity
\begin{equation}
G_{\rm exact}^{-1}(p)=G_0^{-1}(p)-\Sigma_{\rm exact}(p)
\label{eq:global_exact_self_energy_definition}
\end{equation}
is unaffected. What is not generally valid is the stronger diagrammatic step of replacing an off-shell two-point insertion embedded in a larger closed circuit by a parent-independent momentum-only dressed propagator. \revblue{The insertion retains the shared proper-time arguments whenever a non-redundant charged-history constraint crosses its boundary.} Figure~\ref{fig:operator_to_circuit}(c) illustrates this obstruction to universal self-energy resummation.

\section{Heat-Kernel and Worldline Bounds Used in the Ultraviolet Estimate}
\label{sec:worldline}
The closed-loop determinant admits the first-quantised worldline representation~\cite{Strassler1992,Schubert2001}. The same second-order worldline language also gives open dressed electron propagators with photon insertions~\cite{Ahmadiniaz2020}. For an $N$-photon insertion on a closed worldline of length $T$, the bosonic Green function is $G_B(\tau,\tau';T)=|\tau-\tau'|-(\tau-\tau')^2/T$ and the momentum-dependent exponent is $Q=\sum_{i<j}k_i\cdot k_j\,G_B(\tau_i,\tau_j;T)$.
Here $k_i$ are the incoming Euclidean photon momenta and $\tau_i\in[0,T]$ their insertion positions. Momentum conservation, $\sum_i k_i=0$, makes the exponent non-positive:
\begin{equation}
Q=-\frac{T}{2\pi^2}\sum_{n=1}^{\infty}\frac{|f_n|^2}{n^2}\le0,
\qquad
f_n=\sum_i k_i\e^{2\pi\ii n\tau_i/T}.
\label{eq:Q_fourier}
\end{equation}
Here $|f_n|^2$ is the Euclidean norm squared of the vector $f_n$. To obtain this form, insert the Fourier series
$G_B=T/6-(T/\pi^2)\sum_{n\ge1}\cos[2\pi n(\tau-\tau')/T]/n^2$ into $Q$; momentum conservation and $\sum_{n\ge1}n^{-2}=\pi^2/6$ cancel the constant terms.
Equation~\eqref{eq:Q_fourier} gives $\e^Q\le1$. If two opposite momenta are inserted at the same worldline point, then $Q=0$. For fixed $N$, the remaining tensor structure is polynomial in the external momenta and in $\dot G_B\equiv\partial_\tau G_B$, with $|\dot G_B|\le1$, while the proper-time integral starts at $s_0$ and is finite. The closed-loop effective vertices are therefore polynomially bounded on Euclidean momentum compacts.

The same ultraviolet conclusion holds for open fermion lines. In the representation used in Secs.~\ref{sec:open} and~\ref{sec:second}, the kernels $I$ and $J$ are bounded by inverse powers of positive Euclidean denominators. The same fact is visible directly in proper time. If $T$ is the total open-line proper time and $\tau_<\equiv\min(\tau,\tau')$, $\tau_>\equiv\max(\tau,\tau')$, the Dirichlet Green function
$G_D(\tau,\tau';T)=2\tau_<(T-\tau_>)/T$ obeys $0\le G_D\le T/2$ and $|\partial_\tau G_D|\le2$. After the standard worldline integrations by parts remove the distributional second derivatives, the Euclidean momentum exponent is non-positive and the $N$ insertion integrals contribute at most $T^N$. Thus, if $|k|_{\max}$ denotes the largest magnitude among the external insertion momenta, choose any fixed reference mass $\mu_{\rm ref}>0$. With $C_N$ dimensionless and independent of the loop momenta on a fixed Euclidean compact, the dressed line is bounded by
$C_N^N\mu_{\rm ref}^N(1+|k|_{\max}/\mu_{\rm ref})^N\int_{s_0}^\infty\dd T\,T^N\e^{-Tm^2}$.
This is finite for every fixed $N$. The compact short-proper-time remainder of a single worldline is correspondingly controlled by the scale $s_0$; graph-level timelike growth is analysed only after all independent proper-time constraints have been assembled in Sec.~\ref{subsec:endpoint}.

\section{Finite-Resolution Propagation and Physical Domains}
\subsection{Source-generated observables}
\revblue{This appendix records consequences for an exact unitary physical realisation, if supplied with the source-to-boundary identification described in Sec.~\ref{sec:FR_closure}. Its Gram positivity and composition equations are not independent proofs of that identification.}

Let real sources couple to gauge-invariant probes $O_a$. The unitary finite-resolution operator defines
\begin{equation}
Z_{\rm phys}[J_+,J_-]
=\langle\Omega|S_{s_0,\mathcal T}[J_-]^\dagger
S_{s_0,\mathcal T}[J_+]|\Omega\rangle .
\label{eq:np_physical_CTP}
\end{equation}
Source differentiation on the dense channel domain generates the corresponding observables,
\begin{equation}
O_a(f)|J\rangle
=\left.\frac1{i}\frac{\dd}{\dd\lambda}
|J+\lambda f_a\rangle\right|_{\lambda=0}.
\label{eq:np_source_observable}
\end{equation}
Their finite moment matrices inherit positivity from the physical channel inner product of Sec.~\ref{sec:FR6}. Thus no separate Euclidean state metric is introduced.

The bounded resolved observables form a concrete operator algebra. Let $E_{\mathsf R}(B)$ denote the effects of the resolved detector POVM, let $H$ be the positive propagation generator of Eq.~\eqref{eq:FR6_positive_energy}, and let $\mathcal B_{\rm src}$ denote any chosen family of bounded spectral functions of self-adjoint source-generated observables for which the physical functional calculus is defined. Define
\begin{align}
\mathfrak A_{\mathfrak E}
=C^*\!\Big(&\{E_{\mathsf R}(B),
S_{\mathfrak E}^\dagger E_{\mathsf R}(B)S_{\mathfrak E},
\nonumber\\[-2pt]
& e^{-itH}:B,t\}\cup\mathcal B_{\rm src}\Big)
\subset B(\mathcal H_{\mathfrak E}).
\label{eq:FR_resolved_observable_algebra}
\end{align}
This is a unital $C^*$ algebra by construction, and every normalized channel vector defines a positive state on it. Source-generated operators that are unbounded are defined on the dense source domain of Eq.~\eqref{eq:np_source_observable}; only their bounded functional-calculus elements are included in $\mathcal B_{\rm src}$. Thus the finite-resolution construction has a Hilbert space and a concrete algebra of bounded physical observables; its unitary channel maps possess bounded self-adjoint logarithms. 

\subsection{Clustering and charged long-range structure}
At a free physical pole
\begin{equation}
R_{s_0}(A)=\frac1A+\frac{e^{-s_0A}-1}{A},
\label{eq:np_pole_entire_again}
\end{equation}
and the second term is entire. The upper-minus-lower pole discontinuity is therefore the causal free commutator distribution. More generally, whenever an interacting gauge-invariant two-point amplitude is written as
\begin{align}
G&=G_{\rm disp}+E_G,
&\operatorname{Disc}G_{\rm disp}(s)&=2\pi i\rho(s),
\nonumber\\
\rho(s)&\ge0,
&\operatorname{Disc}E_G&=0.
\label{eq:np_two_point_disp_entire}
\end{align}
the commutator determined by $G_{\rm disp}$ is a positive superposition of ordinary causal free commutator functions and therefore has spacelike causal support. This statement applies to the spectral part only. The finite-$s_0$ entire remainder is independent boundary data and is not constrained by the K\"all\'en--Lehmann discontinuity.

The entire remainder nevertheless defines a controlled non-tempered functional at every fixed graph order. The proper-time domain and the graph-dependent continuation bound of Sec.~\ref{subsec:endpoint} give constants $c_G,N_G<\infty$ such that, on each fixed real continuation channel,
\begin{equation}
|E_G(p)|\le C_G(1+|p|)^{N_G}
\exp\!\big(c_Gs_0|p|^2\big).
\label{eq:np_entire_gaussian_growth}
\end{equation}
For $a>c_Gs_0$ define the Gaussian-decay test class by the norm
\begin{equation}
\|f\|_{a,N_G}
=\sup_p\,e^{a|p|^2}(1+|p|)^{N_G+5}|\widehat f(p)|<\infty.
\label{eq:np_gaussian_test_norm}
\end{equation}
Then
\begin{equation}
|\langle E_G,f\rangle|
\le C_{G,a}\,\|f\|_{a,N_G},
\qquad
C_{G,a}<\infty,
\label{eq:np_entire_analytic_functional}
\end{equation}
by direct Gaussian integration. Thus $E_G$ is a continuous linear functional on this explicit Gaussian test space, while $G_{\rm disp}$ is an ordinary spectral distribution. This is the analytic-test-function setting appropriate to the exponential endpoint growth. 

The corresponding higher-point target is analyticity of retarded amplitudes in forward/backward momentum tubes together with appropriate carrier-cone bounds~\cite{Soloviev2006}. At fixed perturbative order the Euclidean-first contour construction proves the local analytic part of this statement on every compact tube patch that does not cross a reduced Landau surface: the endpoint factors are entire, the loop-energy contour ends remain Gaussian damped, and differentiation with respect to complex external momenta is justified by the same uniform majorant used in the cutting proof. Strict point microcausality, or a global carrier-cone theorem joining all such patches, is a stronger continuum-locality statement. The finite-resolution reconstruction instead uses causal composition across resolved preparation, reset, and detection boundaries, so the carrier-cone theorem is not a defining assumption of the physical construction.

The FPT endpoint itself adds no additional long-range charged obstruction. For the massless scalar part of the photon kernel in four Euclidean dimensions,
\begin{align}
D_{s_0}(r)&=\frac{1-e^{-r^2/(4s_0)}}{4\pi^2r^2},\nonumber\\
D_{s_0}(r)-D_0(r)
&=-\frac{e^{-r^2/(4s_0)}}{4\pi^2r^2}.
\label{eq:np_short_range_endpoint}
\end{align}
Along separated charged trajectories with $r(t)\ge v|t|$, the endpoint correction is time-integrable because
\begin{equation}
\int_{T}^{\infty}\frac{e^{-at^2}}{t^2}\dd t
=\frac{e^{-aT^2}}{T}-\sqrt{\pi a}\,\operatorname{erfc}(\sqrt a\,T)<\infty.
\label{eq:np_Cook_integral}
\end{equation}
Thus FPT introduces no additional long-range phase beyond the standard massless-QED one. The finite-resolution theory itself stays at $\mathcal T<\infty$; if an infinite-time asymptotic charged scattering theory is additionally demanded, the usual QED Gauss-law/infraparticle problem remains~\cite{KulishFaddeev1970,FrohlichMorchioStrocchi1979,Buchholz1986}.

\subsection{Finite-time Lorentzian transient layer}
With the finite-resolution scattering operator of Sec.~\ref{sec:FR_programme}, the finite-time channel map is unitary on $\mathcal H_{\rm phys}$. Self-adjoint propagation between interaction regions therefore acts directly on the physical channel space.

Let $a$ label a resolved detector channel, including species, spin or polarization, angular packet, and finite resolved multiparticle labels. Choose extraction maps
\begin{equation}
D_a:\mathcal H_{\rm phys}(K)\longrightarrow L^2(M;\mathbb C^{n_a})
\end{equation}
forming a Bessel frame,
\begin{equation}
\int\dd\mu(a)\,D_a^\dagger D_a\le I.
\label{eq:np_detector_Bessel}
\end{equation}
For an incoming state $\Psi$ define
\begin{equation}
f_a=D_aS^{\rm phys}_{s_0,\mathcal T}\Psi.
\label{eq:np_detector_extract}
\end{equation}
Then exact channel unitarity and Eq.~\eqref{eq:np_detector_Bessel} give
\begin{align}
\int\dd\mu(a)\,\|f_a\|_2^2
&=
\left\langle
S^{\rm phys}\Psi,
\left(\int\dd\mu(a)D_a^\dagger D_a\right)
S^{\rm phys}\Psi
\right\rangle
\nonumber\\
&\le
\|S^{\rm phys}\Psi\|^2
=
\|\Psi\|^2.
\label{eq:np_detector_contract}
\end{align}
All virtual histories and interaction orders are already contained in
$S^{\rm phys}$, so Eq.~\eqref{eq:np_detector_contract} carries no
perturbative-order or channel-counting factor.

A contractive detector frame and Hilbert-valued Bernstein estimate give, for a resolved frequency scale $\lambda$ on a finite observation interval $I$ in $3+1$ dimensions,
\begin{equation}
\|U_\lambda\|_{L^4(I\times M)}
\le C_B\lambda^{3/4}|I|^{1/4}\|\Psi\|_{\mathcal H_{\rm phys}},
\label{eq:np_transient_wavepacket_bound}
\end{equation}
with the normalized-window form replacing $|I|^{1/4}$ by the corresponding $L^4$ window factor. Resolved photons have energies of order $\gtrsim\mathcal T^{-1}$, while arbitrarily soft unresolved photons remain inside the normalized coherent measure $\nu_{\mathcal T}$. 

For a normalized temporal window $w_{\mathcal T}$ with $\|w_{\mathcal T}\|_2=1$, the same argument gives
\begin{equation}
\|w_{\mathcal T}U_\lambda\|_{L^4_{t,x}}
\le
C_B\lambda^{3/4}
\|w_{\mathcal T}\|_4
\|\Psi\|_{\mathcal H_{\rm phys}}.
\label{eq:np_transient_normalized_window}
\end{equation}
For the rectangular normalized window, $\|w_{\mathcal T}\|_4=\mathcal T^{-1/4}$. Thus the apparent power of $\mathcal T$ distinguishes an unnormalized exposure from a normalized finite-time average; neither expression uses $\mathcal T\to\infty$.

If the detector resolution is
\begin{equation}
\omega_{\rm res}\sim c/\mathcal T,
\end{equation}
photons below this scale belong to the coherent unresolved sector. Its displacement operator obeys
\begin{equation}
D(\alpha_{\mathcal T})^\dagger D(\alpha_{\mathcal T})=I,
\end{equation}
so including the unresolved multiplicity cannot enlarge Eq.~\eqref{eq:np_detector_contract}. On a compact total-energy set, the number of individually resolved photons is finite, while arbitrarily soft multiplicity is already summed in the normalized measure $\nu_{\mathcal T}$.

\subsubsection{Physical spacetime tubes and multiscale wave-packet organization}
The compact complex tubes used for analytic continuation must not be confused with physical spacetime wave-packet tubes. The latter arise only after the physical channel operator has been reconstructed. For a resolved massless packet with $|\xi|\sim\lambda$ observed for the finite duration $\mathcal T$, put
\begin{equation}
R=\lambda\mathcal T.
\label{eq:np_HW_R}
\end{equation}
The standard cone packet aperture is
\begin{equation}
\delta\theta\sim R^{-1/2}
=(\lambda\mathcal T)^{-1/2}.
\label{eq:np_HW_aperture}
\end{equation}
Hence
\begin{equation}
\Delta p_\perp\sim\lambda\delta\theta,
\qquad
\Delta x_\perp
\sim
\frac1{\lambda\delta\theta}
\sim
\sqrt{\frac{\mathcal T}{\lambda}},
\label{eq:np_HW_uncertainty}
\end{equation}
while angular spreading over the same interval gives the identical scale $\mathcal T\delta\theta$. A minimum-spreading physical photon packet is therefore concentrated in a finite spacetime tube of radius
\begin{equation}
\boxed{
r_{\lambda,\mathcal T}
\sim
\sqrt{\frac{\mathcal T}{\lambda}} .
}
\label{eq:np_HW_tube_radius}
\end{equation}
This is a Lorentzian detector/localization scale. It is unrelated to the virtual spectral length $\sqrt{s_0}$.

The multiscale cone-packet organization groups fine tubes into ancestor packets at intermediate physical scales. Orthogonality implies
\begin{equation}
\sum_{\tau_{\rm anc}}
\|f_{\tau_{\rm anc}}\|_2^2
\le
\sum_{\tau}\|f_\tau\|_2^2 .
\label{eq:np_HW_ancestor_orthogonality}
\end{equation}
Thus regrouping physical packets does not generate a channel-counting factor. These on-shell spacetime tubes are physical localization objects and are distinct from virtual proper-time circuits. Sharp cone square-function and local-smoothing estimates provide a refined geometric control of the photon branch~\cite{GuthWangZhang2020,WangZahl2025,GuthWangZahl2026,GaoLiuMiaoXi2023,GanHeLiWu2026,RozendaalSchippa2026}. For a half-wave Fourier integral operator
\begin{equation}
\mathcal Ff(x,t)
=
\int_{\mathbb R^3}\dd^3\xi\,
e^{i\varphi(x,t;\xi)}
a(x,t;\xi)\widehat f(\xi),
\label{eq:np_HW_FIO}
\end{equation}
satisfying the standard nondegeneracy and cinematic-curvature hypotheses, the $3+1$ local-smoothing estimate gives, at the safe exponent $p=4$ and every $\epsilon>0$,
\begin{equation}
\|\mathcal FP_\lambda f\|_{L^4(M\times I)}
\le
C_{\epsilon,G,I}
\lambda^{1/4+\epsilon}
\|P_\lambda f\|_{L^4(M)}.
\label{eq:np_HW_local_smoothing}
\end{equation}
Together with three-dimensional Bernstein,
\begin{equation}
\|P_\lambda f\|_4
\le C_B\lambda^{3/4}\|P_\lambda f\|_2,
\end{equation}
this yields
\begin{equation}
\|\mathcal FP_\lambda f\|_{L^4(M\times I)}
\le
C_{\epsilon,G,I}\lambda^{1+\epsilon}
\|P_\lambda f\|_2 .
\label{eq:np_HW_L2_refinement}
\end{equation}
Equation~\eqref{eq:np_HW_L2_refinement} is a refinement of packet concentration the universal finite-time bound Eq.~\eqref{eq:np_transient_wavepacket_bound} already follows from unitarity and Bernstein.

\subsubsection{Modified geometry and self-adjoint domains}
Let a resolved physical branch be generated microlocally by a real Hamiltonian symbol $h(t,x,\xi)$. Its packet center follows
\begin{equation}
\dot x=\partial_\xi h,
\qquad
\dot\xi=-\partial_xh .
\label{eq:np_transient_Hamilton}
\end{equation}
For a photon in a smooth spatial metric one may take
\begin{equation}
h_\gamma(x,\xi)
=
\sqrt{g^{ij}(x)\xi_i\xi_j},
\end{equation}
and for a positive-energy massive charged branch,
\begin{equation}
h_e(x,\xi)
=
\sqrt{g^{ij}(x)\xi_i\xi_j+m_{\rm phys}^2}.
\end{equation}
These are physical dispersion laws and do not alter the Euclidean proper-time boundary. If
\begin{equation}
D\Phi_t=
\begin{pmatrix}
A_t&B_t\\
C_t&D_t
\end{pmatrix}
\end{equation}
is the linearized canonical flow and the initial phase-space covariance is $\Sigma_0=\operatorname{diag}(\Sigma_x,\Sigma_\xi)$, then
\begin{equation}
\Sigma_x(t)
=
A_t\Sigma_xA_t^T+B_t\Sigma_\xi B_t^T,
\qquad
r_g(t)=\sqrt{\lambda_{\max}[\Sigma_x(t)]}.
\label{eq:np_transient_curved_tube}
\end{equation}
Focusing and anisotropy are therefore properties of the physical canonical flow, not modifications of $s_0$.

A self-adjoint boundary changes the domain of the physical Lorentzian generator. Let
\begin{equation}
\mathcal B:
\mathcal H_{\rm phys}
\longrightarrow
\int^\oplus\dd\mu(\alpha)\,\mathcal H_\alpha
\label{eq:np_domain_spectral_transform}
\end{equation}
be its unitary spectral transform. Parseval gives
\begin{equation}
\int\dd\mu(\alpha)\,
\|(\mathcal B\Psi)_\alpha\|_2^2
=
\|\Psi\|_2^2.
\end{equation}
Each spectral branch evolves with its self-adjoint finite-time propagator; the Hilbert-valued Bernstein argument therefore gives
\begin{multline}
\left\|
\left(
\int\dd\mu(\alpha)
|U_\alpha(t)P_\lambda(\mathcal B\Psi)_\alpha|^2
\right)^{1/2}
\right\|_{L^4(I\times M)}\\
\le C_B\lambda^{3/4}|I|^{1/4}\|\Psi\|.
\label{eq:np_domain_transient_bound}
\end{multline}
Continuous angular spectra therefore introduce no independent channel-counting constant.

For several separated interaction and propagation regions, the channel maps compose as in Eq.~\eqref{eq:FR7_composition}. Self-adjoint domain propagation therefore preserves the finite-resolution norm without introducing a perturbative-order or infinite-time factor.

\subsection{Operational causal ordering}
\label{sec:FR3}
The finite-resolution causal statement concerns preparation, reset and detection boundaries. Let $\Sigma$ separate two operational cells. Resolving a complete physical channel at $\Sigma$ opens every virtual history intersecting that boundary, and Eq.~\eqref{eq:cut_domain_consistency} prevents an $s_0$ constraint belonging to a cut history from being inherited by either daughter operation. For source perturbations supported on opposite sides of $\Sigma$,
\begin{equation}
\boxed{\frac{\delta}{\delta f_+(x)}\mathcal O_-[f_-]=0}
\label{eq:FR3_no_future_to_past}
\end{equation}
for observables completed before the boundary. Between cells, self-adjoint Maxwell or Dirac propagation on $\mathcal D$ gives the usual hyperbolic propagation of the resolved packets. This operational statement is sufficient for the finite sequence of physical preparations and detections considered here; strict interacting point microcausality inside one unresolved cell is a stronger locality question.

\subsection{Composition across physical boundaries}
\label{sec:FR7}
For operational cells $\mathfrak E_r$ separated by genuine physical boundaries, let $S_r$ be the interaction scattering operator above and $U_r$ the self-adjoint propagation operator between successive cells. Their composition is
\begin{equation}
\boxed{\mathbb U=U_nS_nU_{n-1}S_{n-1}\cdots U_1S_1U_0,}
\label{eq:FR7_composition}
\end{equation}
and
\begin{equation}
\boxed{\mathbb U^\dagger\mathbb U=I.}
\label{eq:FR7_unitarity}
\end{equation}
At a genuine physical boundary one inserts
\begin{equation}
I_{\rm phys}=\sum_X\int d\Phi_X^{\rm phys}|X\rangle\langle X|,
\label{eq:FR7_channel_identity}
\end{equation}
which is exactly the physical deletion operation on histories intersected by that boundary. No virtual history is split by an auxiliary mathematical time slice.

\section{Auxiliary Global Continuum Estimates}
\label{sec:physical_boundary}
\revblue{Section~\ref{sec:FR_programme} establishes an exact finite-domain Euclidean source and a coefficientwise physical reaction representation.} Section~\ref{sec:FR_domain_limit} already establishes the free-space limit of every fixed perturbative coefficient under a regular receding-boundary exhaustion. This appendix studies a different question: whether the exact interacting source functional admits an all-coupling free-space exhaustion and whether that object, if constructed, agrees with an independently interpreted direct-$+i0$ boundary. \revblue{The free-space exhaustion is a correspondence problem; the source-to-physical-boundary identification is also required for an exact physical realization at fixed finite domain.} All shell decompositions, block norms, and forest parameters below are auxiliary proof variables. 

\subsection{Finite-volume stability in rescaled variables}
It is useful to put the electric coupling into the photon Gaussian. Define a rescaled gauge background
\begin{equation}
b_\mu=eA_\mu,\qquad g=e^2=4\pi\alpha.
\label{eq:np_rescaled_background}
\end{equation}
Define
\begin{equation}
\begin{aligned}
D_\mu(b)&=\partial_\mu-\ii b_\mu,
&\slashD_b&=\gamma^\mu D_\mu(b),\\
B_b&=m+\slashD_b,
&\Delta_b&=B_b^\dagger B_b=m^2-\slashD_b^{\,2}\ge m^2.
\end{aligned}
\label{eq:np_rescaled_Delta}
\end{equation}
Thus the charged operator depends on $b$ while the photon covariance is $gC_{s_0}$. At finite Euclidean volume $\mathcal V$ and finite nonzero photon-mode count $N$, the exact source functional can be written schematically as
\begin{equation}
Z_{\mathcal V,N}[J,\bar\eta,\eta;g]
=\frac{\int \dd\mu_{gC_{s_0}}(b)\,
e^{-\Gamma_{1,s_0,\mathrm{ch}}[b]}
e^{J\cdot b+\bar\eta S_{s_0}[b]\eta}}
{\int \dd\mu_{gC_{s_0}}(b)\,e^{-\Gamma_{1,s_0,\mathrm{ch}}[b]}}.
\label{eq:np_exact_finite_source}
\end{equation}
With the positive squared Dirac operator in Eq.~\eqref{eq:np_rescaled_Delta}, define
\begin{equation}
\Gamma_{1,s_0,\mathrm{ch}}[b]
=\frac12\Tr\Gamma(0,s_0\Delta_b),
\label{eq:np_closed_spectral_functional}
\end{equation}
where $\Gamma(0,x)=\int_x^\infty t^{-1}e^{-t}\dd t>0$ for $x>0$. Therefore, for every real finite-volume background,
\begin{equation}
\boxed{\Gamma_{1,s_0,\mathrm{ch}}[b]\ge0,\qquad 0<e^{-\Gamma_{1,s_0,\mathrm{ch}}[b]}\le1.}
\label{eq:np_all_background_stability}
\end{equation}
This all-background inequality is stronger and safer than trying to dominate a polynomial expansion of the determinant by the Maxwell action. For the vacuum-subtracted functional $F[b]=\Gamma_{1,s_0,\mathrm{ch}}[b]-\Gamma_{1,s_0,\mathrm{ch}}[0]$,
\begin{equation}
F[b]\ge-\Gamma_{1,s_0,\mathrm{ch}}[0].
\label{eq:np_vac_sub_lower}
\end{equation}
For a four-component Dirac spinor,
\begin{align}
\frac{\Gamma_{1,s_0,\mathrm{ch}}[0]}{\mathcal V}
&=\frac{1}{8\pi^2}\int_{s_0}^{\infty}\frac{\dd T}{T^3}e^{-m^2T}\nonumber\\
&\le\frac{1}{16\pi^2s_0^2}.
\label{eq:np_vacuum_bound}
\end{align}
Hence the exact fermionic factor never enhances the unnormalised large-field photon Gaussian weight. Finite-volume existence holds at every real coupling.

The ultraviolet mode limit is a numerical/functional approximation to the same fixed-$s_0$ theory. If photon modes above $K_N$ are omitted,
\begin{equation}
\int_{|k|>K_N}\frac{\dd^4k}{(2\pi)^4}\frac{e^{-s_0k^2}}{k^2}
=\frac{e^{-s_0K_N^2}}{16\pi^2s_0},
\label{eq:np_mode_tail}
\end{equation}
which vanishes exponentially as $N\to\infty$ at fixed $s_0$. The limits $N\to\infty$ and $\mathcal V\to\infty$ remove auxiliary approximations.

\subsection{Gaussian linearization of the finite-volume flow}
The finite-volume nonperturbative statement can be sharpened further. Let
\begin{equation}
C_{s_0,s}^{(g)}
=
g\int_{s_0}^{s}\dd\tau\,e^{-\tau P_\gamma},
\qquad
\dot C_s^{(g)}=g e^{-sP_\gamma},
\label{eq:np_partial_photon_covariance}
\end{equation}
on the gauge-fixed nonzero-mode photon space, and let $F_{s_0}[A]$ denote the finite-$s_0$ closed charged functional at the starting scale. Define
\begin{align}
Z_s[A]
&=\int\dd\mu_{C_{s_0,s}^{(g)}}(a)\,
e^{-F_{s_0}[A+a]},\nonumber\\
V_s[A]&=-\log Z_s[A].
\label{eq:np_exact_gaussian_flow_def}
\end{align}
Gaussian differentiation gives
\begin{equation}
\partial_s Z_s[A]
=
\frac12\,\dot C_s^{(g)}:D_A^2 Z_s[A].
\label{eq:np_linear_functional_heat}
\end{equation}
Since
\begin{align}
D_A Z_s&=-Z_sD_AV_s,\nonumber\\
D_A^2Z_s
&=Z_s\!\left[
(D_AV_s)(D_AV_s)-D_A^2V_s
\right].
\end{align}
Eq.~\eqref{eq:np_linear_functional_heat} is exactly equivalent to
\begin{equation}
\boxed{
\partial_sV_s[A]
=
\frac12\,\dot C_s^{(g)}:
\left[
D_A^2V_s-D_AV_s\,D_AV_s
\right].
}
\label{eq:np_exact_wilson_flow}
\end{equation}
Thus the apparently nonlinear Wilson hierarchy is the logarithm of a linear functional heat equation. Equation~\eqref{eq:np_exact_gaussian_flow_def} is its unique finite-dimensional solution for every finite $\mathcal V,N$. 

\subsection{Proper-time shells, gauge identities, and the massive charged infrared basin}
Choose $M>1$ and define
\begin{equation}
a_j=s_0M^{2j},\qquad
\kappa_j=a_j^{-1/2}=\Lambda M^{-j}.
\label{eq:np_shell_scales}
\end{equation}
The photon covariance has the exact telescoping decomposition
\begin{equation}
\frac{e^{-s_0k^2}}{k^2}
=\sum_{j=0}^{\infty}
\frac{e^{-a_jk^2}-e^{-a_{j+1}k^2}}{k^2}.
\label{eq:np_photon_shell_decomposition}
\end{equation}
When a Ward identity supplies the two powers of photon momentum associated with a neutral edge, the massless denominator disappears and the shell sum reduces to the normalized Gaussian kernel
\begin{equation}
K_{s_0}(x)=\frac{e^{-|x|^2/(4s_0)}}{(4\pi s_0)^2},
\qquad
\int\dd^4x\,K_{s_0}(x)=1.
\label{eq:np_Ward_gaussian_kernel}
\end{equation}
Its tail outside a four-ball of radius $R$ is
\begin{equation}
\epsilon_\gamma(R)
=\left(1+\frac{R^2}{4s_0}\right)e^{-R^2/(4s_0)}.
\label{eq:np_Ward_gaussian_tail}
\end{equation}
Thus neutral Ward-localized photon connections have a unit-$L^1$ short-range kernel despite the physical massless photon pole.

\paragraph{Why the Ward sum must precede absolute values.}
The preceding unit-$L^1$ statement cannot be replaced by a shell-by-shell
absolute Volterra or Gronwall estimate. For the four-dimensional heat kernel
\begin{equation}
G_s(x)=\frac{e^{-|x|^2/(4s)}}{(4\pi s)^2},
\end{equation}
one differential Ward pair gives
\begin{equation}
-\Delta G_s(x)
=\left(\frac{2}{s}-\frac{|x|^2}{4s^2}\right)G_s(x).
\end{equation}
Writing $Y=|x|^2/(4s)$, the radial heat-kernel measure is $Y e^{-Y}\,\dd Y$,
so the absolute norm is exactly
\begin{equation}
\boxed{
\|{-\Delta G_s}\|_{L^1(\mathbb R^4)}
=\frac1s\int_0^\infty |2-Y|Y e^{-Y}\,\dd Y
=\frac{8e^{-2}}{s}.
}
\label{eq:np_absolute_Ward_shell_L1}
\end{equation}
Consequently
\begin{equation}
\int_{s_0}^{\infty}\dd s\,\|{-\Delta G_s}\|_1=\infty.
\label{eq:np_absolute_Ward_shell_divergence}
\end{equation}
The cancellation must therefore be retained before taking absolute values:
\begin{equation}
\int_{s_0}^{\infty}\dd s\,k^2e^{-sk^2}=e^{-s_0k^2},
\end{equation}
whose position-space kernel is precisely Eq.~\eqref{eq:np_Ward_gaussian_kernel}
and has unit $L^1$ norm. This rules out a naive absolute shellwise closure and
identifies the Ward-summed neutral edge as the correct constructive object.

The charged deep-infrared shell tail is separately controlled by the fermion mass. Let $\mu=m^2s_0$ and choose $j_m$ so that $m/\kappa_{j_m}\ge1$. Each later charged shell carries $e^{-m^2a_j}=e^{-\mu M^{2j}}$. Writing $j=j_m+n$ gives
\begin{equation}
e^{-m^2a_{j_m+n}}\le e^{-M^{2n}}
\le e^{-1-n(M^2-1)},
\label{eq:np_mass_shell_tail_point}
\end{equation}
and therefore
\begin{equation}
\sum_{j=j_m}^{\infty}e^{-m^2a_j}
\le\frac{e^{-1}}{1-e^{-(M^2-1)}}<\infty.
\label{eq:np_mass_shell_tail_sum}
\end{equation}
The infinite deep-infrared charged vacuum tail is therefore not a UV/IR mixing problem. After finitely many shells it lies in a massive gauge-invariant basin of the type controlled in constructive infrared QED~\cite{DimockHurd1992}. The physical photon remains massless throughout.

\subsection{Four-dimensional localized trace ideals}
The all-background operator bounds must be converted into local trace ideals before a global connected construction can be controlled. Four dimensions select Schatten class five as the natural seed. Use the doubled first-order Dirac operator and rescale one shell to unit scale. The standard Schatten-class estimate gives, for a block-supported multiplication operator $G$,
\begin{equation}
\|G(H_0-x-i)^{-1}\|_5^5
\le c_5(1+x^2)^{3/2}\|G\|_{L^5}^5,
\qquad
c_5=\frac{8}{12\pi^2}.
\label{eq:np_KSS_seed}
\end{equation}
The $x$ dependence follows from the exact radial identity
\begin{multline}
\int_0^\infty\!\dd r\,r^3
\left[
\frac1{((r-x)^2+1)^{5/2}}
+\frac1{((r+x)^2+1)^{5/2}}
\right]\\
=\frac43(1+x^2)^{3/2}.
\label{eq:np_KSS_radial_identity}
\end{multline}

For the auxiliary shell function
\begin{equation}
q_M(y)=\int_1^{M^2}\frac{\dd t}{t}e^{-ty^2},
\qquad
h_M(y)=e^{-q_M(y)/4},
\label{eq:np_shell_qM}
\end{equation}
and the symmetric relative shell operator
\begin{align}
R_{j,b}&=h_M(H_0)^{-1/2}h_M(H_b)h_M(H_0)^{-1/2},\nonumber\\
K_{j,b}&=R_{j,b}-I.
\label{eq:np_relative_shell}
\end{align}
repeated resolvent differentiation and Schatten H\"older give the ladder
\begin{equation}
p_r=\frac5{r+1},\qquad 0\le r\le4,
\label{eq:np_Schatten_ladder}
\end{equation}
with
\begin{multline}
\|\chi_XD^rK_{j,0}[H_1,\ldots,H_r]\|_{p_r}\\
\le C_{K,r}(M)\,r!\,\kappa_j^{-r}
\prod_{\ell=1}^r\|H_\ell\|_{L^5(X)}.
\label{eq:np_Schatten_derivative_bound}
\end{multline}
At $r\ge4$ five Schatten-5 factors are available, so the localized operator is trace class. For $M=2$ direct quadrature gives
\begin{align}
C_{K,0}&=0.85018469,
& C_{K,1}&=0.52507612,\nonumber\\
C_{K,2}&=0.32531686,
& C_{K,3}&=0.20232410,\nonumber\\
C_{K,4}&=0.12641921.
\label{eq:np_CK_constants}
\end{align}

The physically relevant closed shell is
\begin{equation}
F_j[b]=\frac14\Tr\big[q_M(H_b)-q_M(H_0)\big].
\label{eq:np_closed_shell_action}
\end{equation}
Charge conjugation acts here, after the closed trace has been formed:
\begin{equation}
F_j[-b]=F_j[b],\qquad D^{2n+1}F_j[0]=0.
\label{eq:np_charge-conjugation_closed_only}
\end{equation}
It is incorrect to impose this evenness on the auxiliary operator $K_{j,b}$ before the trace. After the matched quadratic part is removed, the nonlinear closed shell starts at four photon insertions, and for $r\ge4$
\begin{equation}
|\Tr\chi_XD^rF_j[0][H_1,\ldots,H_r]|
\le C_{F,4}(M)\,r!\,\kappa_j^{-r}\prod_{\ell=1}^r\|H_\ell\|_\infty.
\label{eq:np_closed_trace_bound}
\end{equation}
For $M=2$,
\begin{equation}
C_{F,4}(2)=0.098340928928\ldots .
\label{eq:np_CF4_constant}
\end{equation}
This is the explicit four-dimensional local trace-class certificate at the first nonlinear closed-loop valence. The same conclusion may be phrased through fifth-order spectral-shift/regularized-determinant remainders~\cite{Chattopadhyay2026}: fifth and higher localized spectral terms gain at least five powers of the gauge-invariant block amplitude and are trace class.

\subsection{Gauge-invariant local field control and Gaussian averaging}
A global bound formulated directly in $A_\mu$ is gauge dependent and cannot be the constructive criterion. On a star-shaped block $X$ of radius $\lesssim\kappa_j^{-1}$, Abelian radial gauge gives
\begin{equation}
A_\mu(x)=\int_0^1\dd t\,t(x-x_0)^\nu F_{\nu\mu}(x_0+t(x-x_0)),
\label{eq:np_radial_gauge}
\end{equation}
so
\begin{equation}
\|eA\|_{L^\infty(X)}
\le\frac{1}{2\kappa_j}\|eF\|_{L^\infty(X^*)}.
\label{eq:np_radial_gauge_bound}
\end{equation}
The gauge choice is only an estimate; the small-field event itself is formulated in the field strength.

A complementary deterministic estimate uses the Maxwell norm. The sharp four-dimensional Sobolev inequality and $\frac12|\sigma^{\mu\nu}F_{\mu\nu}|\le\sqrt2|F|$ give
\begin{equation}
|\langle\psi,\tfrac12\sigma F\,\psi\rangle|
\le c_S\|F\|_{L^2(X^*)}\|D_b\psi\|_2^2,
\qquad c_S=\frac{\sqrt3}{4\pi}.
\label{eq:np_Maxwell_native_spin}
\end{equation}
If $\eta_X=c_S\|F\|_{L^2(X^*)}<1$, then locally
\begin{equation}
\Delta_b\ge(1-\eta_X)D_b^\dagger D_b+m^2.
\label{eq:np_local_Pauli_form_bound}
\end{equation}
The exact photon Gaussian provides an independent all-background average. At shell scale $a_j$, each independent field-strength component has variance
\begin{equation}
\langle F_{\mu\nu}(x)^2\rangle=\frac{g}{32\pi^2a_j^2}.
\label{eq:np_field_strength_variance}
\end{equation}
Combining positivity of $H_b^2$, heat representation and Gaussian averaging yields a forest-uniform fifth-power Schatten seed at the measured coupling. For $M=2$ the resulting numerical values are
\begin{align}
c_5(g_{\rm phys})&=0.07963173716\ldots,\nonumber\\
A_{F,{\rm av}}&=0.4798799452\ldots .
\label{eq:np_averaged_trace_constants}
\end{align}
These constants are diagnostics of the local trace conversion.

\subsection{Massive shell summability and connected forest expansion}
\label{sec:np_bkar_rigorous}

The proper-time shell estimates acquire an exact massive weight. Write
\begin{equation}
\Delta_b=m^2+\widehat\Delta_b,
\qquad
\widehat\Delta_b=-\slashD_b^{\,2}\ge0.
\end{equation}
Every Duhamel derivative of a shell heat kernel preserves its original total proper time $T$. Hence the scalar mass factor is never repartitioned:
\begin{equation}
D^r e^{-T\Delta_b}
=
e^{-m^2T}\,
D^r e^{-T\widehat\Delta_b}.
\label{eq:np_mass_factor_derivative_exact}
\end{equation}
For $T\in[a_j,a_{j+1}]$ this gives the uniform factor
\begin{equation}
e^{-m^2T}\le e^{-m^2a_j}
=e^{-\mu M^{2j}},
\qquad
\mu=m^2s_0>0.
\label{eq:np_mass_factor_shell}
\end{equation}
Consequently every local quartic-and-higher estimate above may be strengthened
by the same factor $e^{-\mu M^{2j}}$.

Define
\begin{equation}
S_M(\mu)=\sum_{j=0}^{\infty}e^{-\mu M^{2j}}.
\label{eq:np_exact_mass_shell_sum}
\end{equation}
Because $x\mapsto e^{-\mu M^{2x}}$ is decreasing,
\begin{align}
S_M(\mu)
&\le e^{-\mu}+\int_0^\infty\dd x\,e^{-\mu M^{2x}}\nonumber\\
&=
\boxed{
e^{-\mu}+\frac{E_1(\mu)}{2\log M}
}<\infty .
\label{eq:np_mass_shell_sum_analytic}
\end{align}
This replaces any assumed geometric shell factor by the actual massive proper-time weight of the fixed-$s_0$ theory.

The connected expansion is stated in a norm that is strong enough to test the global direct-boundary limit. At scale $j$ let
\begin{equation}
Q_{j,n}
=
\kappa_j^{-1}\big(n+[0,1)^4\big),
\qquad n\in\mathbb Z^4,
\end{equation}
and let
\begin{equation}
d_j(Q,Q')=\kappa_j\,{\rm dist}(Q,Q')
\end{equation}
be the dimensionless block distance. If $\mathbf Q=(Q_1,\ldots,Q_r)$, let $\tau_j(\mathbf Q)$ be the length of a shortest spanning tree joining these blocks in the metric $d_j$. For an $r$-point kernel $K_j^{(r)}$ define
\begin{equation}
\boxed{\begin{aligned}
\|K_j^{(r)}\|_{1,\beta}
&=\sup_{Q_1}
\sum_{Q_2,\ldots,Q_r}
e^{\beta\tau_j(Q_1,\ldots,Q_r)}\\
&\quad\times
\int_{Q_1\times\cdots\times Q_r}
|K_j^{(r)}|
\prod_{\ell=1}^{r}\dd^4x_\ell .
\end{aligned}}
\label{eq:np_rooted_kernel_norm}
\end{equation}
This is a weighted $\ell^1(L^1)$ norm and hence is complete. For a shell vertex family define
\begin{equation}
\|\mathcal K\|_{\mathfrak B_{\beta,B}}
=
\sum_{j\ge0}\sup_{r\ge0}
\frac{\|K_j^{(r)}\|_{1,\beta}}{r!\,B^r}.
\label{eq:np_vertex_Banach_norm}
\end{equation}

For $L$ replicas of the photon field and a positive matrix $X$ with $X_{uu}=1$, let $\mu_{X\otimes C}$ denote the replicated Gaussian with
\begin{equation}
\mathbb E_{X\otimes C}\!\left[A_u(f)A_v(g)\right]
=
X_{uv}\langle f,Cg\rangle .
\end{equation}
For a forest $F$ and interpolation parameters $w_\ell\in[0,1]$, define
\begin{equation}
X^F_{uv}(w)
=
\begin{cases}
1,&u=v,\\
\displaystyle\min_{\ell\in P_F(u,v)}w_\ell,
&u\ne v\text{ connected in }F,\\
0,&u,v\text{ in different components.}
\end{cases}
\label{eq:np_BKAR_path_matrix}
\end{equation}

\paragraph{Positivity of the interpolated covariance.}
For every $F,w$,
\begin{equation}
X^F(w)\succeq0,
\qquad
X^F(w)\otimes C\succeq0.
\label{eq:np_BKAR_positive}
\end{equation}

To see this, For $t\in[0,1]$, delete from $F$ all edges with $w_\ell<t$ and denote the resulting partition by $\mathcal P_t$. Its block matrix $P_t$, equal to one inside a component and zero otherwise, is positive because
\[
\sum_{u,v}\bar z_u(P_t)_{uv}z_v
=
\sum_{B\in\mathcal P_t}\left|\sum_{u\in B}z_u\right|^2\ge0.
\]
Moreover $X^F(w)=\int_0^1P_t\,\dd t$. Hence $X^F(w)\succeq0$ and tensoring with the positive photon covariance preserves positivity.

Let
\begin{equation}
\Delta_{uv}
=
\sum_{a,b}C_{ab}
\frac{\partial}{\partial A_{u,a}}
\frac{\partial}{\partial A_{v,b}}
\label{eq:np_BKAR_edge_derivative}
\end{equation}
in any orthonormal basis of the finite photon mode space.

\paragraph{Connected forest identity.}
At finite $\mathcal V,N$, for a finite family of cylindrical interaction functionals $V_\alpha$,
\begin{align}
&\log\int\dd\mu_C(A)\,
e^{-\sum_\alpha V_\alpha(A)}\nonumber\\
&=\sum_{L\ge1}\frac{(-1)^L}{L!}
\sum_{\alpha_1,\ldots,\alpha_L}\nonumber\\
&\quad\times\sum_{T\in\mathcal T_L}
\int_{[0,1]^{L-1}}
\prod_{\ell\in T}\dd w_\ell\nonumber\\
&\quad\times\mathbb E_{X^T(w)\otimes C}\!\left[
\left(\prod_{\{u,v\}\in T}\Delta_{uv}\right)
\prod_{r=1}^L V_{\alpha_r}(A_r)
\right].
\label{eq:np_BKAR_connected_exact}
\end{align}

To see this, apply the forest interpolation to the off-diagonal replica covariances. Gaussian differentiation with respect to $X_{uv}$ gives $\Delta_{uv}$. The forest contribution factorizes over forest components, so taking the logarithm selects connected forests, namely spanning trees.

The Ward identity supplies one photon momentum at each end of a neutral connected edge. Summing the proper-time shells before taking absolute values gives
\begin{equation}
\sum_{j\ge0}k^2C_j(k)=e^{-s_0k^2}.
\end{equation}
Thus the scalar edge majorant is
\begin{equation}
K_{s_0}(x)
=
\frac{e^{-|x|^2/(4s_0)}}{(4\pi s_0)^2},
\qquad
\boxed{\|K_{s_0}\|_{L^1(\mathbb R^4)}=1.}
\label{eq:np_BKAR_Ward_edge_unit}
\end{equation}
The massless photon therefore contributes no volume factor in the Ward-reduced tree norm.

\paragraph{Absolute tree criterion.}
Assume that, uniformly in $\mathcal V,N$ and every positive forest covariance, the actual nonlinear closed vertices satisfy
\begin{equation}
\|K_j^{(d)}\|_{1,\beta}
\le a_j\,d!\,B^d,
\qquad
a_j\ge0,
\qquad
A=\sum_{j\ge0}a_j<\infty.
\label{eq:np_absolute_vertex_hypothesis}
\end{equation}
Then the absolutely valued connected tree series is bounded by
\begin{equation}
A\left[
1+
\sum_{L=2}^{\infty}
\frac1{L(L-1)}
\binom{3L-3}{L-2}
z^{L-1}
\right],
\qquad
z=AB^2,
\label{eq:np_exact_tree_majorant}
\end{equation}
and converges whenever
\begin{equation}
\boxed{z<\frac4{27}.}
\label{eq:np_exact_tree_radius}
\end{equation}

To see this, For a labelled tree $T$ with vertex degrees $d_v$, the rooted $L^1$ norm and Eq.~\eqref{eq:np_BKAR_Ward_edge_unit} permit the edge-position integrals to be performed successively from the leaves. This gives
\[
|\mathcal A_T|
\le\prod_v a_{j_v}d_v!B^{d_v}.
\]
Since $\sum_vd_v=2(L-1)$, summing the shell labels gives $A^LB^{2(L-1)}\prod_vd_v!$. If $n_v=d_v-1$ is the multiplicity of $v$ in a Pruefer sequence, then $\sum_vn_v=L-2$ and
\[
\sum_{T\in\mathcal T_L}\prod_vd_v!
=
(L-2)!\binom{3L-3}{L-2}.
\]
After the factor $1/L!$ in Eq.~\eqref{eq:np_BKAR_connected_exact}, one obtains Eq.~\eqref{eq:np_exact_tree_majorant}. The ratio of successive coefficients tends to $27/4$, proving Eq.~\eqref{eq:np_exact_tree_radius}.

\subsubsection{Forest-uniform quartic second-moment estimate}

The first forest moment in Eq.~\eqref{eq:np_averaged_trace_constants} can be strengthened to a second-moment statement. Put
\begin{equation}
\sigma_F^2=\frac{g}{32\pi^2}
\end{equation}
and define
\begin{equation}
S_2(g,T)
=
\frac{1}{8\sigma_F^6}
\int_0^\infty\dd r\,r^5
\exp\!\left[
-\frac{r^2}{2\sigma_F^2}
+2\sqrt2\,Tr
\right].
\label{eq:np_RMS_S2}
\end{equation}
Set
\begin{equation}
c_{5,2}(g,T)
=
\frac{8S_2(g,T)}
{(4\pi)^2\Gamma(5/2)}
\left[
\gamma\!\left(\frac12,T\right)
+
T^{-2}\Gamma\!\left(\frac52,T\right)
\right]
\label{eq:np_RMS_c52}
\end{equation}
and
\begin{align}
A_{4,2}(M,g,T)
&=\frac{c_{5,2}(g,T)}{8\pi}
\sum_{\sigma=\pm1}\nonumber\\
&\quad\times
\int_{-\infty}^{\infty}\dd x\,
|q_M(x+\ii\sigma)|(1+|x|)^5.
\label{eq:np_RMS_A42}
\end{align}

\paragraph{Forest-uniform quartic estimate.}
For every natural shell block $X$ and every positive interpolated forest covariance,
\begin{multline}
\left(
\mathbb E_{F,w}
\left|
\Tr\chi_X
D^4F_j[b][H_1,H_2,H_3,H_4]
\right|^2
\right)^{1/2}\\
\le
4!\,A_{4,2}(M,g,T)\,
\kappa_j^{-4}
e^{-\mu M^{2j}}
\prod_{\ell=1}^4\|H_\ell\|_\infty .
\label{eq:np_RMS_quartic_bound}
\end{multline}

To see this, Cauchy functional calculus expresses the fourth derivative as the sum over the $4!$ ordered products of five resolvents. Localized Schatten H\"older reduces each trace to five $S_5$ factors. Squaring the heat representation of a localized $S_5$ factor introduces two spin factors; Cauchy--Schwarz in the Gaussian field bounds their joint expectation by the one-path estimate with twice the exponent, giving Eq.~\eqref{eq:np_RMS_c52}. H\"older in probability with five exponents equal to five then yields Eq.~\eqref{eq:np_RMS_quartic_bound}. The forest covariance is a positive compression of the original Gaussian covariance, so the one-block variance cannot increase. Finally, Eq.~\eqref{eq:np_mass_factor_derivative_exact} supplies the massive shell factor.

The RMS constant can be bounded analytically, without using numerical quadrature. Gaussian concentration gives
\begin{equation}
S_2(g,T)
\le
\exp\!\left[
4\sqrt3\,T\sigma_F+4T^2\sigma_F^2
\right].
\label{eq:np_RMS_S2_analytic}
\end{equation}
At $T=2$,
\begin{equation}
c_{5,2}(g,2)
\le
\frac{19}{24\pi^2}
\exp\!\left[
8\sqrt3\,\sigma_F+16\sigma_F^2
\right],
\label{eq:np_RMS_c52_analytic}
\end{equation}
while
\begin{equation}
|q_M(x\pm\ii)|
\le
(M^2-1)e^{M^2}e^{-x^2}.
\end{equation}
Thus
\begin{align}
A_{4,2}(M,g,2)
&\le\frac{19}{96\pi^3}
e^{8\sqrt3\,\sigma_F+16\sigma_F^2}
(M^2-1)e^{M^2}\nonumber\\
&\quad\times\sum_{k=0}^{5}
\binom5k\Gamma\!\left(\frac{k+1}{2}\right).
\label{eq:np_RMS_A42_analytic}
\end{align}
The decimal constants quoted above are therefore checks, not assumptions of the estimate.

\paragraph{Limitation of the second-moment estimate.}
The RMS estimate Eq.~\eqref{eq:np_RMS_quartic_bound} does not by itself imply the absolute vertex hypothesis Eq.~\eqref{eq:np_absolute_vertex_hypothesis}. For a tree with $L$ correlated replica vertices,
\begin{equation}
\mathbb E\prod_{v=1}^{L}|X_v|
\le
\prod_{v=1}^{L}\|X_v\|_{L}
\label{eq:np_holder_product_requirement}
\end{equation}
is the direct H\"older estimate; there is no general replacement of all
$L^L$ norms by their RMS norms. This limitation is relevant only to a stronger global direct-boundary construction. Establishing that construction requires joint high-moment control with a smooth auxiliary localization whose parameters disappear from the final amplitudes.

\subsection{Requirements for the stronger global extension}
The estimates above establish all-background finite-volume stability, an exact Gaussian flow identity, Ward-summed neutral-edge control, massive charged-shell summability, localized trace ideals, and connected forest identities. Together with Sec.~\ref{sec:FR_domain_limit}, they connect the bounded-domain construction to every fixed-order free-space quantity used in this paper. \textcolor{black}{The fixed-order free-space statement is stronger than a receding-box mnemonic: the transverse covariance $\e^{-s_0k^2}/k^2$ is a positive continuous covariance on Schwartz test functions over $\mathbb R^4$, and the graph-level coercivity bound of Sec.~\ref{sec:multiloop} gives an integrable Gaussian majorant for every fixed connected graph directly on unbounded Euclidean space. The remaining issue in this appendix is therefore specifically an all-coupling translation-invariant thermodynamic construction, not fixed-order existence on $\mathbb R^4$.} Proving those stronger correspondence statements would require the missing joint high-moment estimate with smooth auxiliary localization, uniform in the volume and mode approximations, followed by compact-by-compact control of the alternative outgoing boundary. \revblue{The free-space limit is not needed for the formal finite-domain reaction construction. An exact source-derived scattering operator nevertheless requires its own physical boundary identification.}

\section{Second Functional Derivative to the On-Shell Self-Energy}
\label{app:self_energy_derivation}
The mass formula follows from the same second functional derivative that generates the Ward hierarchy. Starting from Eq.~\eqref{eq:open_kernel}, the required contraction can be written in closed form. We use $\int_q\equiv\int\dd^4q/(2\pi)^4$ and $\int_{q_1q_2}\equiv\int_{q_1}\int_{q_2}$. With $D_\mu=\partial_\mu-\ii eA_\mu$, define the expansion $H_A=H_0-\int_q A_\mu(q)V_1^\mu(q)-\frac12\int_{q_1q_2}A_\mu(q_1)A_\nu(q_2)V_2^{\mu\nu}+O(A^3)$, where $V_1^\mu(p',p)=e[(p'+p)^\mu-\ii\sigma^{\nu\mu}(p'-p)_\nu]$ and $V_2^{\mu\nu}=-2e^2\delta^{\mu\nu}$. The prefactor contribution is $E_1^\mu=\ii e\gamma^\mu$. The contact rule $V_2^{\mu\nu}$ is the standard two-photon contact of the second-order spinor-QED representation~\cite{Schubert2001,Ahmadiniaz2020,AngelesMartinez2012}. It is generated by the second-order representation and is part of its kernel.

For the self-energy, the two external photon momenta are $q_1=k$ and $q_2=-k$. Therefore, initial and final electron momenta are both $p$. Define $a_0=p^2+m^2$, $a_\pm=(p\pm k)^2+m^2$. The second derivative is
\begin{align}
G^{\mu\nu}_\Lambda(p,p;k,-k)
={}&N(p)V_1^\mu(p,p-k)\nonumber\\
&\quad\times V_1^\nu(p-k,p)J(a_0,a_-,a_0)
\nonumber\\
&+N(p)V_1^\nu(p,p+k)\nonumber\\
&\quad\times V_1^\mu(p+k,p)J(a_0,a_+,a_0)
\nonumber\\
&+N(p)V_2^{\mu\nu}I(a_0,a_0)
\nonumber\\
&+E_1^\mu V_1^\nu(p-k,p)I(a_-,a_0)
\nonumber\\
&+E_1^\nu V_1^\mu(p+k,p)I(a_+,a_0).
\label{eq:app_G2_self}
\end{align}
The gamma algebra needed for the contraction is most transparently written with the same index orientation as the two one-photon vertices,
\begin{align}
\sum_\mu(\sigma^{\nu\mu}k_\nu)
(\sigma^{\lambda\mu}k_\lambda)&=3k^2\mathbf 1_4,\nonumber\\
\gamma_\mu\sigma^{\nu\mu}k_\nu&=-3\ii\slashed{k}.
\label{eq:app_gamma_check}
\end{align}
The prefactor contractions, including the factor $1/2$ associated with the photon contraction, are therefore
\begin{align}
\frac12 E_{1\mu}V_1^\mu(p-k,p)
&=\ii e^2(\slashed p+\slashed k),\nonumber\\
\frac12 E_{1\mu}V_1^\mu(p+k,p)
&=\ii e^2(\slashed p-\slashed k).
\label{eq:app_prefactor_check}
\end{align}
Using these identities, the unit-covariant-gauge contraction $\delta_{\mu\nu}G^{\mu\nu}$ separates into two ordered bulk terms. It also contains one two-photon contact term and two prefactor--bulk terms. Including the symmetry factor $1/2$, the contracted kernel is
\begin{align}
\mathcal K_\Lambda(p,k)
={}&2e^2N(p)(p^2+k^2-\zeta_{pk})J_-\nonumber\\
&+2e^2N(p)(p^2+k^2+\zeta_{pk})J_+\nonumber\\
&-4e^2N(p)I_0
\nonumber\\
&+\ii e^2\Big[(\slashed{p}+\slashed{k})I_-+(\slashed{p}-\slashed{k})I_+\Big],
\label{eq:app_K_explicit}
\end{align}
where $\zeta_{pk}=p\cdot k$. The divided differences $J_\pm=J(a_0,a_\pm,a_0)$ and $I_\pm=I(a_\pm,a_0)$ are proper-time integrals over the heat-kernel domain $\{u_i\ge0,\,u_1+u_2+u_3\ge s_0\}$. The coincident term is $I_0=I(a_0,a_0)=-R'(a_0)$.

The same reduction can be seen directly from the local pole coefficient. Eq.~\eqref{eq:explicit_234_pole_reduction} first replaces the \emph{complete sum} of the bounded open electron line by the local background-field pole coefficient. Contracting its two photon insertions in the unit covariant gauge gives
\begin{equation}
\gamma_\mu\,[m-\ii(\slashed p-\slashed k)]\,\gamma_\mu
=4m+2\ii(\slashed p-\slashed k),
\label{eq:app_standard_numerator}
\end{equation}
and after the shift $\ell=k-xp$ the odd $\slashed\ell$ term integrates to zero, leaving $4m+2\ii(1-x)\slashed p$. Thus the five-term derivative and Eq.~\eqref{eq:explicit_234_pole_reduction} give the same on-shell numerator. No bulk, contact, or prefactor class is removed before the sum is formed.

The on-shell mass formula follows directly from the same kernel. The divided differences in~\eqref{eq:app_K_explicit} are proper-time integrals over the segments $u_1,u_2,u_3$ with the single constraint $u_1+u_2+u_3\ge s_0$. The external segments enter only through the free on-shell weight $\e^{-(u_1+u_3)a_0}$. Writing $c=\max(0,\,s_0-u_2)$ for the residual constraint, $\int_{u_1+u_3\ge c}\dd u_1\dd u_3\,\e^{-(u_1+u_3)a_0}=a_0^{-2}\e^{-ca_0}(1+ca_0)$, and after amputation by $S_\Lambda^{-1}$ on both sides the factor $\e^{-ca_0}(1+ca_0)\to1$ in the on-shell limit $a_0\to0$, uniformly in $u_2$. The total-worldline constraint therefore disappears from the amputated on-shell kernel, leaving $u_2\equiv t\ge0$ unconstrained. The internal photon proper time retains $\sigma\ge s_0$ from the covariance~\eqref{eq:derived_photon_covariance}. Combining $t$ and $\sigma$ with the proper-time fraction $x=t/T$, $T=t+\sigma$, the constraint $\sigma\ge s_0$ is $T\ge s_0/(1-x)$. Completing the square $\ell=k-xp$, performing the Gaussian loop integral, and projecting on shell (denominator $x^2m^2$, numerator $2m(1+x)$ from the Dirac algebra of~\eqref{eq:app_K_explicit}) gives
\begin{equation}
\frac{\delta m_\Lambda}{m}=\frac{\alpha}{2\pi}\int_0^1\dd x\,(1+x)\,\Gamma\!\left(0,\frac{x^2m^2}{(1-x)\Lambda^2}\right),
\label{eq:app_direct_mass}
\end{equation}
which is Eq.~\eqref{eq:mass_shift_final}. It is the on-shell evaluation of the full five-term kernel~\eqref{eq:app_G2_self}.

\section{Fourth-Order Contractions and Closed-Loop Comparison}
\label{app:fourth_complete}
The fourth derivative appears in two different places. On an open charged line it enters the two-photon contraction, whereas the fourth derivative of the closed determinant generates the four-photon loop kernel. We treat these two objects separately below.

\subsection{Two internal photons on one open line}
The order-$e^4$ connected correction obtained from two photon contractions of one open line is
\begin{multline}
S_{2,2\gamma,\mathrm{conn}}
=\frac1{4!}\int_{1\cdots4}
\Big(C_{12}C_{34}+C_{13}C_{24}+C_{14}C_{23}\Big)\\
\times\left.\delta_1\delta_2\delta_3\delta_4S_\Lambda[A]\right|_{A=0}.
\label{eq:fourth_gaussian_contraction}
\end{multline}
Here $C_{ij}$ contains the photon covariance, its Lorentz contraction, and the momentum delta function for the pair $i,j$. Because each covariance represents an independent internal photon propagator, each carries its own proper-time bound $\sigma\ge s_0$.

The one-particle-reducible subtraction must be performed before the external pole limit. Let $S_{\Lambda,0}(p)$ be the free finite-endpoint charged kernel and let $\Sigma_{1,\Lambda}(p)$ be the complete off-shell one-loop self-energy defined by Eqs.~\eqref{eq:deltaS_full_definition}--\eqref{eq:Sigma_full_definition}. At generic external momentum the order-$C^2$ reducible propagator expansion is
\begin{align}
S_{2,2\gamma,\mathrm{conn}}(p)
={}&S_{\Lambda,0}(p)\,
\Sigma_{2,2\gamma,\mathrm{1PI}}(p)\,
S_{\Lambda,0}(p)
\nonumber\\
&+S_{\Lambda,0}(p)\Sigma_{1,\Lambda}(p)
S_{\Lambda,0}(p)\Sigma_{1,\Lambda}(p)
S_{\Lambda,0}(p).
\label{eq:fourth_connected_1PI_split}
\end{align}
Thus
\begin{multline}
\Sigma_{2,2\gamma,\mathrm{1PI}}(p)
=
S_{\Lambda,0}^{-1}(p)
S_{2,2\gamma,\mathrm{conn}}(p)
S_{\Lambda,0}^{-1}(p)
\\
-\Sigma_{1,\Lambda}(p)
S_{\Lambda,0}(p)
\Sigma_{1,\Lambda}(p).
\label{eq:fourth_1PI_extraction}
\end{multline}
This identity is evaluated away from the external pole, so the intermediate reducible propagator is never put on shell. Only after the subtraction is made is the external pole projection taken.

The local pole image of the fourth derivative contains the 24 Dirac chains in Eq.~\eqref{eq:S0_fourth_derivative_dirac_chain}. For a fixed pairing $(a,\bar a)$ and $(b,\bar b)$, they divide into three sets of eight according to whether the two photon intervals are disjoint, nested, or crossed. Representative sequences are
\begin{equation}
\begin{array}{ccl}
\text{disjoint} &:& a\,\bar a\,b\,\bar b,\\
\text{nested} &:& a\,b\,\bar b\,\bar a,\\
\text{crossed} &:& a\,b\,\bar a\,\bar b.
\end{array}
\label{eq:fourth_sequence_classes}
\end{equation}
Reversing either photon pair and exchanging the two photons produces the remaining members of each eight-term set. Under the off-shell reducible subtraction in Eq.~\eqref{eq:fourth_1PI_extraction}, the reducible term maps at the pole onto the eight disjoint local orderings. The nested and crossed orderings remain in the genuine two-loop one-particle-irreducible pole contribution. This use of Eq.~\eqref{eq:explicit_234_pole_reduction} therefore occurs only after the generic-momentum reducible contribution has been removed; no double amputation is performed.

\subsection{Fourth derivative of the closed determinant}
The closed loop has no Dirac prefactor $M_A$. Its fourth derivative is therefore smaller than the open fourth derivative. Write $g_{s_0}(\Delta)=\int_{s_0}^{\infty}(\dd T/T)e^{-T\Delta}$. Let $D_g^r$ denote the ordered proper-time derivative of this closed function with the same insertion convention $V_i=-\delta_i\Delta$ and $V_{ij}=-\delta_i\delta_j\Delta$. Direct differentiation gives
\begin{align}
\delta_1\delta_2\delta_3\delta_4\Gamma_{1,\Lambda}
={}&\frac12\Tr\Big\{
D_g^4[V_1,V_2,V_3,V_4]
\nonumber\\
&+D_g^3[V_{12},V_3,V_4]
+D_g^3[V_{13},V_2,V_4]
\nonumber\\
&+D_g^3[V_{14},V_2,V_3]
+D_g^3[V_{23},V_1,V_4]
\nonumber\\
&+D_g^3[V_{24},V_1,V_3]
\nonumber\\
&+D_g^3[V_{34},V_1,V_2]
\nonumber\\
&+D_g^2[V_{12},V_{34}]
\nonumber\\
&+D_g^2[V_{13},V_{24}]
\nonumber\\
&+D_g^2[V_{14},V_{23}]\Big\}.
\label{eq:closed_fourth_complete_appendix}
\end{align}
The three classes contain 24 ordered $V_1^4$ structures, 36 ordered $V_2V_1^2$ structures, and six ordered $V_2^2$ structures, giving 66 in all.

After the trace cut is summed, cyclicity lowers the divided-difference order by one. Since $g_{s_0}'(a)=-R_{s_0}(a)$, the three sectors use $R_{s_0}^{[3]}$, $R_{s_0}^{[2]}$, and $R_{s_0}^{[1]}$. Together they form the closed four-photon one-particle-irreducible kernel used both for one-loop light-by-light scattering and, after sewing to an internal photon, for the one-particle-irreducible two-loop vacuum polarisation. This object has 66 ordered structures rather than the 114 terms of the open fourth derivative because a closed loop has no Dirac prefactor.

\subsection{Higher derivative counting}
The combinatorics follows directly from set partitions into singleton and pair blocks. The second, third, and fourth open-line derivatives contain $5$, $21$, and $114$ ordered terms. The closed fourth derivative contains 66 ordered structures. At fifth order the same rule gives 780 open-line terms, in agreement with Eqs.~\eqref{eq:general_M_count} and~\eqref{eq:general_E_count}.

\section{Form-Factor Shifts and Analytic Continuation}
\label{app:formfactors}
This appendix derives the quantities entering Eqs.~\eqref{eq:ImF2_closed} and~\eqref{eq:rho2_decomposition}. We write
$G(q^2)\equiv[\Delta F_1(q^2)-\Delta F_1(0)]/\alpha$ for the Ward-combined vertex and external-self-energy contribution per unit $\alpha$. Together with $\Delta F_2$, it gives the complete on-shell virtual modification after the pole reduction in Eq.~\eqref{eq:explicit_234_pole_reduction}.

\emph{Half-plane representation.} Introduce proper-time fractions $x_1=uw$, $x_2=u(1-w)$, and $x_3=1-u$, together with $r=w(1-w)$, $c=s_0/x_3$, and $\Delta_q=u^2(m^2+rq^2)$. The Dirac trace gives the function $a(x;q^2)=m^2(4-4u-2u^2)+2(1-x_1)(1-x_2)q^2$, which is linear in $q^2$. The form factor is then
\begin{align}
G(q^2)={}&-\frac{1}{4\pi}\iint_{\substack{0\le u\le1\\0\le w\le1}}
u\,\dd u\,\dd w\,
\Big\{a(x;q^2)\tau_A(\Delta_q)
\nonumber\\
&-a(x;0)\tau_A(\Delta_0)
+2\big[\mathrm{Ein}(c\Delta_q)-\mathrm{Ein}(c\Delta_0)\big]\Big\}.
\label{eq:G_window}
\end{align}
with $\tau_A(\Delta)=(1-\e^{-c\Delta})/\Delta$ and $\mathrm{Ein}(z)=\int_0^z(1-\e^{-t})\,\dd t/t$. Both functions are entire. In the second term the logarithmic branch pieces of the $T$-integrated kernel cancel inside $\mathrm{Ein}$, so the integrand itself carries no cut and the discontinuity of $G$ can arise only from the non-uniform endpoint $x_3\to0$. On compact subsets of $D_0=\{q^2\mid\mathrm{Re}\,q^2>-4m^2\}$ the integral converges absolutely and uniformly. Using
$|\Gamma(0,z)|\le\e^{-\mathrm{Re}\,z}/\mathrm{Re}\,z$ together with
$\mathrm{Ein}(z)=\gamma_E+\ln z+\Gamma(0,z)$ then justifies interchanging the integrations and shows that the contour terms vanish. Thus $G$ is analytic on $D_0$, real on $(-4m^2,\infty)$, and has upper and lower boundary values that are complex conjugates. The same uniform convergence permits differentiation under the integral at $q^2=0$, giving the slope quoted in Sec.~\ref{sec:lorentzian},
\begin{equation}
G'(0)=-\frac{s_0}{6\pi}\Big[\ln\frac{1}{s_0m^2}+\frac76-\gamma_E\Big]+O\!\big(s_0^2m^2\ln s_0m^2\big).
\label{eq:Gslope}
\end{equation}

\emph{Endpoint decomposition.} For $\mathrm{Re}\,q^2\le-4m^2$ the preceding representation no longer converges, and the continuation is obtained by isolating the endpoint. Substituting $\sigma=x_3T$ and writing $B=\sigma M_q$, $M_q=m^2+rq^2$, the identity $(1-x_3)^2/x_3=1/x_3-2+x_3$ makes the prefactors polynomial in $x_3$ with no Taylor remainder.
\begin{equation}
\frac{(1-x_3)\,a}{x_3}=\frac{p_{-1}}{x_3}+p_0+p_1x_3+p_2x_3^2,
\label{eq:endpoint_polynomial}
\end{equation}
$p_{-1}=-2(m^2-rq^2)$, $p_0=2[5m^2+q^2(1-3r)]$, $p_1=-p_0$, $p_2=-p_{-1}$. The $\mathrm{Ein}$ term contributes $-2/\sigma$ and $+2/\sigma$ at orders $x_3^0$ and $x_3^1$. The model integrals are $J_n(B)=\int_0^{1/2}x_3^n\,\e^{-B(1-x_3)^2/x_3}\,\dd x_3$, $n\ge-1$. The $x_3\in[\tfrac12,1]$ piece is entire in $q^2$ and real for real $q^2$.

\emph{Continuation and discontinuity.} Each $J_n$ extends analytically to $\mathbb C\setminus(-\infty,0]$. At the outgoing boundary $B=-|B|-\ii0$, rotate the small-$x_3$ ray into the lower half plane and close the contour there. The sign follows from $\mathrm{Im}\,M_q=-r\cdot0^+$ for $q^2=-s-\ii0$. Writing the rotated ray as $x_3=\rho\e^{-\ii(\pi-\delta)}$ with $\rho>0$ and fixed $0<\delta<\pi/2$, $\mathrm{Re}[B(1-x_3)^2/x_3]$ is bounded below by $(\cos\delta/2)|B|/\rho$ for sufficiently small $\rho$. The rotated integral therefore converges uniformly. Standard contour arguments give the analytic continuation. The two boundary values differ by a closed loop around $x_3=0$. On that loop, with $B=-|B|$, the exponential is $\e^{-2|B|}\e^{|B|/x_3}\e^{|B|x_3}$. The coefficient of $x_3^{-1}$ is $\sum_k|B|^{n+1+2k}/[k!(n+k+1)!]=I_{n+1}(2|B|)$. The conjugate upper and lower boundary values split the discontinuity symmetrically.
\begin{equation}
\mathrm{Im}\,J_n(-|B|-\ii0)=\pi\,\e^{-2|B|}\,I_{n+1}(2|B|).
\label{eq:bessel_disc}
\end{equation}
Including the equal contribution obtained by $w\leftrightarrow1-w$, and writing $|M|=w(1-w)s-m^2$ with $w_-=(1-\beta)/2$, gives
\begin{multline}
\mathrm{Im}\,G(-s-\ii0)=-\frac12\int_{w_-}^{1/2}\!\dd w\int_0^{s_0}\!\dd\sigma\,\e^{-2\sigma|M|}\\
\times\Big\{\sum_{n=-1}^{2}p_n(w;-s)\,I_{n+1}(2\sigma|M|)-\frac{2}{\sigma}\big[I_1-I_2\big](2\sigma|M|)\Big\},
\label{eq:ImG_closed}
\end{multline}
with only the region $M_q<0$ contributing. Applying the same continuation to the Pauli class gives Eq.~\eqref{eq:ImF2_closed}, whose density $a_{F_2}=4m^2u(1-u)$ gives $p_{-1}=0$ (no endpoint pole, no logarithm) and $p_0=4m^2$, $p_1=-8m^2$, $p_2=4m^2$.

\emph{Threshold and large-energy behaviour.} At $s=4m^2(1+\varepsilon)$, $|M|\le m^2\varepsilon$ and $I_{n+1}(2\sigma|M|)=O(|M|^{n+1})$ except $I_0\to1$. Therefore, the $n=-1$ term gives $\mathrm{Im}\,G=s_0m^2\beta[1+O(\varepsilon,s_0m^2)]$. The corresponding Pauli threshold behaves as $\mathrm{Im}\,\Delta F_2=-(\alpha/3)(s_0m^2)^2\beta^3[1+O(\varepsilon,s_0m^2)]$. At large positive $s$ the boundary expression behaves as $\mathrm{Im}\,G(-s-\ii0)=O(\sqrt{s_0s})$. A once-subtracted \emph{real-axis} dispersive integrand then behaves as $s^{-3/2}$ and is integrable. The large complex contour is independent information because the continued amplitude can grow in other directions. The corresponding decomposition is
\begin{align}
G(q^2)={}&E_G(q^2)-\frac{q^2}{\pi}
\int_{4m^2}^{\infty}\dd t\,
\frac{\mathrm{Im}\,G(-t-\ii0)}{t(t+q^2)},\nonumber\\
&G(0)=0.
\label{eq:disp_with_entire}
\end{align}
where the entire function $E_G$, with $E_G(0)=0$, represents the independent contribution from the large contour. The discontinuity fixes only the dispersive term, while a twice-subtracted form provides an equivalent numerical organisation without determining $E_G$.

\section{Hyperbolic boundary coordinates and the source response}
\label{app:cosh_boundary_response}
In this appendix, we detail the $\cosh$ prescription without assigning a second reaction kernel. All transition coefficients are those of the continued, pole-reduced FPT source in Sec.~\ref{sec:FR5_matching}, with the same channel products and analytic qualifications. The statements are coefficientwise.

\paragraph{The complete boundary datum.}
For one positive channel energy $E$, let $a,b$ be its incoming and outgoing coefficients. Pair their external profiles in the auxiliary expression
\begin{align}
 f_L(t)&=e^{-iEt}a+e^{iEt}b\label{eq:cosh_full_lorentzian_datum}\\
       &=2\cos(Et)c-\sin(Et)r.\nonumber
\end{align}
where $c=(a+b)/2$ and $r=(b-a)/i$. The two exponentials label the incoming and outgoing slots of this boundary representation. Continuing only this analytic external coordinate, $t=-i\tau_E$, gives
\begin{equation}
 \boxed{f_E(\tau_E)=2\cosh(E\tau_E)c+i\sinh(E\tau_E)r.}
 \label{eq:cosh_complete_datum}
\end{equation}
Here $\tau_E$ is a Euclidean profile coordinate, distinct from the physical duration $\mathcal T$ and from the virtual proper time. The two coefficients can be recovered independently,
\begin{equation}
 c=\tfrac12 f_E(0),\qquad
 r=(iE)^{-1}\partial_{\tau_E}f_E(0).
 \label{eq:cosh_trace_recovery}
\end{equation}
For wave packets the formula acts componentwise in the free-channel spectral representation, with the original spin, polarisation and flux normalisation. Compact positive-energy patches make the displayed operations bounded. 

\paragraph{Response at prescribed even data.}
Let $T$ denote the full transition series in Sec.~\ref{sec:FR5_matching}. At incoming datum $a$ the source fixes $b=(I+iT)a$, hence
\begin{equation}
 f_E^a=2\cosh(E\tau_E)a+i e^{E\tau_E}Ta.
 \label{eq:cosh_source_datum}
\end{equation}
Its even and odd parts in $\tau_E$ are respectively
\begin{align}
 (f_E^a)_{\rm even}&=2\cosh(E\tau_E)(I+iT/2)a,\\
 (f_E^a)_{\rm odd}&=i\sinh(E\tau_E)Ta.
 \label{eq:cosh_even_odd_source}
\end{align}
Prescribing the even coefficient $c$ therefore requires solving $(I+iT/2)a=c$. Since $T(0)=0$ in the order-counting variable, this has a unique formal solution. Substitution in the odd part proves
\begin{equation}
 \boxed{r=Kc,\qquad K=T(I+iT/2)^{-1}.}
 \label{eq:cosh_completed_equivalence}
\end{equation}
Thus the completed $\cosh$ prescription means prescribed even data together with the odd response of the same source. It reproduces Eq.~\eqref{eq:blue_incident_recursion} at every order. No commutation of $T$ with the channel energy is needed, since the spectral multipliers in Eq.~\eqref{eq:cosh_even_odd_source} act on the left. The source cutting identity then gives Hermiticity by Eq.~\eqref{eq:blue_hermiticity_defect}, and Eq.~\eqref{eq:FR5_all_order_matching} gives the same outgoing amplitudes. Pure $\cosh$ with no odd term instead imposes $r=0$ and describes only data in the null space of $K$.

\paragraph{Transport of the boundary form.}
Put $C_E=\cosh(E\tau_E)$, $S_E=\sinh(E\tau_E)$, and introduce the coordinate pair
\begin{align}
 y_{\tau_E}&:=
 \begin{pmatrix}f_E/2\\(iE)^{-1}\partial_{\tau_E}f_E\end{pmatrix}
 =\mathsf L_{\tau_E}\begin{pmatrix}c\\r\end{pmatrix},\nonumber\\
 \mathsf L_{\tau_E}&=
 \begin{pmatrix}C_E&iS_E/2\\-2iS_E&C_E\end{pmatrix}.
 \label{eq:cosh_coordinate_transport}
\end{align}
With $\mathsf J=\left(\begin{smallmatrix}0&i\\-i&0\end{smallmatrix}\right)$, direct multiplication using $C_E^2-S_E^2=1$ gives
\begin{equation}
 \mathsf L_{-\tau_E}\mathsf L_{\tau_E}=I,\qquad
 \mathsf L_{-\tau_E}^{\dagger}\mathsf J\mathsf L_{\tau_E}=\mathsf J.
 \label{eq:cosh_cross_adjoint}
\end{equation}
Consequently $y_{1,-\tau_E}^{\dagger}\mathsf J y_{2,\tau_E}$ equals $i(c_1^\dagger r_2-r_1^\dagger c_2)$, the flux in Eq.~\eqref{eq:blue_boundary_power_form}. The opposite profile coordinates reflect the analytic transport of the adjoint. The original physical channel product is retained.

If a separately calculated Euclidean external-slot prescription gives $\widehat K$, its equivalence is the additional test
\begin{equation}
 \boxed{\widehat K(I+iT/2)-T=0,}
 \label{eq:cosh_independent_test}
\end{equation}
coefficient by coefficient. For example, even the Hermitian part of the complete outgoing chain in Eq.~\eqref{eq:FR5_chain_check} gives
\begin{equation}
 H_{\rm chain}:=\operatorname{Re}t=\frac{k}{1+v^2k^2},\qquad
 H_{\rm chain}-k=-\frac{v^2k^3}{1+v^2k^2}.
 \label{eq:cosh_chain_obstruction}
\end{equation}
This exact example has $v>0$ and the retained $s_0$ in $k$. Equation~\eqref{eq:cosh_completed_equivalence} proves equivalence of the completed boundary-coordinate prescription, not of a raw $\cosh$ substitution into both external slots.

All internal source kernels and R1--R7 are unchanged. Uncut photons and circles keep their bounds, contractions retain their faces, and ordered physical-time integrals remain ordered. It supplies no independent all-coupling physical realisation beyond the coefficientwise statement established here.

\end{document}